\documentclass{aastex62}

\usepackage{amsmath}
\usepackage{gensymb}
\graphicspath{{./}{figures/}}
\usepackage{graphicx}
\usepackage{multirow}

\begin{document}

\title{Kilonova Emission From Black Hole-Neutron Star Mergers. I. Viewing-Angle-Dependent Lightcurves}

\author[0000-0002-9195-4904]{Jin-Ping Zhu}
\altaffiliation{zhujp@pku.edu.cn}
\affil{Department of Astronomy, School of Physics, Peking University, Beijing 100871, China}

\author[0000-0001-6374-8313]{Yuan-Pei Yang}
\affiliation{South-Western Institute for Astronomy Research, Yunnan University, Kunming, Yunnan, China}

\author{Liang-Duan Liu}
\affiliation{Department of Astronomy, Beijing Normal University, Beijing 100875, China}

\author{Yan Huang}
\affiliation{Department of Astronomy, School of Physics, Peking University, Beijing 100871, China}

\author[0000-0002-9725-2524]{Bing Zhang}
\altaffiliation{zhang@physics.unlv.edu}
\affiliation{Department of Physics and Astronomy, University of Nevada, Las Vegas, NV 89154, USA}

\author{Zhuo Li}
\altaffiliation{zhuo.li@pku.edu.cn}
\affiliation{Department of Astronomy, School of Physics, Peking University, Beijing 100871, China}
\affiliation{Kavli Institute for Astronomy and Astrophysics, Peking University, Beijing 100871, China}

\author[0000-0002-1067-1911]{Yun-Wei Yu}
\affiliation{Institute of Astrophysics, Central China Normal University, Wuhan 430079, China}

\author[0000-0002-3100-6558]{He Gao}
\affiliation{Department of Astronomy, Beijing Normal University, Beijing 100875, China}

\begin{abstract}

As the new era of gravitational-wave-led multi-messenger astronomy is ushered in, kilonovae from black hole-neutron star (BH-NS) mergers powered by $r$-process radioactivity are expected to be detected in the future. Recent numerical simulations revealed that  tidal dynamical ejecta and disk wind outflows of the BH-NS mergers are highly anisotropic. Exploring the viewing angle effect on kilonova lightcurves is of great interest for further understanding the BH-NS merger kilonova physics. In this paper, we present a numerical method to study the predicted lightcurves as a function of viewing angle. We extrapolate the fitting formulae for the mass and velocity of tidal dynamical ejecta across a wide range of mass ratio validated with 66 simulations and use them in the calculations of the kilonova lightcurves. The calculated peak luminosity of a black hole-neutron star (BH-NS) merger kilonova is typically about a few times $10^{41}\,{\rm erg\,s^{-1}}$, which is always $\lesssim4.5\times10^{41}\,{\rm erg\,s^{-1}}$. This corresponds to the AB absolute magnitudes fainter than $\sim -15\,{\rm mag}$ in optical and $\sim -16\,{\rm mag}$ in infrared. Since the projected photosphere area of the dynamical ejecta is much larger than that of the disk wind outflows, the dynamical ejecta usually contribute to the majority of the kilonova emission from BH-NS mergers. The fitted blackbody temperature and the shape of the observed multi-band lightcurves are insensitive to the line of sight. The peak time of the observed multi-band lightcurves, affected by the light propagation effect, is related to the relative motion direction between the dynamical ejecta and the observer. The observed luminosity varies with the projected photosphere area determined by the viewing angles. However, the predicted peak luminosity only varies by a factor of $\sim (2 - 3)$ (or by $\sim1\,{\rm mag}$) for different viewing angles. When the short-duration gamma-ray burst afterglow is taken into account, for an on-axis geometry, the kilonova emission is usually outshone by the afterglow emission and can be only observed in the redder bands, especially in the $K$-band at late times. Compared with GW170817/AT2017gfo, the BH-NS merger kilonovae are optically dim but possibly infrared bright. At the same epoch after the merger, the blackbody fitting temperature of the BH-NS merger kilonovae is lower than that of GW170817/AT2017gfo.

\end{abstract}


\section{Introduction}
It has long been proposed that mergers of binary neutron stars (BNS) and black hole-neutron star (BH-NS) are progenitors of short-duration gamma-ray bursts (sGRBs)\citep{Paczynski1986,Paczynski1991,Eichler1989,Narayan1992}. Rapid accretion of a centrifugally supported disk/torus by the merger remnant (likely a BH or a rapidly rotating, highly magnetized NS) would drive a pair of collimated relativistic jets, which can be observed as a sGRB if the jet points towards Earth \citep{Rezzolla2011,Paschalidis2015,Ruiz2016}\footnote{A more exotic way of making a sGRB or short electromagnetic transient from BH-BH merger systems (e.g. GW150914-GBM, \citealt{Connaughton2016}) would be to invoke a large enough charge in at least one of the BHs \citep{Zhang2016}.}. The interaction of the relativistic jets with the surrounding interstellar medium would generate afterglows with emission ranging from radio to X-rays \citep{Rees1992,Meszaros1993,Paczynski1993,Meszaros1997,Sari1998}, which have been confirmed observationally \citep{Berger2005,Fox2005,Hjorth2005}.

Besides collimated relativistic jets, BNS and BH-NS mergers are expected to release an amount of neutron-rich matter \citep{Lattimer1974,Lattimer1976,Symbalisty1982}, which can synthesize the elements heavier than iron via the rapid neutron-capture process ($r$-process). \cite{Li1998} first predicted a type of transient event powered by the radioactive decays of $r$-process nuclei following a BNS or BH-NS merger. The more detailed research by \cite{Metzger2010} showed that the peak luminosity of such an optical-infrared transient is around several times $10^{41}\,{\rm erg\,s^{-1}}$ and hence, named it a ``kilonova'' (i.e. the luminosity is a factor $\sim 10^3$ higher than a typical nova). Later, the characteristics of kilonova emission have been widely investigated theoretically \citep{Kulkarni2005,Roberts2011,Kasen2013,Tanaka2013,Barnes2013,Yu2013,Metzger2014,Grossman2014,MetzgerF2014,Perego2014,Wanajo2014,Just2015,Martin2015,Kasen2015,Kasen2017,Metzger2017}. In the past a few years, several kilonova candidates following sGRBs have been claimed from the optical-infrared emission in excess to the afterglow emission \citep{Tanvir2013,Berger2013,Yang2015, Jin2015,Jin2016,Jin2020,Gao2015,Gao2017,Gompertz2018,Ascenzi2019,Rossi2020}. However, due to the scarce and ambiguous observational data and the lack of smoking-gun evidence that sGRBs are indeed from BNS or BH-NS mergers, these cases cannot be firmly confirmed. In general, for sGRB-related kilonovae (presumably on-axis mergers), the putative kilonova emission could be outshone by the luminous afterglow. Also, the usual emission timescale of kilonovae is only about a few days to a few weeks. Therefore, for blind optical transient surveys, it is easy to miss the optimal observational time window of searching kilonovae. On the other hand, BNS and BH-NS mergers are gravitational wave (GW) events which could be detected by the Advanced LIGO out to a distance $\sim300$ and $\sim 650\,{\rm Mpc}$, respectively \citep{Cutler2002}. An optimal searching strategy for kilonovae is to take advantage of GW triggers to make electromagnetic (EM) follow-up observations \citep[e.g.,][]{Metzger2012,Cowperthwaite2015,Gehrels2016}.

On 17 August, 2017, the first BNS merger gravitational wave source GW170817 was detected by LIGO/Virgo Collaborations \citep{Abbott2017a}. At $\Delta t \sim 1.7\,{\rm s}$ after the merger, an sGRB lasting $\sim2\,{\rm s}$, GRB170817A, triggered the {\em Fermi} Gamma-ray Burst Monitor (GBM) \citep{Abbott2017b,Goldstein2017,Zhang2018}. About 11 hours later, an associated ultraviolet-optical-infrared kilonova transient, named AT2017gfo, was discovered in a galaxy NGC4993 $\sim40\,{\rm Mpc}$ away \citep{Abbott2017c,Arcavi2017,Andreoni2017,Coulter2017,Cowperthwaite2017,Covino2017,Chornock2017,Diaz2017,Drout2017,Hu2017,Kasliwal2017,Kilpatrick2017,Lipunov2017,McCully2017,Nicholl2017,Pian2017,Pozanenko2018,Shappee2017,Smartt2017,Soares-Santos2017,Tanvir2017,Troja2017,Utsumi2017,Valenti2017}. At the location coincident with this kilonova, a broad-band afterglow source from radio to X-rays was detected, which is consistent with the standard synchrotron afterglow model with an off-axis viewing angle \citep{Alexander2017,D'Avanzo2018,Dobie2018,Ghirlanda2019,Haggard2017,Hallinan2017,Lazzati2018,Lyman2018,Margutti2017,Troja2017}. The discovery by LIGO/Virgo of this GW event from a BNS merger and the subsequent observations of EM counterparts set a milestone for the era of GW-led multi-messenger astronomy. The associations between GW170817, GRB 170817A and AT2017gfo provided the smoking-gun evidence for the BNS merger origin of sGRBs and confirmed the kilonova theoretical prediction. 

The comprehensive observations of AT2017gfo showed that its early- and late-stage lightcurve cannot be explained by one single radioactivity-powered component \citep[however, see][]{Waxman2018}. In the literature, the data are widely interpreted by invoking two or even three different radioactivity-powered components \citep{Cowperthwaite2017,Kawaguchi2018,Perego2017,Tanaka2017,Villar2017,Wanajo2018,Wu2019}. More specifically, the early-stage emission of AT2017gfo was explained by a lanthanide-free ``blue" component (with low opacity) while he late-stage emission is interpreted by a lanthanide-rich ``red" component (with high opacity). Introducing both components could account for the AT2017gfo emission evolving from a blue-component-dominated stage to a red-component-dominated stage. In addition, some authors (e.g. \citealt{Perego2017} and \citealt{Villar2017}) found that an intermediate opacity ``purple" component may be needed to fully account for the broad-band kilonova emission data of AT2017gfo. Theoretically, the lanthanide-free ``blue" component is thought to be produced due to heating by the shocks at the contact interface between two merging NSs  \citep[e.g.,][]{Oechslin2006,Radice2016,Sekiguchi2016,Wanajo2014} or by neutrino irradiation from the remnant hypermassive NS \citep[e.g.,][]{MetzgerF2014,Perego2014,Yu2018}, while the tidal dynamical ejecta is usually interpreted as the ``red" component. For the ``purple" component, viscous heating and angular momentum transport of the remnant disk could form such intermediate opacity ejecta \citep[e.g.,][]{Fernandez2013,Fujibayashi2020,Just2015,Siegel2017}. One issue of this main-stream interpretation is that the blue component is too bright and too early unless an unreasonably small opacity is introduced \citep{Li2018}. \cite{Li2018} (see also \citealt{Yu2018,Ren2019}) argued that this may point toward a long-lived central engine that continuously injects energy into the ejecta \citep{Yu2013}.

{Besides BNS mergers, BH-NS mergers could be another type of sources to produce kilonova transients that astronomers expect to detect.} During the third observing run (O3) of the LIGO-Virgo Collaboration, several GW events from BH-NS or MassGap merger candidates with low false alarm rates have been discovered, e.g., S190814bv, S190924h, S190930s, and S200115j \citep{LIGO2019a,LIGO2019b,LIGO2019c,LIGO2020}. Among them, the BH-NS merger candidate S190814bv has been widely discussed \citep[e.g.,][]{Ackley2020,Andreoni2019,Dobie2019,Gomez2019,Kawaguchi2020a}. Besides, there was a BNS merger event (GW190425) with a total mass $\sim3.4\ M_\odot$. The possibility that the system is BH-NS merger cannot be ruled out from gravitational-wave data \citep{TheLIGO2020,Han2020,Kyutoku2020}. Mostly because of the incomplete coverage of the error boxes of the GW events, so far follow-up observations did not detect any EM counterpart, in particular, the associated kilonovae emission from any of these events\footnote{A low-significance association between a sub-threshold GRB candidate GBM-190816 and a sub-threshold LIGO/Virgo GW event candidate was reported \citep{LIGO2019,Goldstein2019,Yang2019}}. 

There are two types for BH-NS mergers: one is the case that the NS directly plunges into the BH without being tidally disrupted \citep[e.g.,][]{Shibata2009} while the other is that the NS undergoes tidal disruption before the merger so that an amount of matter will remain outside the BH after the merger. No sGRB or kilonova is expected in the former case\footnote{These events will make brief, weak EM signals due to the non-negligible charge of the NS or BH \citep{Zhang2019,Dai2019}.}. Only the latter case (relative a small mass ratio between BH and the NS) is interesting for sGRB and kilonova follow-up observations. Numerical relativity (NR) simulations revealed that the dynamical ejecta from these system, caused by the tidal forces, are highly anisotropic, showing a crescent shape \citep{Kyutoku2013,Kyutoku2015,Kawaguchi2015,Brege2018}. Moreover, the wind outflows from the disk around the remnant BH are directional \citep[e.g.,][]{Just2015,Wu2016,Siegel2017}. As a result, the outflowing materials that power the kilonova emission are highly anisotropic in the BH-NS cases. Therefore, it is of great interest to explore the viewing angle effect on the kilonova lightcurves for these systems. On the other hand, due to the lack of shock heating and neutrino irradiation during or shortly after the merger, only a small fraction (e.g. a few percent \citep{Just2015}) of remnant disk can be transformed into lanthanide-free ``blue" component ejecta. The kilonovae from the BH-NS mergers would be obviously different from those of BNS mergers (e.g., AT2017gfo). 

In this paper, we study the lightcurves of BH-NS merger kilonovae in detail, paying special attention on the viewing angle effect. This paper is organized as follows. In Section \ref{sec:2}, we collect 66 results from recent NR simulations for BH-NS mergers and extrapolate fitting formulae for tidal dynamical ejecta mass and velocity across a wide range of mass ratio. In Section \ref{sec:3}, we model the dynamics and temperature evolution of each radioactivity-powered component for a BH-NS merger. In Section \ref{sec:4}, we present our simulation results for the BH-NS merger kilonova emission and compare them with the data of AT2017gfo. In Section \ref{sec:5}, we briefly discuss the relationship between kilonovae and GRB afterglows for the on-axis cases. The discussion and the conclusions are presented in Section \ref{sec:6}. In Appendix \ref{app:A0}, we summarizes the definitions of frequently used variables. In Appendix \ref{app:A}, we provide a numerical method to model photosphere evolution and the predicted lightcurves as seen by observers from different lines of sight. In Appendix \ref{app:B}, we briefly introduce the sGRB afterglow model.

\section{Remnant Disk and Dynamical Ejecta Mass}\label{sec:2}

In this section, we collect 66 simulation results of BH-NS mergers from the published literature and extrapolate the fitting formulae for the mass of tidal dynamical ejecta across a wide range of mass ratios. We also unify the velocity of the dynamical ejecta from different papers and give a new fitting formula covering the range from the near-equal-mass regime to the large mass ratio regime. The fitting formulae for the mass and velocity of dynamical ejecta are used to discuss viewing-angle-dependent kilonovae lightcurves later.

In a BH-NS merger system, whether the NS is directly plunged into the BH or tidally disrupted by the BH can be described by the relative positions of the radius of the innermost stable circular orbit (ISCO) $R_{\rm ISCO}$ and the location of the radius at which the tidal disruption occurs $R_{\rm tidal}$ \citep[see][for a review]{Shibata2011,Shibata2019}. The normalized ISCO radius $\widetilde{R}_{\rm {ISCO}} = c^2R_{\rm {ISCO}} / GM_{\rm BH}$, determined by the BH mass and the spin of BH, is given by \cite{Bardeen1972}, i.e.,
\begin{equation}
    \widetilde{R}_{\rm {ISCO}} = 3 + Z_2 - {\rm {sign}}(\chi_{\rm {BH}})\sqrt{(3 - Z_1)(3 + Z_1 + 2Z_2)},
\end{equation}
with 
\begin{equation}
\begin{split}
    &Z_1 = 1 + (1 - \chi_{\rm {BH}}^2) ^ {1 / 3} [(1 + \chi_{\rm {BH}})^{1 / 3} + (1 - \chi_{\rm {BH}})^{1 / 3}], \\
    &Z_2 = \sqrt{3 \chi_{\rm {BH}}^2 + Z_1^2},
\end{split}
\end{equation}
where $M_{\rm BH}$ is the BH mass and $\chi_{\rm BH}$ is the dimensionless spin parameter of the BH. On the other hand, the radius $R_{\rm tidal}$ at which tidal disruption occurs can be estimated by balancing the tidal force and the self-gravitational force at the surface of the NS. In the Newtonian theory, it reads
\begin{equation}
    R_{\rm tidal} \sim R_{\rm NS}\left( \frac{3M_{\rm BH}}{M_{\rm NS}} \right) ^ {1 / 3}.
\end{equation}
If $R_{\rm tidal}\lesssim R_{\rm ISCO}$, the NS would plunge into the BH without material outside the remaining BH; if $R_{\rm tidal}\gtrsim R_{\rm ISCO}$, the NS 
{would} be tidally disrupted by the BH while forming an accretion disk around the BH and an unbound tidal dynamical ejecta.

\cite{Foucart2018} presented a nonlinear model that the mass outside of the BH is determined by the relative location of $R_{\rm ISCO}$ and $R_{\rm tidal}$. By considering 75 numerical relativity (NR) simulation results, the fitting model of the total remnant mass outside the remnant BH given by \cite{Foucart2018} is
\begin{equation}
\label{Eq. TotalEjectaMassFunction}
    \frac{M_{\rm total,fit}}{M^{\rm b}_{\rm NS}} = \max\left[\left(0.406\frac{1 - 2C_{\rm NS}}{\eta^{1 / 3}} - 0.139 \widetilde{R}_{\rm {ISCO}}\frac{C_{\rm NS}}{\eta} + 0.255 \right) ^ {1.761} , 0\right],
\end{equation}
where $\eta = Q / (1 + Q) ^ 2$, $ M^{\rm b}_{\rm NS}$ is the baryonic mass of the NS, $Q = M_{\rm BH} / M_{\rm NS}$ is the mass ratio between the BH mass and the NS mass, and $C_{\rm NS} = GM_{\rm NS}/c^2R_{\rm NS}$ is the compactness of the NS. This formula covers the range $Q \in [1 , 7]$, $\chi_{\rm BH} \in [-0.5 , 0.9]$, and $C_{\rm NS} \in [0.13 , 0.182]$.  

\begin{figure}[htbp]
    \centering
    \includegraphics[width=0.70\textwidth, trim=50 45 50 50, clip]{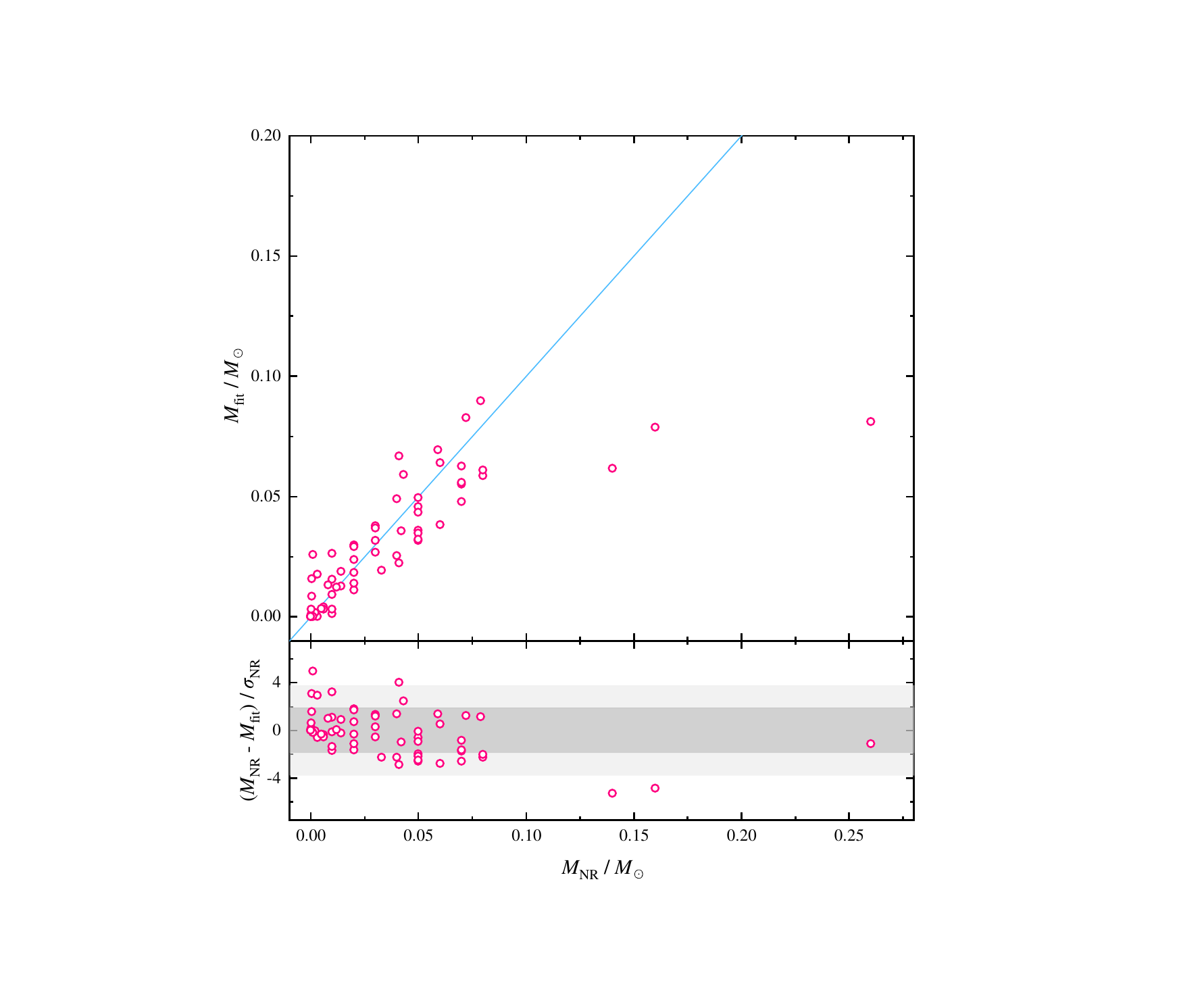}
    \caption{The top panel shows the NR data listed in Table \ref{tab:list} (horizontal axis) and the fitting results (vertical axis) of the dynamical ejecta mass. The blue line represents the position that the NR data equal to the fitting results. The bottom panel shows the normalized residuals between the NR data and the fitting results. The shaded regions include the $1-\sigma$ and $2-\sigma$ confidence intervals.}
    \label{fig:DynamicalEjecta1}
\end{figure}

As for the dynamical ejecta mass, \cite{Kawaguchi2016} presented a fitting model (by referring to \citealt{Foucart2012})
\begin{equation}
    \frac{M_{\rm d,fit}}{M^{\rm b}_{\rm NS}} = \max\left[4.464\times10^{-2}Q^{0.250}\frac{1 - 2C_{\rm NS}}{C_{\rm NS}} - 2.269 \times 10^{-3} Q^{1.352} \widetilde{R}_{\rm {ISCO}} + 2.431 \left( 1 - \frac{M_{\rm NS}}{M^{\rm b}_{\rm NS}} \right) - 0.4159 , 0\right],
\end{equation}
considering 45 NR simulation results. The range of parameters are $Q \in [3 , 7]$, $\chi_{\rm BH} \in [0 , 0.9]$, and $C_{\rm NS} \in [0.13 , 0.18]$. Recently, some new NR simulation results of BH-NS mergers including the case with near-equal-mass regime \citep[e.g.,][]{ Foucart2019} and high-spin BH regime \citep{Lovelace2013} have been published. Here we attempt to  fit 66 NR simulation results we collected (see Table \ref{tab:list}) using the similar form of formula in Equation (\ref{Eq. TotalEjectaMassFunction}). The numerical models cover the parameter range with $Q \in [1 , 7]$, $\chi_{\rm BH} \in [0 , 0.97]$, and $C_{\rm NS} \in [0.108 , 0.18]$. We assume the resulting error estimate combines a $10\%$ relative error and $0.005M_\odot$ absolute error similar to \cite{Kawaguchi2016}\footnote{The simulation result of \cite{Lovelace2013} indicated that the merger of a BH-NS with an initial BH spin parameter of $\chi_{\rm BH} = 0.97$ has a relative larger dynamical ejecta mass with $M_{\rm d,NR} = 0.26M_\odot \pm 0.16\,M_\odot$. For this simulation, we directly use the estimation error $\sigma_{\rm d,NR}$ as reported.}:
\begin{equation}
    \sigma_{\rm d,NR} = \sqrt{(0.1M_{\rm d,NR})^2 + (0.005M_\odot)^2}.
\end{equation}
The best least-square method fitting result is
\begin{equation}
    \frac{M_{\rm d,fit}}{M^{\rm b}_{\rm NS}} = \max\left[ \left(  0.273\frac{1 - 2C_{\rm NS}}{\eta^{1 / 3}} - 0.035 \widetilde{R}_{\rm {ISCO}}\frac{C_{\rm NS}}{\eta} -0.153 \right) ^ {1.491} , 0 \right].
\end{equation}

\begin{figure}[tbp]
    \centering
    \includegraphics[width=0.70\textwidth, trim=50 45 50 50, clip]{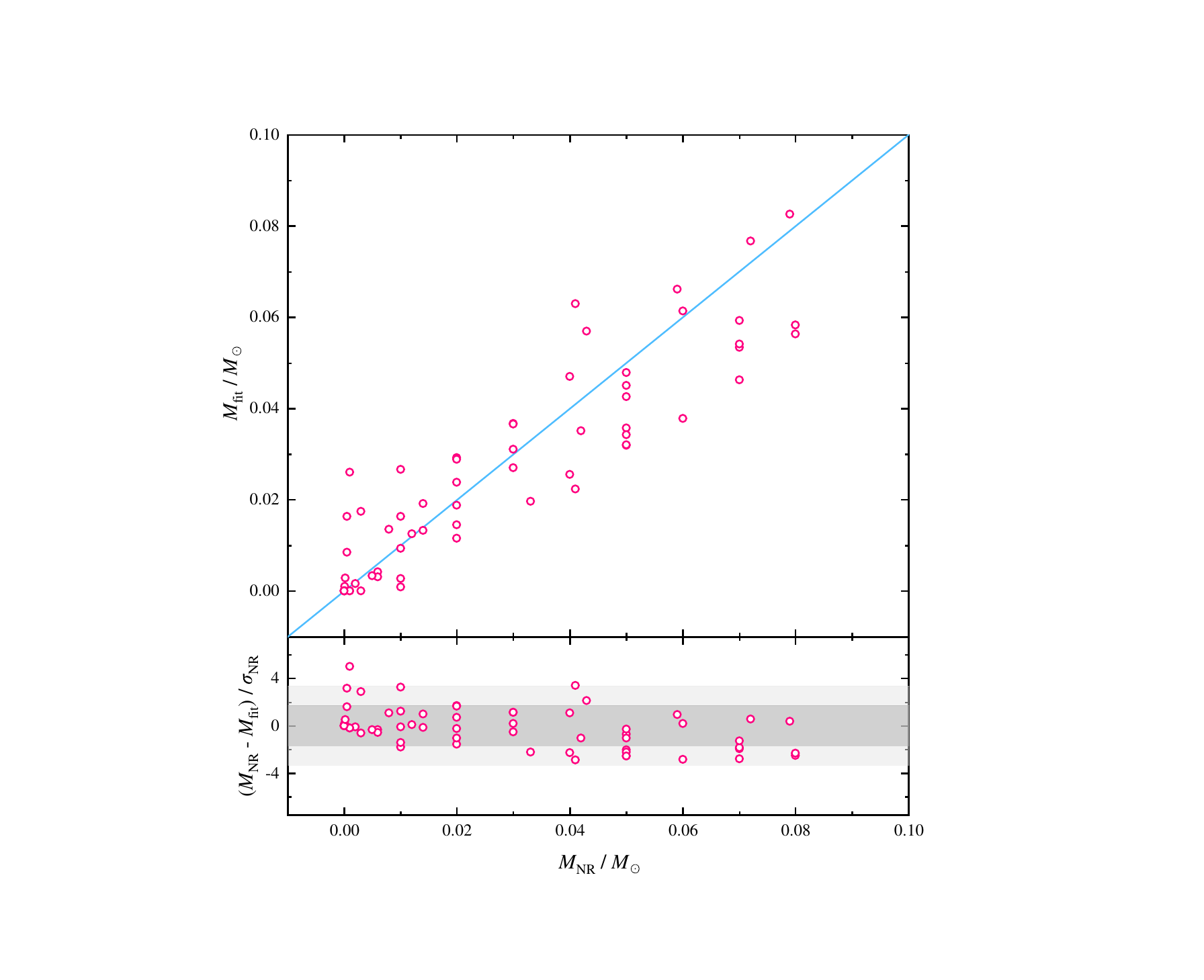}
    \caption{Similar to Figure \ref{fig:DynamicalEjecta1}, with two outlier points removed from the fitting.}
    \label{fig:DynamicalEjecta2}
\end{figure}

Figure \ref{fig:DynamicalEjecta1} shows the NR simulation data of the dynamical ejecta mass versus our fitting results and the normalized residuals. One can see that three points have relatively large dynamical ejecta mass deviating from the  fitting result. The deviated data with the largest dynamical ejecta mass (No.21 in Table \ref{tab:list}) is from \cite{Lovelace2013}. Due to its large estimated numerical error, it is still located within the region of $1-\sigma$ confidence interval.
The other two points of deviation (No.53 and No.54 in Table \ref{tab:list}) are from \cite{Foucart2014}. We note that similar initial conditions have been also studied by \cite{Kyutoku2015} and \cite{Brege2018}, who obtained results consistent with our fitting results. Removing these two points to refit the NR simulations data of the dynamical ejecta, we replot the NR data of the dynamical ejecta versus fitting results in Figure \ref{fig:DynamicalEjecta2} from $0$ to $0.1\,M_\odot$. The best least-square method fitting result now reads
\begin{equation}
\label{Eq. DynamicalEjctaFunciton}
    \frac{M_{\rm d,fit}}{M^{\rm b}_{\rm NS}} = \max\left[ \left(  0.218\frac{1 - 2C_{\rm NS}}{\eta^{1 / 3}} - 0.028 \widetilde{R}_{\rm {ISCO}}\frac{C_{\rm NS}}{\eta} -0.122 \right) ^ {1.358} , 0 \right].
\end{equation}
This fitting formula applies to dynamical ejecta mass for binaries in the following parameter ranges $Q \in [1 , 7]$, $\chi_{\rm BH} \in [0 , 0.9]$\footnote{Our fitting has covered the range $\chi_{\rm BH}\in[0 , 0.97]$. However, there is only one NR simulation result \citep{Lovelace2013} that has the extremely high spin ($\chi_{\rm BH} \sim 0.97$). We believe that the fitting formula is most reliable for the range $\chi_{\rm BH}\in[0 , 0.9]$. 
}, and $C_{\rm NS} \in [0.108 , 0.18]$. The 
{relative error of} numerical data for $0.05 \lesssim M_{\rm d,NR} \lesssim 0.1\, M_\odot$, $M_{\rm d,NR}\simeq 0.04\, M_\odot$ and $M_{\rm d,NR}\simeq 0.02\, M_\odot$ are fitted within $\sim 25\%$, $\sim 30\%$ and $\sim 40\%$. 

Additionally, we compare our dynamical ejecta fitting formula with the fitting model from \cite{Kawaguchi2016}. The reduced $\chi^2$ for our fit is 
\begin{equation}
    \chi^2 = \frac{1}{N_{\rm d,NR} - N_{\rm p} - 1}\sum_{i = 1}^{N_{\rm d,NR}}\left( \frac{M_{\rm d,NR} - M_{\rm d,fit}}{\sigma_{\rm d,NR}} \right)^2,
\end{equation}
where $N_{\rm d,NR} = 64$ and $N_{\rm p} = 4$ are the numbers of the NR data points and the fitting parameters. The number of parameters changes to $N_{\rm p} = 6$ when we use the fitting model of \cite{Kawaguchi2016}. The $\chi^2$ values of our result and that of \cite{Kawaguchi2016} are 2.926 and 3.443, respectively, suggesting a comparable fitting results between the two fitting formulae.

The fitting formula of the total remnant mass ($M_{\rm total,fit}$) and the dynamical ejecta mass ($M_{\rm d,fit}$) are fitted with independent data. Within a certain range of parameters, we find that $M_{\rm d,fit}$ could be approximately to, or even larger than $M_{\rm total,fit}$. However, the mass of the dynamical ejecta cannot exceed a few tens percent of the total remnant mass outside the remnant BH. Therefore, we assume an upper limit on how much total remnant mass can become unbound dynamical ejecta, i.e.,
\begin{equation}
    M_{\rm d,max} = f_{\rm max}M_{\rm total,fit},
\end{equation}
where $f_{\rm max}$ is the maximum fractional factor. The largest unbound component based on the NR results we collected is $\sim 50\%$ {(the model MS1-Q7a5 in \citealt{Kyutoku2015}, i.e., No.61 in Table \ref{tab:list}).} We therefore roughly define the factor is $f_{\rm max} = 0.5$. The mass of dynamical ejecta can be then expressed as
\begin{equation}
    M_{\rm d} = \min(M_{\rm d,fit} , f_{\rm max} M_{\rm total,fit}),
\end{equation}
while the remnant disk mass around BH can be estimated as
\begin{equation}
    M_{\rm disk} = M_{\rm total,fit} - M_{\rm d}.
\end{equation}

\begin{figure}
    \centering
    \includegraphics[width = 0.5\linewidth , trim = 70 30 100 60, clip]{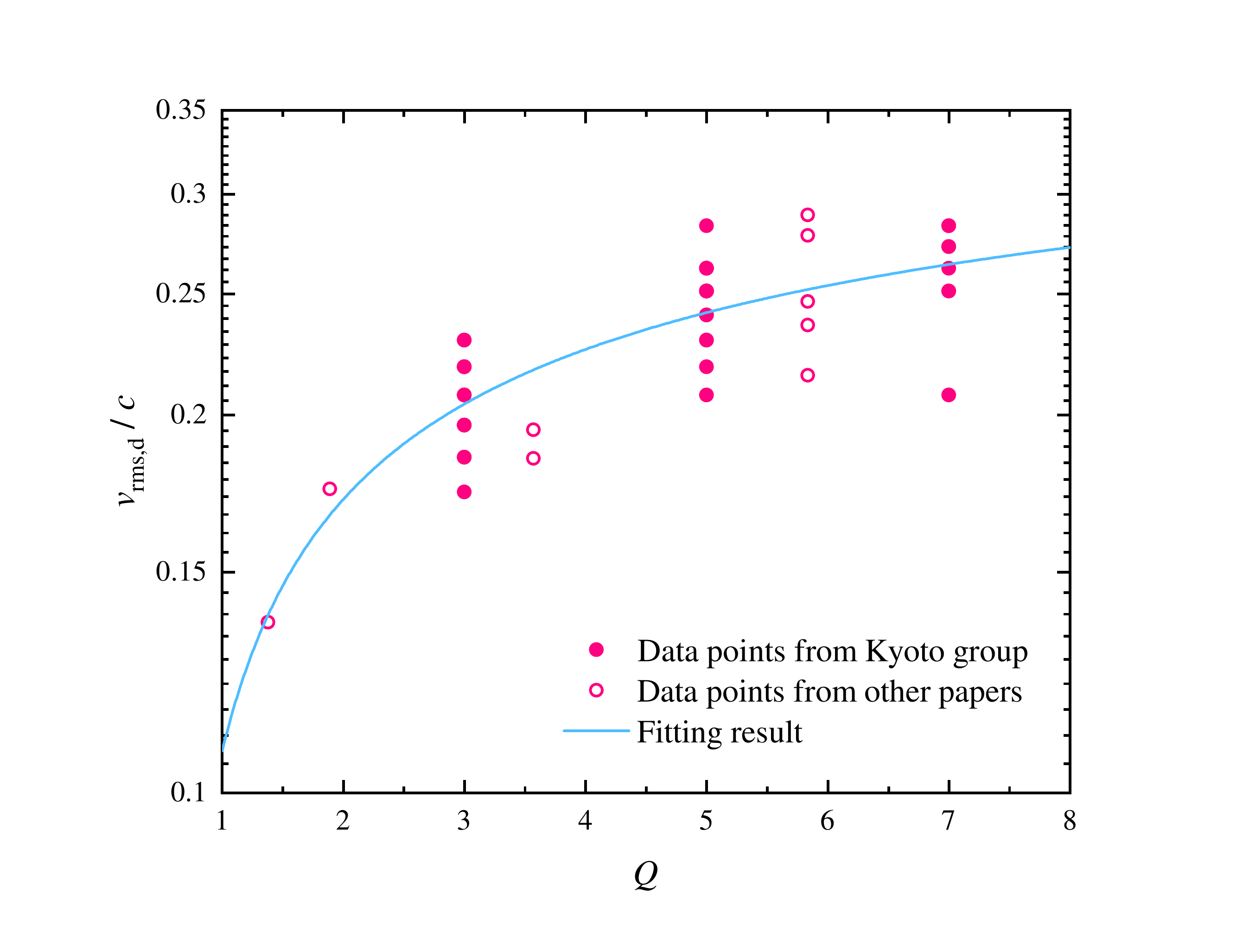}
    \caption{Comparison of the rms velocity fitting formula with the results of NR simulations. The blue line represents our rms velocity fitting formula. The pink solid points are taken from the Kyoto group after unification, while the pink open points are collected from the simulation results of \cite{Foucart2014,Foucart2017,Foucart2019} and \cite{Brege2018} after unification.}
    \label{fig:FigureRMSVelocity}
\end{figure}

We also give a fitting model for dynamical ejecta average velocity with 60 valid NR data points. In the literature, the definitions of the average velocity are inconsistent. The Kyoto group \citep[e.g.,][]{Kawaguchi2016,Kyutoku2015} defines the average velocity as the root-mean-square (rms) velocity, whereas the linear average velocity is adopted in \cite{Foucart2014,Foucart2017,Foucart2019} and \cite{Brege2018}. We transform the linear average velocity into the rms velocity by simply assuming a ratio of $\sim 1.11$ between the two \citep{Foucart2017}. In addition, as discussed in \cite{Foucart2017}, there is a $\Delta\epsilon = 7.6\,{\rm MeV/nuc}$ difference in the internal energy of the ejecta matter between their NR simulation results and the Kyoto group's simulation results due to different assumptions of the equations of state (EOSs). \cite{Foucart2017} also mentioned that the ejecta will have a kinetic energy $\sim4.6\,{\rm MeV/nuc}$ lower than predicted due to the energy lost during $r$-process nucleosynthesis \citep{Metzger2010}. Following this discussion, we unify our collected data points which are summarized in Table \ref{tab:list}. As a simple function of the mass ratio, the fitting model (see Figure \ref{fig:FigureRMSVelocity}) for the rms velocity of the dynamcial ejecta reads
\begin{equation}
    v_{\rm rms,d} = (-0.441Q^{-0.224} + 0.549)\,c.
\end{equation}
This fitting formula can achieve a good fitting in a wide $Q$ range (from 1 to 7) within $\sim15\%$ of the relative errors.

\startlongtable
\begin{deluxetable*}{cccrcccccc}
\tablecaption{List of the Numerical Relativity Simulation Results and References \label{tab:list}}
\tablecolumns{8}
\tablenum{1}
\tablewidth{0pt}
\tablehead{
\colhead{ID} & 
\colhead{$Q$} & 
\colhead{$M_{\rm BH}/M_\odot$} &
\colhead{$\chi_{\rm BH}$} & 
\colhead{$M_{\rm NS}/M_\odot$} & 
\colhead{$M^{\rm b}_{\rm NS}/M_\odot$} & 
\colhead{$C_{\rm NS}$} & 
\colhead{$M_{\rm d,NR} / M_\odot$} &  
\colhead{$v_{\rm rms,d} / c$} & 
\colhead{Reference}
}
\startdata
1 & 1.0 & 1.44 & 0.00 & 1.44 & 1.57 & 0.160 & $<1\times10^{-3}$ & \nodata & (1) \\ 
2 & 1.2 & 1.44 & 0.00 & 1.20 & 1.29 & 0.134 & $<1\times10^{-3}$ & \nodata & (1) \\ 
3 & 1.4 & 1.60 & 0.00 & 1.16 & 1.24 & 0.121 & 0.001 & 0.14 & (1) \\ 
4 & 1.9 & 1.89 & 0.15 & 1.00 & 1.06 & 0.108 & 0.03 & 0.17 & (1) \\ 
5 & 3.0 & 4.05 & -0.04 & 1.35 & 1.50 & 0.180 & $<1\times10^{-3}$ & 0.19 & (2) \\
6 & 3.0 & 4.05 & -0.04 & 1.35 & 1.47 & 0.147 & 0.006 & 0.20 & (2) \\ 
7 & 3.0 & 4.05 & 0.00 & 1.35 & 1.50 & 0.180 & $<1\times10^{-3}$ & 0.17 & (2,3) \\ 
8 & 3.0 & 4.05 & 0.00 & 1.35 & 1.48 & 0.161 & 0.003 & 0.20 & (2,3) \\ 
9 & 3.0 & 4.05 & 0.00 & 1.35 & 1.47 & 0.147 & 0.006 & 0.20 & (2,3) \\ 
10 & 3.0 & 4.05 & 0.00 & 1.35 & 1.45 & 0.138 & 0.02 & 0.21 & (2,3) \\ 
11 & 3.0 & 4.05 & 0.35 & 1.35 & 1.47 & 0.147 & 0.02 & 0.22 & (2) \\ 
12 & 3.0 & 4.05 & 0.50 & 1.35 & 1.50 & 0.180 & 0.002 & 0.19 & (2,3) \\ 
13 & 3.0 & 4.05 & 0.50 & 1.35 & 1.48 & 0.161 & 0.02 & 0.22 & (2,3) \\ 
14 & 3.0 & 4.05 & 0.50 & 1.35 & 1.47 & 0.147 & 0.03 & 0.21 & (2,3) \\ 
15 & 3.0 & 4.05 & 0.50 & 1.35 & 1.45 & 0.138 & 0.05 & 0.22 & (2,3) \\ 
16 & 3.0 & 4.05 & 0.64 & 1.35 & 1.47 & 0.147 & 0.03 & 0.20 & (2) \\ 
17 & 3.0 & 4.05 & 0.75 & 1.35 & 1.50 & 0.180 & 0.01 & 0.21 & (2,3) \\ 
18 & 3.0 & 4.05 & 0.75 & 1.35 & 1.48 & 0.161 & 0.05 & 0.23 & (2,3) \\ 
19 & 3.0 & 4.05 & 0.75 & 1.35 & 1.47 & 0.147 & 0.05 & 0.22 & (2,3) \\ 
20 & 3.0 & 4.05 & 0.75 & 1.35 & 1.45 & 0.138 & 0.07 & 0.23 & (2,3) \\ 
21 & 3.0 & 4.20 & 0.97 & 1.40 & 1.51 & 0.144 & 0.26 & \nodata & (4) \\ 
22 & 3.6 & 5.00 & 0.35 & 1.40 & 1.53 & 0.152 & 0.014 & 0.20 & (5) \\ 
23 & 3.6 & 5.00 & 0.45 & 1.40 & 1.53 & 0.152 & 0.014 & 0.19 & (5) \\ 
24 & 4.0 & 5.40 & 0.75 & 1.35 & 1.48 & 0.167 & 0.01 & \nodata & (6) \\
25 & 4.0 & 5.40 & 0.75 & 1.35 & 1.47 & 0.151 & 0.05 & \nodata & (6) \\
26 & 4.0 & 5.40 & 0.75 & 1.35 & 1.45 & 0.138 & 0.08 & \nodata & (6) \\
27 & 5.0 & 6.75 & -0.05 & 1.35 & 1.50 & 0.180 & $<1\times10^{-3}$ & 0.22 & (2,7) \\ 
28 & 5.0 & 6.75 & -0.05 & 1.35 & 1.48 & 0.161 & $<1\times10^{-3}$ & 0.24 & (2,7) \\ 
29 & 5.0 & 6.75 & -0.05 & 1.35 & 1.47 & 0.147 & 0.001 & 0.26 & (2,7) \\ 
30 & 5.0 & 6.75 & -0.04 & 1.35 & 1.45 & 0.138 & 0.01 & 0.25 & (2,7) \\ 
31 & 5.0 & 6.75 & 0.34 & 1.35 & 1.50 & 0.180 & 0.001 & 0.25 & (2,7) \\ 
32 & 5.0 & 6.75 & 0.34 & 1.35 & 1.48 & 0.161 & 0.01 & 0.26 & (2,7) \\ 
33 & 5.0 & 6.75 & 0.34 & 1.35 & 1.47 & 0.147 & 0.012 & 0.23 & (2,7) \\ 
34 & 5.0 & 6.75 & 0.34 & 1.35 & 1.45 & 0.138 & 0.041 & 0.25 & (2,7) \\ 
35 & 5.0 & 6.75 & 0.50 & 1.35 & 1.50 & 0.180 & $<1\times10^{-3}$ & 0.21 & (2,3) \\ 
36 & 5.0 & 6.75 & 0.50 & 1.35 & 1.48 & 0.161 & 0.01 & 0.25 & (2,3) \\
37 & 5.0 & 6.75 & 0.50 & 1.35 & 1.47 & 0.147 & 0.02 & 0.24 & (2,3) \\
38 & 5.0 & 6.75 & 0.50 & 1.35 & 1.45 & 0.138 & 0.05 & 0.25 & (2,3) \\ 
39 & 5.0 & 6.75 & 0.63 & 1.35 & 1.50 & 0.180 & 0.005 & 0.28 & (2,7) \\ 
40 & 5.0 & 6.75 & 0.63 & 1.35 & 1.48 & 0.161 & 0.033 & 0.25 & (2,7) \\ 
41 & 5.0 & 6.75 & 0.63 & 1.35 & 1.47 & 0.147 & 0.042 & 0.25 & (2,7) \\ 
42 & 5.0 & 6.75 & 0.64 & 1.35 & 1.45 & 0.138 & 0.07 & 0.26 & (2,7) \\ 
43 & 5.0 & 6.75 & 0.75 & 1.35 & 1.50 & 0.180 & 0.008 & 0.23 & (2,3) \\ 
44 & 5.0 & 6.75 & 0.75 & 1.35 & 1.48 & 0.161 & 0.05 & 0.26 & (2,3) \\ 
45 & 5.0 & 6.75 & 0.75 & 1.35 & 1.47 & 0.147 & 0.05 & 0.25 & (2,3) \\ 
46 & 5.0 & 6.75 & 0.75 & 1.35 & 1.45 & 0.138 & 0.08 & 0.26 & (2,3) \\ 
47 & 5.0 & 7.00 & 0.70 & 1.40 & 1.55 & 0.163 & 0.04 & 0.24 & (8) \\ 
48 & 5.0 & 7.00 & 0.80 & 1.40 & 1.55 & 0.163 & 0.06 & 0.22 & (8) \\ 
49 & 5.0 & 7.00 & 0.85 & 1.40 & 1.53 & 0.152 & 0.043 & 0.24 & (5) \\ 
50 & 5.0 & 7.00 & 0.90 & 1.40 & 1.55 & 0.163 & 0.07 & 0.22 & (8) \\ 
51 & 5.0 & 7.00 & 0.90 & 1.40 & 1.53 & 0.156 & 0.06 & 0.24 & (9) \\ 
52 & 5.0 & 7.00 & 0.90 & 1.40 & 1.53 & 0.152 & 0.059 & 0.23 & (9) \\ 
53 & 5.8 & 7.00 & 0.80 & 1.20 & 1.31 & 0.139 & 0.14 & 0.28 & (8) \\ 
54 & 5.8 & 7.00 & 0.90 & 1.20 & 1.31 & 0.139 & 0.16 & 0.29 & (8) \\ 
55 & 5.8 & 7.00 & 0.90 & 1.20 & 1.29 & 0.135 & 0.072 & 0.25 & (9) \\ 
56 & 5.8 & 7.00 & 0.90 & 1.20 & 1.29 & 0.130 & 0.079 & 0.24 & (9) \\ 
57 & 5.8 & 7.00 & 0.90 & 1.20 & 1.31 & 0.148 & 0.041 & 0.22 & (9) \\
58 & 7.0 & 9.45 & 0.50 & 1.35 & 1.50 & 0.180 & $<1\times10^{-3}$ & 0.21 & (2,3) \\
59 & 7.0 & 9.45 & 0.50 & 1.35 & 1.48 & 0.161 & $<1\times10^{-3}$ & 0.25 & (2,3) \\ 
60 & 7.0 & 9.45 & 0.50 & 1.35 & 1.47 & 0.147 & 0.003 & 0.27 & (2,3) \\ 
61 & 7.0 & 9.45 & 0.50 & 1.35 & 1.45 & 0.138 & 0.02 & 0.28 & (2,3) \\ 
62 & 7.0 & 9.45 & 0.63 & 1.35 & 1.47 & 0.147 & 0.03 & 0.26 & (2) \\ 
63 & 7.0 & 9.45 & 0.75 & 1.35 & 1.50 & 0.180 & $<1\times10^{-3}$ & 0.25 & (2,3) \\ 
64 & 7.0 & 9.45 & 0.75 & 1.35 & 1.48 & 0.161 & 0.02 & 0.27 & (2,3) \\ 
65 & 7.0 & 9.45 & 0.75 & 1.35 & 1.47 & 0.147 & 0.04 & 0.27 & (2,3) \\ 
66 & 7.0 & 9.45 & 0.75 & 1.35 & 1.45 & 0.138 & 0.07 & 0.28 & (2,3) \\
\enddata
\tablecomments{We list the mass ratio $Q = M_{\rm BH} / M_{\rm NS}$, the BH mass $M_{\rm BH}$, the dimensionless spin of the BH $\chi_{\rm BH}$, the NS mass $M_{\rm NS}$, the baryon mass of the NS $M^{\rm b}_{\rm NS}$, compactness of the NS $C_{\rm NS}$, the dynamical ejecta mass $M_{\rm d,NR}$, the mass-weighted root-mean-square velocity $v_{\rm rms,d}$, and references of these simulation results. We note that $\chi_{\rm BH}$ here represents the dimensionless spin of the BH in the direction of the orbital angular momentum. We calculate the mass-weighted root-meen-square velocity $v_{\rm rms,d}$ based on the comments from \cite{Foucart2017}. The references include: (1)\cite{Foucart2019}; (2)\cite{Kawaguchi2016}; (3)\cite{Kyutoku2015}; (4)\cite{Lovelace2013}; (5)\cite{Foucart2017}; (6)\cite{Kyutoku2018}; (7)\cite{Kawaguchi2015}; (8)\cite{Foucart2014}; (9)\cite{Brege2018}.}
\end{deluxetable*}

\section{Dynamics and Temperature Evolution of Mass Ejection}\label{sec:3}

\begin{figure}[htbp]
    \centering
    \includegraphics[width=0.60\textwidth, trim = 170 260 170 100, clip]{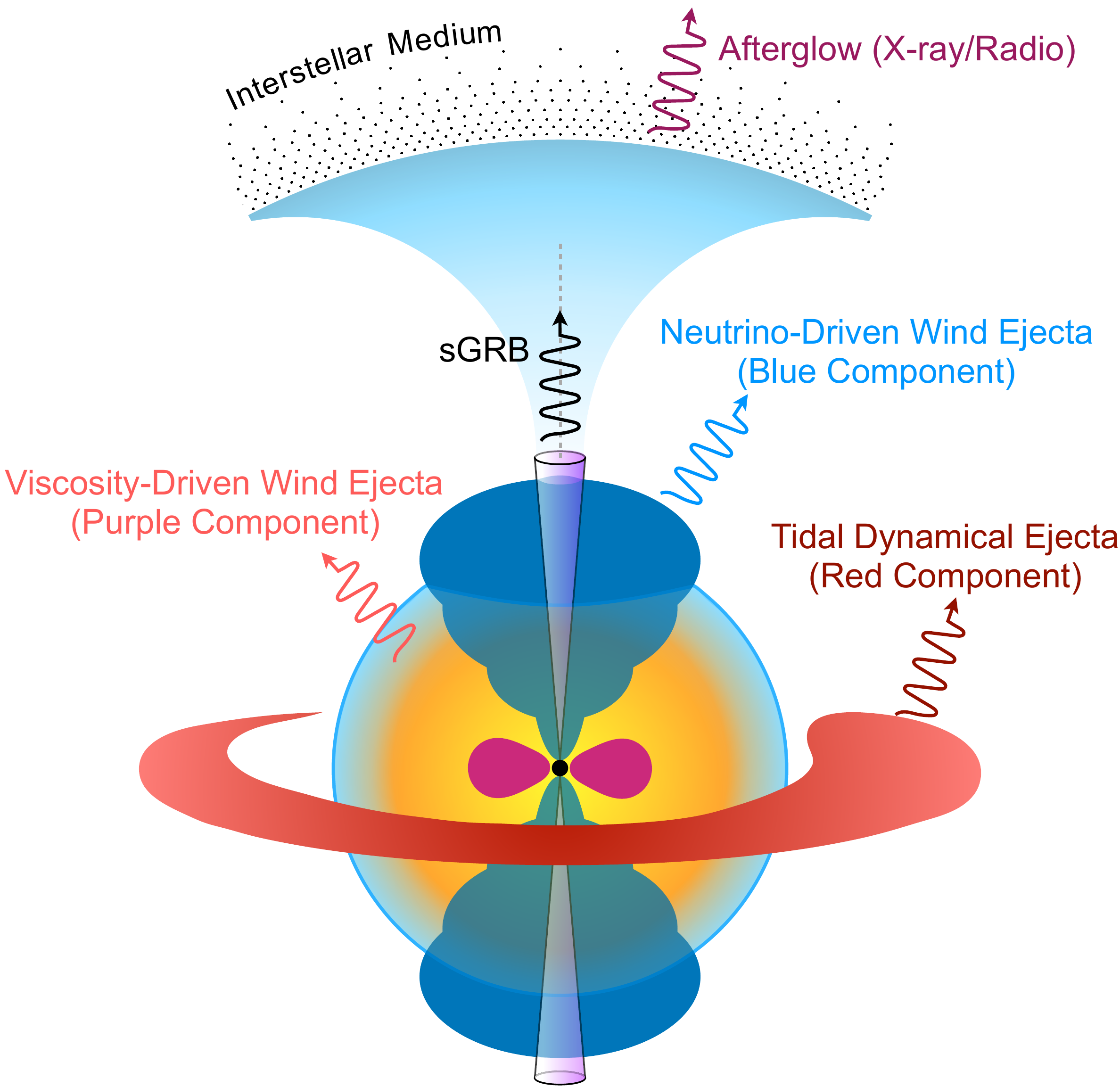}
    \caption{A Cartoon for several EM counterpart emission components after the BH-NS mergers. The unbound tidal dynamical ejecta (the ``red'' component, denoted in red color) and bound disk (magenta color) around the remnant BH are distributed in the equatorial plane. Near the polar axis, there might be a sGRB jet formed by accretion of the remnant BH via the Blandford-Znajek effect \citep{Blandford1977}. The jet would expand into the interstellar medium to power a GRB afterglow. During the evolution of the BH accretion disk, additional matter ejections are expected: The neutrino-driven wind ejecta (the ``blue'' component, denoted in blue color) launched through neutrino-matter interactions and magnetic pressure \citep{Fernandez2013,Just2015} mainly points towards the polar direction. The viscosity-driven wind ejecta (the ``purple'' component, denoted in orange-light-blue color) 
    would be launched more isotropically than the red-component but is a preferable direction in the equatorial plane.}
    \label{fig:Cartoon}
\end{figure}

For BH-NS mergers with NS tidal disruption, there are three possible ejecta components. First, a fraction of neutron-rich matter is tidally ejected. Lacking of weak-interaction processes such as neutrino irradiation, such {\em dynamical ejecta} should be lanthanide-rich which has a relatively low electron fraction $Y_e \lesssim 0.2$ (e.g., \citealt{Foucart2012,Foucart2014,Kyutoku2018}). 

Second, an accretion disk is formed around the remnant BH. At early times after disk forming, mass loss is driven by neutrino heating thanks to the high temperature in the disk. Due to neutrino illumination, such a {\em neutrino-driven wind ejecta} is less neutron rich, with a high electron fraction $Y_e \gtrsim 0.4$. Since the neutrino luminosity decreases rapidly with time \citep{Fernandez2013,Just2015}, such a wind, directing in the polar direction, only lasts for a short period of time so that only a few percent of disk material is ejected in this form. Third, in the later stages, viscous heating and angular momentum transport play important roles in the mass loss of the disk. The electron fraction of this equatorial dominated {\em viscosity-driven wind ejecta} is low, lying in the range of $Y_e\sim0.15-0.25$ \citep[e.g.,][]{Fernandez2013,Just2015,Siegel2017,Metzger2019}.

\cite{Tanaka2019} found that the average effective the gray opacities of the mixture of $r$-processes are $\kappa\sim20-30\,{\rm cm}^2\,{\rm g}^{-1}$ for $Y_{\rm e} \leq 0.20$, $\kappa\sim3-5\,{\rm cm}^2\,{\rm g}^{-1}$ for $Y_{\rm e} \approx 0.25 - 0.35$, and $\kappa\sim1\,{\rm cm}^2\,{\rm g}^{-1}$ for $Y_{\rm e} = 0.40$ at the temperature $T = 5-10\times10^3\,{\rm K}$. At lower temperature these opacities can decrease steeply. In order to simplify the model, we set the gray opacity of dynamical ejecta, neutrino-driven ejecta, and viscosity-driven ejecta as $\kappa_{\rm d} = 20\,{\rm cm}^2\,{\rm g}^{-1}$, $\kappa_{\rm n} = 1\,{\rm cm}^2\,{\rm g}^{-1}$ and $\kappa_{\rm v} = 5\,{\rm cm}^2\,{\rm g}^{-1}$, respectively.

Figure \ref{fig:Cartoon} summarizes the EM counterparts of a BH-NS merger and its mass ejection distribution. In this section, we model the dynamics and temperature evolution of each ejecta component. Since we assume that the three-component ejecta have a homologous expansion of the mass shells for each velocity, i.e., $r = vt$, one can substitute $(r , t)$ by $(v , t)$.

\subsection{Tidal Dynamical Ejecta}

According to NR simulations of BH-NS mergers \citep[e.g.,][]{Kyutoku2013,Kyutoku2015,Kawaguchi2015}, the mass distribution of dynamical ejecta is highly anisotropic, with the mass mainly distributed around the equatorial plane and shaped like a crescent. \cite{Kyutoku2013,Kyutoku2015} and \cite{Kawaguchi2016} indicated that the distribution  angle range of the ejecta in the longitudinal direction is almost $\varphi_{\rm d} \approx \pi$, while the half opening angle in the latitudinal direction is typically in the range of $\theta_{\rm d} \approx 10^\circ-20^\circ$. Hereafter, we assume that the half opening angle in the latitudinal direction is a constant with $\theta_{\rm d} = 15^\circ$. We assume that the dynamical ejecta of the BH-NS mergers have a relatively flat distribution velocity profile with $v_{\min , {\rm d}}<v<v_{\max , {\rm d}}$ \citep{Kyutoku2015,Kawaguchi2016}, i.e., ${\rm d}M_{\rm d}/{\rm d}v \approx {\rm const}$. The mass normalization is determined by 
\begin{equation}
\label{Eq. Normalization Dynamical ejecta 1}
\int_{v_{\min,{\rm d}}}^{v_{\max,{\rm d}}} \frac{{\rm d}M_{\rm d}}{{\rm d}v}{\rm d}v = M_{\rm {d}}.
\end{equation}
Similarly, the normalization can be also determined by the mass distribution of the dynamical ejecta in the latitudinal direction. Following \cite{Kawaguchi2016}, we simply assume that the mass distribution of ejecta in the latitudinal direction is homogeneous\footnote{\cite{Huang2018} considered the inhomogeneous mass distribution in the latitudinal direction and found that it has little effect on lightcurves.}. The volume element of the dynamical ejecta is ${\rm d}V = r^2\cos\theta{\rm d}r{\rm d}\theta{\rm d}\varphi = v^2t^3\cos\theta{\rm d}v{\rm d}\theta{\rm d}\varphi$. Therefore, another normalization of the dynamical ejecta mass is
\begin{equation}
    \int_{v_{\min , {\rm d}}}^{v_{\max , {\rm d}}}\int_{0}^{\theta_{\rm {d}}}\int_{-\varphi_{\rm {d}}/2}^{\varphi_{\rm d}/2}2\rho_{\rm d}(v,t)v^2t^3\cos\theta{\rm d}v{\rm d}\theta{\rm d}\varphi = M_{\rm {d}},
\end{equation}
where $\varphi_{\rm d}$ and $\theta_{\rm d}$ are the longitudinal half opening angle and the latitudinal opening angle, respectively. Expanding the above integral, we get
\begin{equation}
    2\varphi_{\rm {d}}\theta_{\rm d}t^3\int_{v_{\min , {\rm d}}}^{v_{\max , {\rm d}}}\rho(v , t)v^2{\rm d}v = M_{\rm d}
\end{equation}
The dynamical ejecta in the latitudinal direction is approximately geometrically thin, i.e., $\sin\theta_{\rm d} \approx \theta_{\rm d} \ll 1$. The ejecta profile is approximately
\begin{equation}
\label{Eq. Normalization Dynamical ejecta 2}
    \frac{{\rm d}M_{\rm d}}{{\rm d}v} = 2\varphi_{\rm d}\theta_{\rm d}\rho_{\rm d}(v,t)v^2t^3.
\end{equation}
Combining Equation (\ref{Eq. Normalization Dynamical ejecta 1}) and Equation (\ref{Eq. Normalization Dynamical ejecta 2}), one gets the density in the form of
\begin{equation}
    \rho_{\rm d}(v , t) = \frac{M_{\rm d}}{2\varphi_{\rm d}\theta_{\rm d}(v_{\max , {\rm d}} - v_{\min , {\rm d}})}v^{- 2}t^{-3}.
\end{equation}

In Section \ref{sec:2}, we give a fitting model of the rms velocity of the dynamical ejecta. The kinetic energy of the dynamical ejecta is
\begin{equation}
\label{Eq. KineticEnergy1}
    E_{\rm K,d} = \frac{1}{2}M_{\rm d}v_{\rm rms,d}^2,
\end{equation}
which can be also expressed as
\begin{equation}
\label{Eq. KineticEnergy2}
    E_{\rm K,d} = \int_{v_{\rm min,d}} ^ {v_{\rm max,d}}\frac{1}{2}v^2\frac{{\rm d}M_{\rm d}}{{\rm d}v}{\rm d}v.
\end{equation}
Based on the NR results \citep[e.g.,][]{Kyutoku2015,Brege2018}, we find the lower positions at the half maximum of ${{\rm d}M_{\rm d}}/{{\rm d}v}$ are nearly to $0.1\,c$ so that we set the minimum velocity of dynamical ejecta is $v_{\rm min,d} \simeq 0.1\, c$\footnote{In principle, there is a small amount of mass below $0.1 \,c$. However, including it does not noticeably affect the calculation results.}. Combining Equation (\ref{Eq. KineticEnergy1}) and Equation (\ref{Eq. KineticEnergy2}), one can express the maximum velocity as
\begin{equation}
    v_{\rm max,d} = \sqrt{3v_{\rm rms,d}^2 - \frac{3}{4}v_{\rm min,d}^2} - \frac{1}{2}v_{\rm min,d}.
\end{equation}

\begin{figure}[tbp]
    \centering
    \includegraphics[width=0.60\textwidth, trim=50 210 145 60, clip]{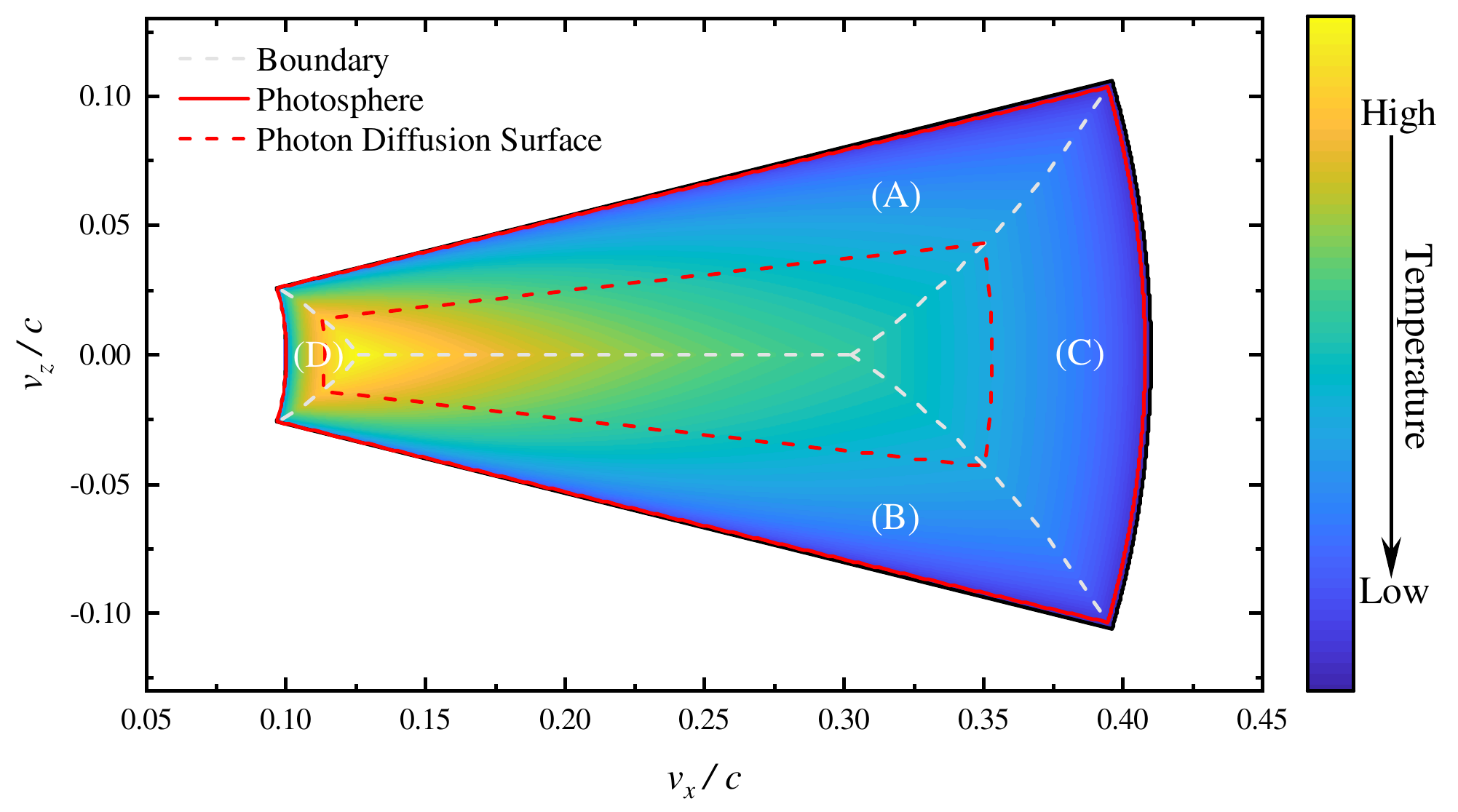}
    \caption{A schematic diagram of the dynamical ejecta that is divided into different regions. It shows a sectional drawing of the $v_y - v_z$ plane. The inner color scheme qualitatively depicts the temperature distribution of the dynamical ejecta. The light gray dashed lines divide the dynamical ejecta into four region. The red solid lines and dashed lines represent the photosphere and photon diffusion surface, respectively. For Regions A and B, the temperature gradient is along the latitudinal direction. For Region C and region D, it is along radial and anti-radial directions, respectively. }
    \label{fig:DynamicalEjectaRegion}
\end{figure}

{Figure \ref{fig:DynamicalEjectaRegion} shows a schematic diagram of the dynamical ejecta. We define a critical surface where $\tau \approx c / v$ as the {\em photon diffusion surface} (red dashed lines) 
below which  photons are trapped in the ejecta because photon diffusion is slower than ejecta expansion. Above this critical surface, the photon diffusion velocity is larger than thye ejecta expansion velocity so that they can escape (even though photon energy can be changed). We also define the {\em photosphere} (red solid lines) from which photons are last scattered. This is at an optical depth of $\tau \approx 2 / 3$.} 

For a given point in the dynamical ejecta, photons diffuse in all directions. However, for the four directions denoted A, B, C, and D, the shortest diffusion times should be the latitudinal directions, radial direction, and the anti-radial direction, respectivelty. Considering random walk of the photons, the diffusion time can be expressed as $t_{\rm diff}\sim(R/\lambda)^2(\lambda / c)$, where $R$ is the distance between the position of the given point and the surface, and $\lambda = (\kappa_{\rm d}\rho_{\rm d})^{-1}$ is the mean free path {of photons}. Therefore, the diffusion times in the latitudinal, radial, and anti-radial directions can be estimated as
\begin{equation}
    t_{\rm diff,lat}(\theta, t)\sim\frac{(\theta_{\rm d} - \theta)^2\kappa_{\rm d}M_{\rm d}}{2c\varphi_{\rm d}\theta_{\rm d}(v_{\rm max,d} - v_{\rm min,d})t},
\end{equation}
\begin{equation}
    t_{\rm diff,rad}(v , t)\sim\frac{(v_{\rm max,d} - v)^2\kappa_{\rm d}M_{\rm d}}{2c\varphi_{\rm d}\theta_{\rm d}(v_{\rm max,d} - v_{\rm min,d})v^2t},
\end{equation}
and
\begin{equation}
    t_{\rm diff,-rad}(v , t)\sim\frac{(v - v_{\rm min,d})^2\kappa_{\rm d}M_{\rm d}}{2c\varphi_{\rm d}\theta_{\rm d}(v_{\rm max,d} - v_{\rm min,d})v^2t}.
\end{equation}
By comparing these three diffusion times, we can find two boundaries which are shown by the light gray dashed lines in Figure \ref{fig:DynamicalEjectaRegion}, i.e.,
\begin{equation}
\begin{split}
\label{Eq. Bounday}
    &v_{\rm bou1} = v_{\rm min,d}(\theta - \theta_{\rm d} + 1)^{-1} \approx v_{\rm min,d}(\theta_{\rm d} - \theta + 1), \\
    &v_{\rm bou2} = v_{\rm max,d}(\theta_{\rm d} - \theta + 1)^{-1} \approx v_{\rm max,d}(\theta - \theta_{\rm d} + 1). \\
\end{split}
\end{equation}
The latter approximations of the above two equations are due to the thin half opening angle of the dynamical ejecta in the latitudinal directions. These two boundaries divide the dynamical ejecta into four different regions. For Regions A and B in the Figure \ref{fig:DynamicalEjectaRegion}, the directions of the temperature gradient is along the latitudinal directions; for Region C, it is along the radial direction; and for Region D, it is along the anti-radial direction. The photon diffusion directions are oppose to the directions of the temperature gradients.

One can also calculate the optical depths in each of the three directions. The optical depths in the latitudinal, radial, and anti-radial directions are
\begin{equation}
    \tau_{\rm d,lat}(v , \theta , t) \approx \int_{\theta}^{\theta_{\rm d}}\kappa_{\rm d}\rho_{\rm d}(v,t) vt{\rm d}\theta = \frac{\kappa_{\rm d}M_{\rm d}(\theta_{\rm d} - \theta)}{2\varphi_{\rm d}\theta_{\rm d}(v_{{\max},{\rm d}} - v_{{\min},{\rm d}})vt^2},
\end{equation}
\begin{equation}
    \tau_{\rm d,rad}(v , t) \approx \int_{v}^{v_{\rm max,d}}\kappa_{\rm d}\rho_{\rm d}(v,t) t{\rm d}v = \frac{\kappa_{\rm d}M_{\rm d}(v^{-1} - v_{\rm max,d}^{-1})}{2\varphi_{\rm d}\theta_{\rm d}(v_{{\max},{\rm d}} - v_{{\min},{\rm d}})t^2},
\end{equation}
and
\begin{equation}
    \tau_{\rm d,-rad}(v , t) \approx \int_{v_{\rm min,d}}^{v}\kappa_{\rm d}\rho_{\rm d}(v,t) t{\rm d}v = \frac{\kappa_{\rm d}M_{\rm d}(v_{\rm min,d}^{-1} - v^{-1})}{2\varphi_{\rm d}\theta_{\rm d}(v_{{\max},{\rm d}} - v_{{\min},{\rm d}})t^2},
\end{equation}
respectively. By comparing these optical depths, one can derive two boundaries $v_{\rm bou1} = v_{\rm min,d}(\theta - \theta_{\rm d} + 1)$ and $v_{\rm bou2} = v_{\rm max,d}(\theta - \theta_{\rm d} + 1)$, which are equal to those derived in Equation (\ref{Eq. Bounday}). The fact that the boundaries defined by the photon diffusion surface and by the photosphere are consistent with each makes sense, since the optical depths at the two surfaces are connected through the same mathematical conditions. It also suggests that the physical conditions in adjacent regions (e.g. A vs. C, A vs. D, B. vs. C, and B. vs. D) are continuous.

In the following, we will use an approximation method to model the temperature gradient of the dynamical ejecta for different regions. First, we consider the temperature gradient in the latitudinal direction. Due to the thin half opening angle of the dynamical ejecta in the latitudinal direction, we assume that all the photons escape the ejecta along the $v_z-$axis. The skin depth angle $\theta_{\rm diff,d}$ (below the ejecta surface) where the photons can diffuse vertically out of the ejecta surface within the dynamical time is given by $(\theta_{\rm d} - \theta_{\rm diff,d})vt\approx(c/\tau_{\rm d,lat}) t$ and $\tau_{\rm d,lat}\approx\kappa_{\rm {d}}\rho_{\rm d} (\theta_{\rm d} - \theta_{\rm diff,d})vt$, i.e.,
\begin{equation}
  \theta_{\rm diff,d}(t) = \theta_{\rm {d}}\left(1 - \frac{t}{t_{\rm c}}\right),
\end{equation}
where we set the critical diffuse timescale that all ejecta can be seen for each velocity $v$ as
\begin{equation}
    t_{\rm c} = \sqrt{\frac{\kappa_{\rm d} M_{\rm d}\theta_{\rm d}}{2c\varphi_{\rm d}(v_{\max,{\rm d}} - v_{\min,{\rm d}})}}.
\end{equation}

Consider the luminosity ${\rm d}L_{\rm d}$ from the part of ejecta in a velocity bin of $v$ to $v + {\rm d}v$. We assume the specific energy injection rate due to radioactive decay is $\dot{\epsilon}(t) = \epsilon_{Y_e}\dot{\epsilon}_0 (t / {\rm day})^{-s}$ with $\dot{\epsilon}_0 \approx 1.58\times 10^{10}{\rm {erg}}\,{\rm g}^{-1}\,{\rm s}^{-1} $ and $s\approx1.3$, where  $\epsilon_{Y_e}$ is an electron-fraction-dependent term which takes into account extremely neutron-rich ejecta with a decay half-life of a few hours  \citep{Perego2017}, i.e.
\begin{equation}
    \epsilon_{Y_e} = \left\{\begin{matrix}
0.5 + 2.5[1 + e^{4(t/{\rm day} - 1)}] ^ {-1},\ &{\rm if}\ Y_e\ge0.25, \\ 
1,\ &{\rm otherwise}
\end{matrix}\right.
\end{equation}
At $t < t_{\rm c}$, only the photons at a depth smaller than $\theta_{\rm d}$ can escape within a dynamical time and hence contribute to the emission; at $t \ge t_{\rm c}$ all the photons can escape. Therefore, the luminosity per unit velocity is
\begin{equation}
    \frac{{\rm d}L_{\rm {d}}}{{\rm d}v} = \frac{1}{2}\epsilon_{\rm {th}}\epsilon_{Y_e}\dot{\epsilon}_0\frac{{\rm d}M_{\rm d}}{{\rm d}v}
    \left\{ \begin{matrix}
    \frac{t}{t_{\rm c}}\left(\frac{t}{\rm day}\right)^{-s} &, t<t_{\rm c}, \\
    \left(\frac{t}{\rm day}\right)^{-s} &, t \ge t_{\rm c} .
    \end{matrix}\right.
\end{equation}
where the efficiency of thermalization is taken as  $\epsilon_{\rm {th}}=0.5$ \citep{Metzger2010} and
the factor of $1 / 2$ accounts for the two surfaces of the ejecta in the latitudinal directions.

The expression for ${\rm d}L_{\rm d}/{\rm d}v$ enables us to calculate the local emissivity per unit area, $D_{\rm d}(v,t)$, which is approximately a perfect blackbody. Because ${\rm d}L_{\rm d}$ is released over an area of $\varphi_{\rm d}r{\rm d}r$, we get
\begin{equation}
    D_{\rm d}(v , t) = \frac{{\rm d}L_{\rm d}}{\varphi_{\rm d}r{\rm d}r} \propto \left\{ \begin{matrix}
    v^{-1}t^{-(s + 1)} & ,t<t_{\rm c}, \\
    v^{-1}t^{-(s + 2)} & ,t \ge t_{\rm c}.
    \end{matrix}\right.
\end{equation}
Therefore, one can derive the effective temperature of each velocity at a given time
\begin{equation}
    T_{\rm eff,d}(v,t) = \left(\frac{D_{\rm d}(v,t)}{\sigma_{\rm SB}}\right)^{1 / 4},
\end{equation}
where $\sigma_{\rm SB}$ is Steffan-Boltzmann constant. We then use the Eddington approximation \citep{Mihalas1970,Rybicki1979} to describe thermal temperature of each point in the velocity space. The internal thermal temperature can be written as
\begin{equation}
    T_{\rm d}(v , \theta , t) = T_{\rm eff,d}(v , t)\left[ \frac{3}{4}\left(\tau_{\rm d,lat}(v , \theta , t) + \frac{2}{3} \right) \right]^{1 / 4}.
\end{equation}

We have considered the temperature gradient if all the photons escape the ejecta along the $v_z-$direction which have an equal gradient directions in Regions A and B. In the following, we model the direction of temperature gradient in Regions C and D. If all the photons escape from the radial direction, as for the dynamical ejecta with a relatively flat velocity distribution profile in the radial direction, the temperature of each velocity $v$ would be equal. Since the physical conditions at the boundaries between adjacent regions (A vs. C, A. vs. D, B. vs. C, and B. vs. D) are continuous, one can simply map the velocity gradient (i.e. temperature gradient) in Regions C and D by that in Regions A and B through the adjacent boundaries. The temperature distributions of the dynamical ejecta in all four regions are qualitatively depicted with the color scheme in Figure \ref{fig:DynamicalEjectaRegion}.

\subsection{Black Hole Disk Outflows}

\subsubsection{Neutrino-Driven Wind Ejecta}
Different from BNS mergers that may have significant neutrino-driven ejecta due to neutrino heating by a remnant supramassive or hypermassive NS that can release copious thermal neutrinos \cite[e.g.,][]{Dessart2009,Perego2014,Martin2015}, BH-NS mergers only have a limited amount of neutrino-driven ejecta. \cite{Just2015} presented that there is only $\sim 1\%$ disk mass around the remnant BH that can contribute to the neutrino-driven ejecta. In our calculation, we assume that the neutrino-driven ejecta is a constant fraction of the remnant disk mass, i.e., $M_{\rm n} = \xi_{\rm n}M_{\rm disk}$ with $\xi_{\rm n}\sim 0.01$. \cite{Martin2015} indicated that the neutrino-driven ejecta is a polar emission which has a rather uniform mass distribution, i.e., $F_{\rm n}(\theta) \approx {\rm const}$, ${\rm for}\ \theta \le\theta_{\rm n}$, {where $\theta_{\rm n}$ is the half opening angle of the neutrino-driven ejecta in the latitudinal direction}. The normalization of $M_{\rm n}$ is
\begin{equation}
\label{Eq. Normalization Neutrino-driven ejecta 1}
    M_{\rm n} = 2\int_0^{\theta_{\rm n}}\frac{M_{\rm n}}{2(1 - \cos\theta_{\rm n})}\sin\theta{\rm d}\theta.
\end{equation}
We note that the latitudinal angle $\theta$ of the dynamical ejecta and the latitudinal angle $\theta$ for the neutrino-driven ejecta and viscosity-driven ejecta are defined differently. The former is defined as the angle with respect to the $v_{x}-v_{y}$ plane, while the latter is defined as the angle from the $\boldsymbol{v_{z}}$ direction.

\cite{Wollaeger2018} indicated that for a spherically symmetric radiation-dominated outflow with {adiabatic index} $\Gamma = 4/ 3$, its expansion profile is ${\rm d}m/{\rm d}v \propto (1 - (v / v_{\max}) ^ 2) ^ 3$. We then simply define the neutrino-driven ejecta mass distribution in the radial direction according to the similar  profile, i.e., ${\rm d}m_{\rm n}/{\rm d}v \propto (1 - (v / v_{{\max},{\rm n}}) ^ 2) ^ 3$, where ${\rm d}m_{\rm n}$ is the neutrino-driven ejecta mass of the material with a certain $\theta$ and $\varphi$, and $v_{\max , {\rm n}}$ is its maximum velocity. Also, ${\rm d}m_{\rm n}/{\rm d}v$ can be expressed as
\begin{equation}
\label{Eq. Normalization Neutrino-driven ejecta 2}
    2\pi\int_0^{v_{\max , {\rm n}}}\frac{{\rm d}m_{\rm n}}{{\rm d}v}{\rm d}v = \frac{{\rm d}M_{\rm n}}{\sin\theta{\rm d}\theta}.
\end{equation}
Combining Equation (\ref{Eq. Normalization Neutrino-driven ejecta 1}) and Equation (\ref{Eq. Normalization Neutrino-driven ejecta 2}), one gets the mass per velocity with a certain $\theta$ and $\varphi$ 
\begin{equation}
\label{Eq. Normalization Neutrino-driven ejecta 3}
    \frac{{\rm d}m_{\rm n}}{{\rm d}v} = \frac{35M_{\rm n}}{64\pi(1 - \cos\theta_{\rm n})v_{{\max} , \rm n}}\left[ 1 - \left( \frac{v}{v_{{\max} , \rm n}} \right) ^2 \right] ^3.
\end{equation}

Another normalization of $M_{\rm n}$ is
\begin{equation}
\label{Eq. Normalization Neutrino-driven ejecta 4}
    2\times2\pi\int_{0}^{\theta_{\rm n}}\int_{0}^{v_{{\max} , \rm n}}\rho_{\rm n}(v , t) v^2t^3\sin\theta{\rm d}v{\rm d}\theta = M_{\rm n},
\end{equation}
where $\rho_{\rm n}$ is the density of the neutrino-driven ejecta and ${\rm d}V = r^2\sin\theta{\rm d}r{\rm d}\theta{\rm d}\varphi$ is the volume element. The density can be calculated by combining Equation (\ref{Eq. Normalization Neutrino-driven ejecta 2}), Equation (\ref{Eq. Normalization Neutrino-driven ejecta 3}) and Equation (\ref{Eq. Normalization Neutrino-driven ejecta 4}):
\begin{equation}
    \rho_{\rm n}(v , \theta , t) = \frac{{\rm d}m_{\rm n}}{v^2t^3{\rm d}v} = \frac{35M_{\rm n}}{64\pi(1 - \cos\theta_{\rm n})v_{{\max} , \rm n}v^2t^3}\left[ 1 - \left( \frac{v}{v_{{\max} , \rm n}} \right) ^2 \right] ^3.
\end{equation}

It is easy to calculate the root-mean-square (rms) velocity:
\begin{equation}
    v_{\rm rms,n} = \sqrt{\frac{2}{M_{\rm n}}\int\frac{1}{2}\rho_{\rm n}v^2{\rm d}V} = \frac{1}{3}v_{\max , {\rm n}}.
\end{equation}
Hereafter we define $v_{\max , {\rm n}} = 3v_{\rm rms,n}$. The simulation results of \cite{Martin2015} presented the rms velocity of neutrino-driven ejecta to be mainly in the range of $v_{\rm rms,n}\simeq0.055-0.075\,c$, while the best fitting results for the lightcurve of AT2017gfo from \cite{Perego2017} showed the rms velocity as $v_{\rm rms,n}\approx0.0667\,c$. Based on these results, we directly set $v_{\rm rms,n} = 0.0667\,c$ in our calculations.

\subsubsection{Viscosity-Driven Wind Ejecta}
Simulation results indicated that the fraction mass of the viscosity-driven ejecta in the remnant disk mass is determined by the spin of the remnant BH. It can range from $\sim5\%$ for a low-spin BH to $\sim30\%$ for a high-spin BH with the dimensionless spin parameter $\chi_{\rm BH} \simeq 0.95$  \citep{Just2015,Fernandez2015,Fujibayashi2020}. As for the BH-NS mergers, the spin of the remnant BH is mainly dependent on the spin of the initial BH and the mass ratio between the BH and the NS \citep{Kyutoku2011,Zappa2019}. An NS can easily plunge into an anti-aligned BH which can hardly form a remnant disk and tidal dynamical ejecta. \cite{Pannarale2013} and \cite{Zappa2019} found that the spin distribution of the remnant BH is in the range of $\chi_{\rm BH} \simeq 0.6-0.9$ if the spin of the initial BH is aligned with (i.e., $\chi_{\rm BH} \geq 0$) the orbital angular momentum\footnote{This result is similar to the case of remnant BHs in NS-NS mergers \citep{Kiuchi2009}.}. We therefore simply set the viscosity-driven ejecta mass as a constant fraction of the disk mass $M_{\rm v} = \xi_{\rm v}M_{\rm disk}$, where $\xi_{\rm v} = 0.2$ \citep{Fernandez2015,Fujibayashi2020,Just2015,Siegel2017}. \cite{Perego2017} assumed an equatorial-dominated outflow of the viscosity-driven ejecta $F_{\rm v}(\theta) = \sin^2\theta$ based on simulations \citep{Wu2016,Lippuner2017,Siegel2017}. We still use this mass distribution to model the viscosity-driven ejecta. For the total mass, the normalization with $\theta$ is
\begin{equation}
      M_{\rm v}\propto\int_0^{\pi/2}F_{\rm v}(\theta)\sin\theta{\rm d}\theta,
\end{equation}
Therefore, one has
\begin{equation}
    \frac{{\rm d}M_{\rm v}}{{\rm d}\theta} = \frac{3}{4}M_{\rm v}\sin^3\theta.
\end{equation}
Similar to the neutrino-driven ejecta, we define ${\rm d}m_{\rm v}$ as the mass of the material with a certain $\theta$ and $\varphi$. The mass per velocity at certain $\theta$ and $\varphi$ is ${\rm d}m_{\rm v}/{\rm d}v \propto (1 - (v / v_{{\max} , \rm v}) ^ 2) ^ 3$, where $v_{{\max} , \rm v} = 3v_{\rm rms , v}$. The simulation results from \cite{Just2015} showed that the rms velocity of the viscosity-driven ejecta ejected from the BH remnant disk mainly lies in the range of $v_{\rm rms,v}\sim0.03-0.04\,c$. Hereafter, we directly assume  $v_{\rm rms , v} = 0.03\,c$. The normalization can be also expressed as
\begin{equation}
\label{Eq. Normalization Vis ejecta 2}
    2\pi\int_0^{v_{{\max} , \rm v}} \frac{{\rm d}m_{\rm v}}{{\rm d}v}{\rm d}v = \frac{{\rm d}M_{\rm v}}{\sin\theta{\rm d}\theta}.
\end{equation}
Therefore, we have
\begin{equation}
    \frac{{\rm d}m_{\rm v}}{{\rm d}v} = \frac{105M_{\rm v}\sin^2\theta}{128\pi v_{{\max} , \rm v}}\left[ 1 - \left( \frac{v}{v_{{\max} , \rm v}} \right) ^2 \right] ^3.
\end{equation}
The density can be easily calculated as
\begin{equation}
    \rho_{\rm v}(v , \theta , t) = \frac{{\rm d}m_{\rm v}}{v^2t^3{\rm d}v} = \frac{105M_{\rm v}\sin^2\theta}{128\pi v_{{\max} , \rm v}v^2t^3}\left[ 1 - \left( \frac{v}{v_{{\max} , \rm v}} \right) ^2 \right] ^3.
\end{equation}

\subsubsection{Temperature Evolution Model}

\begin{figure}[tbp]
    \centering
    \includegraphics[width=0.60\textwidth, trim=50 70 145 60, clip]{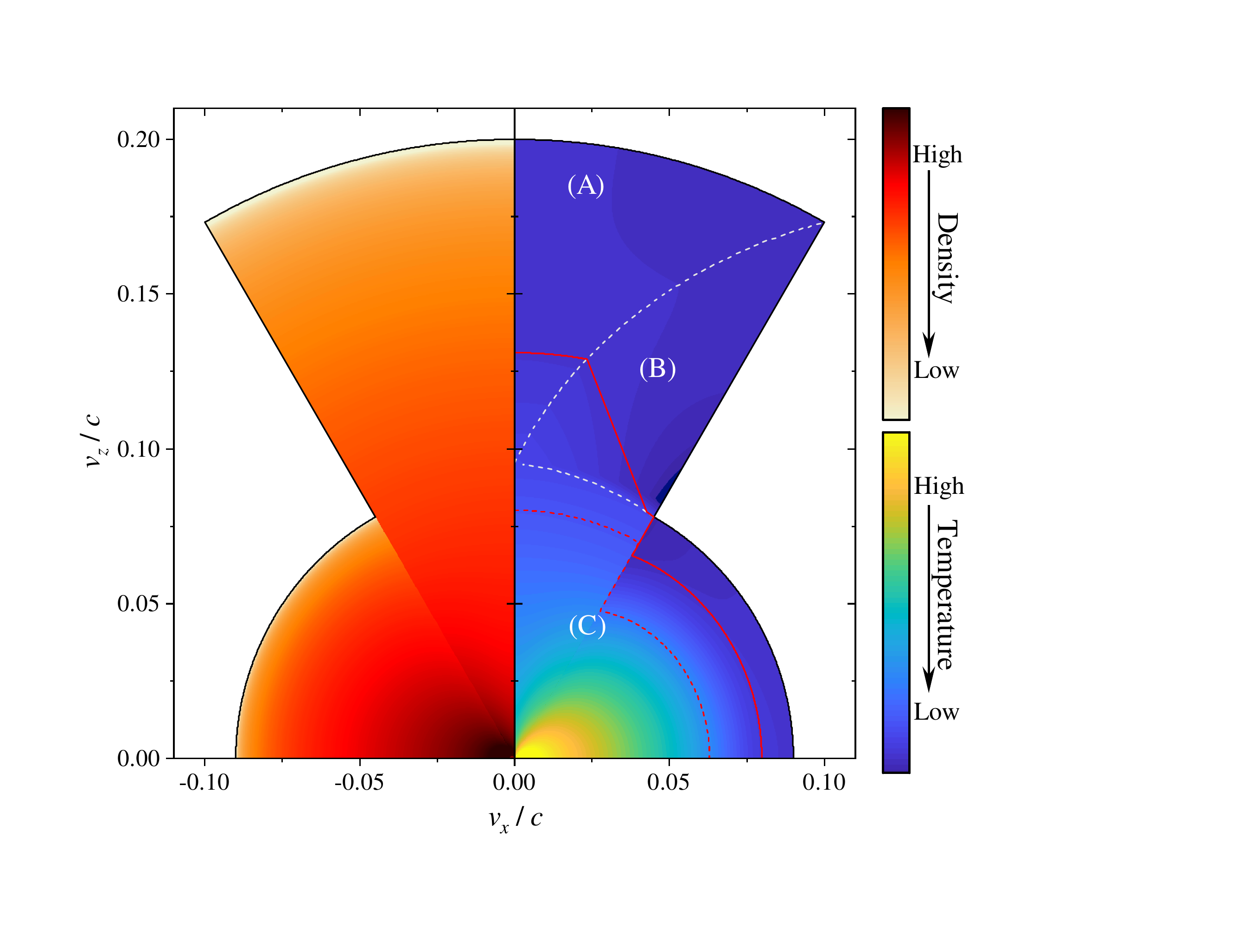}
    \caption{Schematic diagram of the wind ejecta, showing a sectional drawing in the $v_x-v_z$ plane. The color scheme in the left panel qualitatively depicts the density of the wind ejecta, while that in the right panel depicts the thermal temperature. The gray dashed lines divide the wind ejecta into Three regions. The red solid and dashed lines respectively represent the photosphere and photon diffusion surface in the early evolution stage. For Regions A and C, the temperature gradient is along the radial direction. For Region B, the temperature gradient in the direction perpendicular to the edge of the wind ejecta.}
    \label{fig:WindEjectaRegion}
\end{figure}

In the following, we present a similar method as the dynamical ejecta model to delinate the temperature distribution of the two-component wind ejecta.

First, we assume that the radiation direction and temperature gradient are along  the radial direction for the two-component wind ejecta. For the neutrino-driven ejecta, the optical depth along the radial direction is $\tau_{\rm n}(v) \approx \int_v^{v_{\max,{\rm n}}}\kappa_{\rm n}\rho_{\rm n} t{\rm d}v$. Photons will escape if $(v_{\max , {\rm n}} - v_{\rm diff,n})t \approx ct/\tau_{\rm n}(v_{\rm diff,n})$, where $\tau_{\rm n}(v_{\rm diff,n}) \approx \int_{v_{\rm diff,n}}^{v_{\max,{\rm n}}}\kappa_{\rm n}\rho_{\rm n} t{\rm d}v$, and $v_{\rm diff,n}$ is the diffusion velocity for  photons to escape. One can then obtain $v_{\rm diff,n}(t)$ at a given time. Therefore, the radial comoving luminosity is
\begin{equation}
    {\rm d}L'_{\rm n} = \frac{{\rm d}L_{\rm n}}{\mathcal{D} ^ 2} = \dot{\epsilon}(t){\rm d}M_{\rm n},
\end{equation}
where ${\rm d}M_{\rm n} = \rho_{\rm n}v^2t^3\sin\theta{\rm d}v{\rm d}\theta{\rm d}\varphi$, $\mathcal{D} = 1 / [\Gamma(1 - \beta)]$ is Doppler factor and $\Gamma = 1/\sqrt{1 - \beta ^ 2}$. Similar to dynamical ejecta, one can calculate the total emissivity per unit area $D_{\rm n}$. Since ${\rm d}L_{\rm n}$ is released over an area of $v_{\rm phot,n}^2t^2\sin\theta{\rm d}\theta{\rm d}\varphi$, where the photosphere position $\tau_{\rm n}(v_{\rm phot,n}) = 2 / 3$, one gets
\begin{equation}
    D_{\rm n}(t) = \frac{\dot{\epsilon}(t)t\int_{v_{\rm diff,n}}^{{v_{\max,{\rm n}}}}\mathcal{D} ^ 2\rho_{\rm n}v^2{\rm d}v}{v_{\rm phot,n}^2}.
\end{equation}
The effective temperature {of the neutrino-driven ejecta} is
\begin{equation}
    T_{\rm eff,n}(t) = \left(\frac{D_{\rm n}(t)}{\sigma_{\rm SB}}\right) ^ {1 / 4}.
\end{equation}
With the Eddington approximation \citep{Mihalas1970,Rybicki1979}, the thermal temperature of each point in the velocity space of the neutrino-driven ejecta is given by
\begin{equation}
\label{Eq.Boundary2}
    T_{\rm n}(v , t) = T_{\rm eff,n}(t)\left[ \frac{3}{4}\left(\tau_{\rm n}(v , t) + \frac{2}{3} \right) \right]^{1 / 4}.
\end{equation}

The temperature gradient of the viscosity-driven ejecta is along the radial direction similar to the neutrino-driven ejecta. The optical depth of the viscosity-driven ejecta is $\tau_{\rm v}(v) \approx \int_v^{v_{\max,{\rm v}}}\kappa_{\rm v}\rho_{\rm v} t{\rm d}v$. Photons will escape if $(v_{\max , {\rm v}} - v_{\rm diff,v})t \approx ct/\tau_{\rm v}(v_{\rm diff,v})$, where $\tau_{\rm v}(v_{\rm diff,v}) \approx \int_{v_{\rm diff,v}}^{v_{\max,{\rm v}}}\kappa_{\rm v}\rho_{\rm v} t{\rm d}v$. One can obtain $v_{\rm diff,v}(\theta , t)$ as a function of time. We define the total emissivity per unit area $D_{\rm v}$ as
\begin{equation}
    D_{\rm v}(t) = \frac{\dot{\epsilon}(t)t\int_{v_{\rm diff,v}}^{{v_{\max,{\rm v}}}}\mathcal{D} ^ 2\rho_{\rm v}v^2{\rm d}v}{v_{\rm phot,v}^2},
\end{equation}
where $v_{\rm phot,v}$ is for photosphere position where $\tau_{\rm v}(v_{\rm phot, v}) = 2 / 3$.
The effective temperature of the viscosity-driven ejecta is
\begin{equation}
    T_{\rm eff,v}(\theta , t) = \left(\frac{D_{\rm v}(\theta , t)}{\sigma_{\rm SB}}\right) ^ {1 / 4}.
\end{equation}
The thermal temperature of each point in the velocity space of the viscosity-driven ejecta can be also derived by
\begin{equation}
    T_{\rm v}(v , \theta , t) = T_{\rm eff,v}(\theta , t)\left[ \frac{3}{4}\left(\tau_{\rm v}(v , \theta , t) + \frac{2}{3} \right) \right]^{1 / 4}.
\end{equation}

Based on the above assumptions, there is an overlapping region between the neutrino-driven ejecta and the viscosity-driven ejecta in the parameter regime of $v \leq v_{\rm max,v}$ and $0 \leq \theta \leq \theta_{\rm n}$ or $(\pi / 2 - \theta_{\rm n}) \leq \theta \leq \pi / 2$. For simplicity, we directly overlay the density of the two-component wind ejecta in this region, but consider that each point of this region has a single temperature. The viscosity-driven ejecta is equator-dominated so that the mass of the overlapping portion accounts for only a small part of the viscosity-driven ejecta. Therefore, this assumption has a slight effect on our final results. A schematic diagram of the density and temperature is present in Figure \ref{fig:WindEjectaRegion}.

So far we have assumed that the temperature gradient direction is along the radial direction for both the neutrino-driven ejecta and viscosity-driven ejecta. However, photons can also escape from the side of the neutrino-driven ejecta. As shown in Figure \ref{fig:WindEjectaRegion}, we divide the wind ejecta into three regions. We define the temperature gradient of Region A and Region C as along the radial direction, while the temperature gradient of Region B along the direction perpendicular to the edge of the neutrino-driven wind. The boundaries of two light gray dashed lines are, respectively,
\begin{equation}
\begin{split}
     v_{\rm bou1} &\approx v_{\rm max,n}(\theta - \theta_{\rm n} + 1), \\
     v_{\rm bou2} = v_{\rm max,v}&\left( \frac{\theta}{\theta_{\rm n}} \right)  + v_{\rm max,n}(1 - \theta_{\rm n})\left( 1 - \frac{\theta}{\theta_{\rm n}} \right).
\end{split}
\end{equation}
The upper boundary can be derived by comparing the diffusion time similar to the dynamical ejecta. The lower boundary is our hypothetical boundary. Both the photon diffusion surface and the photosphere are nearly continuous at the boundaries.

Using the similar method to  deal with the temperature evolution of the dynamical ejecta, we can calculate the thermal temperature, i.e., $T_{\rm n,bou1}(v_{\rm bou1} , \theta , t)$ and $T_{\rm n,bou2}(v_{\rm bou2} , \theta , t)$, located at the two boundaries by Equation (\ref{Eq.Boundary2}). The optical depth in the direction perpendicular to the edge of the wind is $\tau_{\rm n,side}(v,\theta,t) = \kappa_{\rm n}\rho_{\rm n}(\theta_{\rm n} - \theta)vt$. Using the Eddington approximation, one can obtain the thermal temperature of Region B of the wind ejecta.

\section{Results\label{sec:4}}

\begin{figure}[tbp]
    \centering
    \includegraphics[width=0.40\textwidth, trim=270 160 270 150, clip]{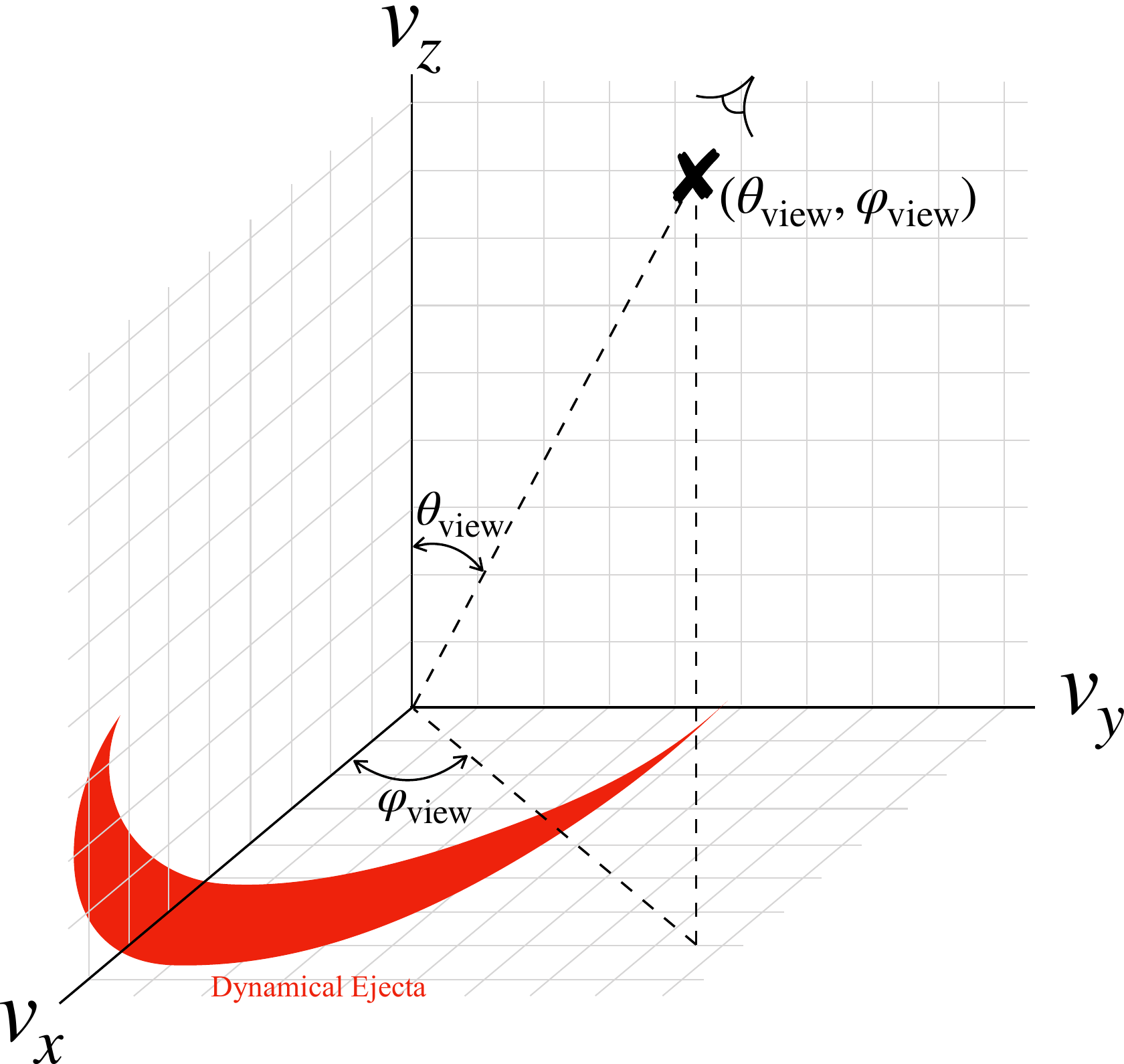}
    \caption{The local coordinate system of the ejecta components. The red crecsent represents the profile of the dynamical ejecta. We set the $v_x-v_z$ plane to divide the dynamical ejecta equally and symmetrically, with the $v_z-$axis aligned with the jet axis. The line of sight forms a latitudinal viewing angle $\theta_{\rm view}$ with respect to the $v_z-$axis and a longitudinal angle $\varphi_{\rm view}$ with respect to the $v_x-v_z$ plane.}
    \label{fig:CoordinateSystem}
\end{figure}

With the above preparation, in this section we investigate the  evolution of the temperature profile of the ejecta and the observed photosphere emission. We will calculate the evolution of the emergent spectra and lightcurves as a function of viewing angle using a spatial discretization model as described in detail in Appendix \ref{app:A}. In order to discuss the viewing angle effect, we define a spherical coordinate system with the $v_z-$axis aligned with the jet axis and the $v_x-v_z$ plane divide the dynamical ejecta equally and symmetrically. As shown in Figure \ref{fig:CoordinateSystem}, the line of sight forms a latitudinal viewing angle $\theta_{\rm view}$ with respect to the $v_z-$axis and a longitudinal angle $\varphi_{\rm view}$ with respect to the $v_x-v_z$ plane. The range of these two viewing angles are $\theta_{\rm view} \in [0 , \pi / 2]$ and $\varphi_{\rm view} \in [0 , \pi]$, respectively.

As discussed in Section \ref{sec:2}, the remnant disk mass, the dynamical ejecta mass and the rms velocity of the dynamical ejecta depend on the following parameters: the mass ratio (or the BH mass), the dimensionless spin of the BH, the NS (gravitational) mass, the NS baryonic mass, and the compactness of the NS. The NS baryonic and gravitational masses are related through the compactness parameter \cite[e.g.][]{Coughlin2017,Gao2020} \begin{equation}
    M^{\rm b}_{\rm NS} = M_{\rm NS}\left( 1 + 0.8858C_{\rm NS}^{1.2082}\right),
\end{equation}
so that there are only four independent parameters. The masses of the neutrino-driven ejecta and the viscosity-driven ejecta can be described as  constant fractions of the total mass of the remnant disk. Therefore, by setting four parameters, one can calculate all the input parameters (the neutrino-driven ejecta mass, the viscosity-driven ejecta mass, the dynamical ejecta mass, and the rms velocity of the dynamical ejecta) needed to calculate kilonova emission. In the following, we give two example cases including a small mass remnant and a large mass remnant outside of the BH. The relevant four parameters in the two cases are summarized in Table \ref{tab:list2}. Hereafter, we mark Case I and Case II corresponding to the small remnant mass case and the large remnant mass case, respectively. At the end of this section, we also extend our results to different mass ratio regimes.

\begin{deluxetable*}{cccccccccc}
\tablecaption{Input Parameters for the Two Example Cases \label{tab:list2}}
\tablecolumns{10}
\tablenum{2}
\tablewidth{0pt}
\tablehead{
\colhead{Case} & 
\colhead{Q} &
\colhead{$M_{\rm BH}/M_\odot$} &
\colhead{$\chi_{\rm BH}$} & 
\colhead{$M_{\rm NS}/M_\odot$} & 
\colhead{$C_{\rm NS}$} & 
\colhead{$M_{\rm d} / M_\odot$} &  
\colhead{$M_{\rm n} / M_\odot$} &
\colhead{$M_{\rm v} / M_\odot$} &
\colhead{$v_{\rm rms,d} / c$}
}
\startdata
{\rm I} & 5 & 6.75 & 0.75 & 1.35 & 0.180 & 0.014 & $6.53\times10^{-4}$ & 0.013 & 0.24 \\ 
{\rm II} & 5 & 6.75 & 0.75 & 1.35 & 0.130 & 0.069 & $2.52\times10^{-3}$ & 0.050 & 0.24 \\ 
\enddata
\tablecomments{We list the mass ratio $Q$, the BH mass $M_{\rm BH}$, the dimensionless spin of the BH $\chi_{\rm BH}$, the NS (gravitational) mass $M_{\rm NS}$, the compactness of the NS $C_{\rm NS}$, the dynamical ejecta mass $M_{\rm d}$, the neutrino-driven ejecta mass $M_{\rm n}$, the viscosity-driven ejecta mass $M_{\rm v}$ and the rms velocity of the dynamical ejecta $v_{\rm rms,d}$. The two cases I and II correspond to the small remnant mass case and the large remnant mass case, respectively.}
\end{deluxetable*}

\subsection{Temperature profile evolution}\label{sec:4.1}

\begin{figure}[htbp]
\centering
\includegraphics[width = 0.49\linewidth]{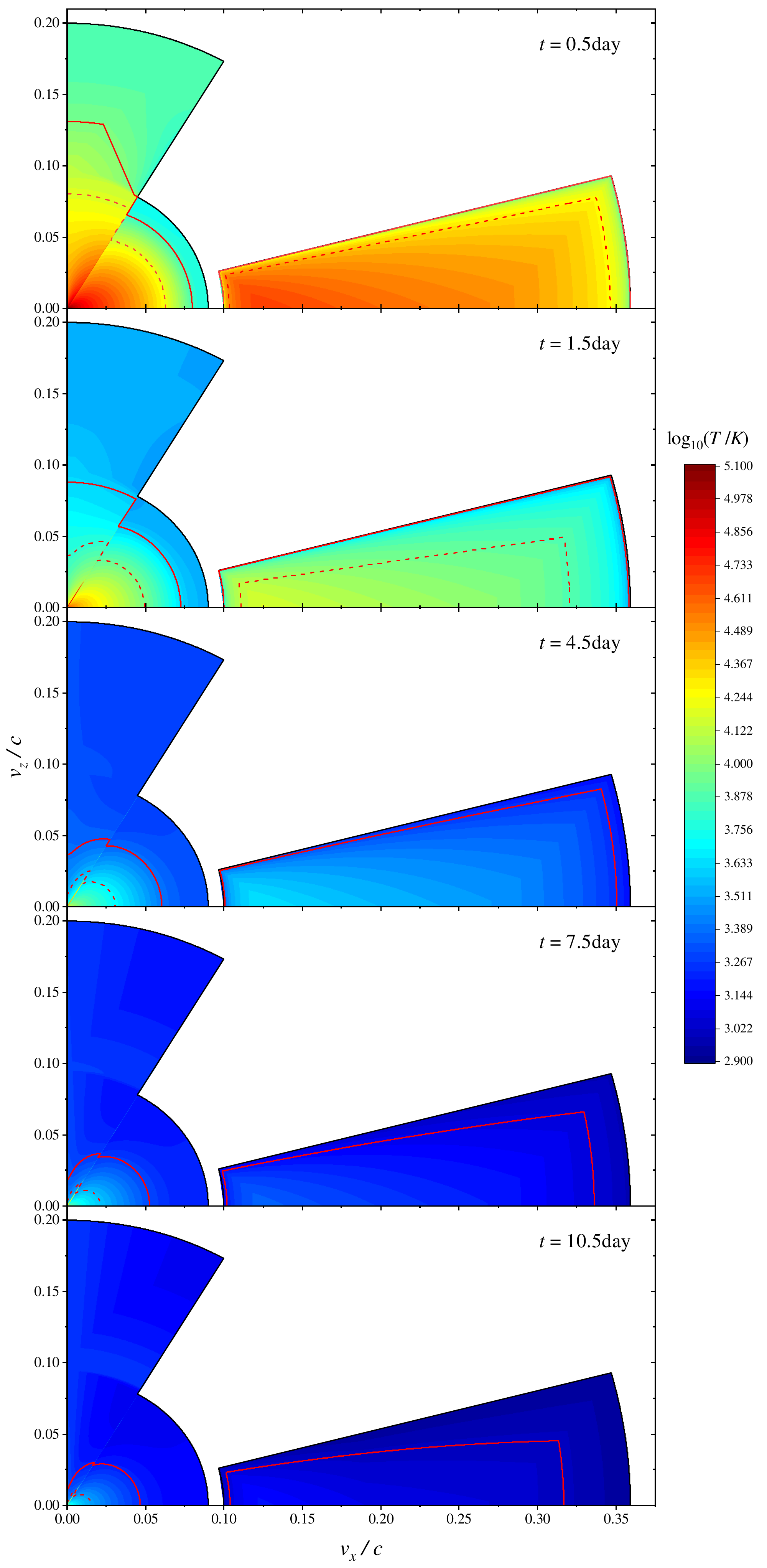}\quad\includegraphics[width = 0.49\linewidth]{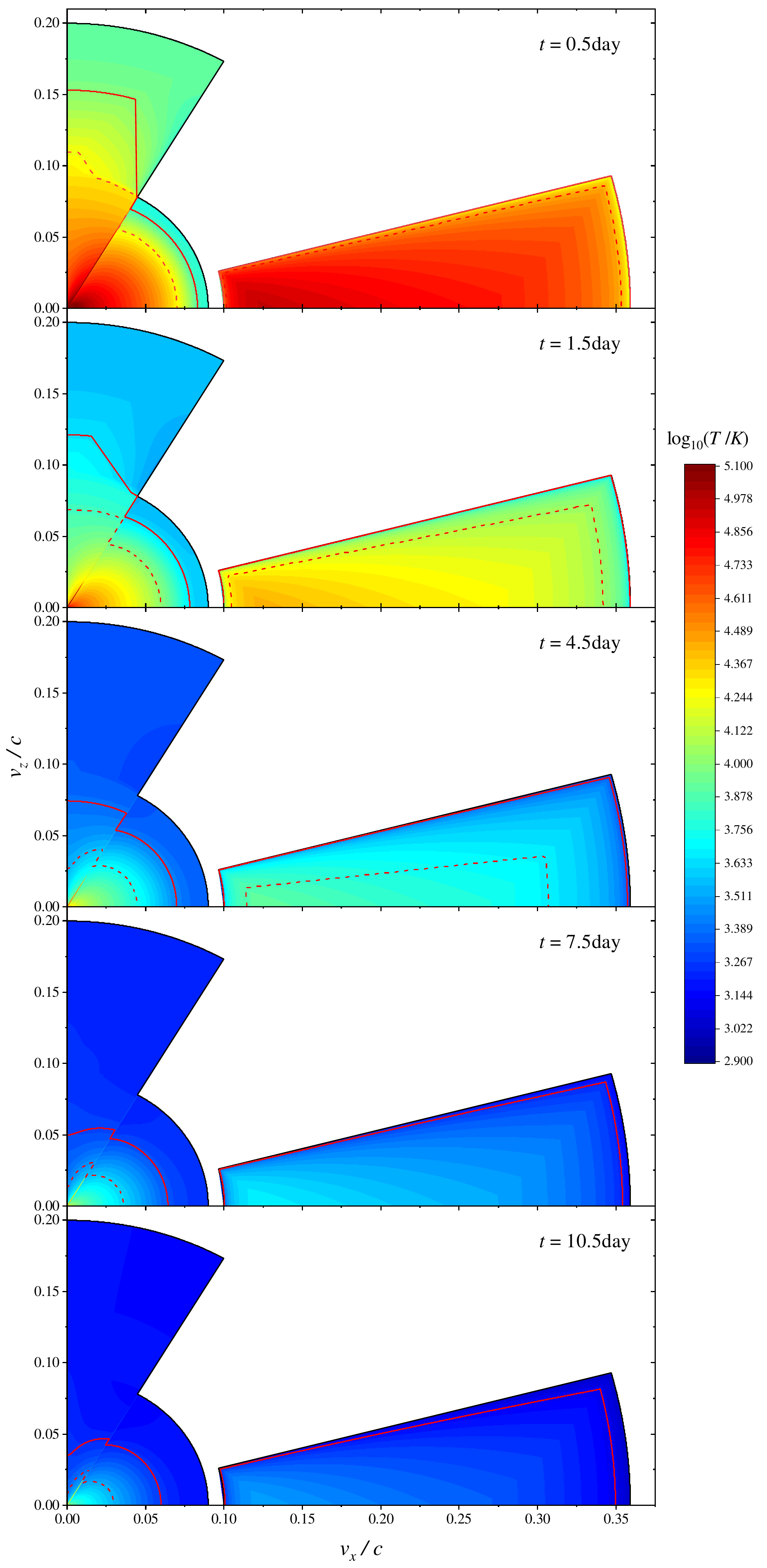}
\caption{Sectional drawings of the temperature profile evolution for Case I (left panels) and Case II (right panels) at $t = 0.5$, $1.5$, $4.5$, $7.5$ and $10.5$ days after the merger. The red solid lines and red dashed lines represent photosphere and the photon diffusion surface in the rest frame of the source, respectively.\label{fig:TemperatureEvolution}} 
\end{figure}

As discussed in Section \ref{sec:3}, we assume a homologous expansion of the mass shells for each velocity and specific mass distribution and model the dynamics evolution of each component ejecta. We also assume that the kilonovae of BH-NS mergers are powered only by the radioactive decay of $r$-process nuclei. In Figure \ref{fig:TemperatureEvolution}, we show the results of the evolution of the ejecta temperature profiles Case I and Case II, respectively. The evolution of the photosphere (red solid lines) and photon diffusion surface (red dashed lines) in the rest frame are shown in both cases. 

Consistent with intuition, Case II has a larger mass outside of the remnant BH and therefore has a higher temperature profile compared with Case I. In both cases, at the beginning of the merger, the ejecta cool rapidly due to the rapid expansion of the ejecta (the size of the ejecta increases by orders of magnitude in a short duration). At later times, the cooling slows down since the relative expansion is smaller (the size only increases linearly with time). This can be clearly seen in Figure \ref{fig:TemperatureEvolution}.

For the two-component BH disk wind ejecta, the matter is mainly concentrated in the region where the velocity is smaller than the rms velocity. The photosphere and the photon diffusion surface evolve quickly at the beginning of the emission due to the density distribution and rapid density change caused by the initial expansion. Compared with Case II, the photosphere and photon diffusion surface of Case I can penetrate deeper into the ejecta at the same time after the merger. As these two surfaces pass through the low density part of the ejecta, their evolution rates gradually slow down and they are close to no evolution at $\sim10\ {\rm days}$. The photosphere cannot penetrate into the central matter of BH disk wind ejecta,  indicating that it cannot become completely optical thin. In addition to the photosphere, the photon diffusion surface cannot penetrate the entire ejecta even for the small mass remnant case. The emission of a large portion of matter below the photon diffusion surface cannot contribute to the luminosity of the kilonova. 

Different from the BH disk wind ejecta, the density distribution of the dynamical ejecta is relatively homogeneous. As a result, the evolution rate of the photosphere and photon diffusion surface does not show the tendency of slowing down. The photon diffusion surface spends $\sim 4\,$days and $\sim 7\,$days passing through the entire dynamical ejecta for the Case I and II, respectively. After this, the energy deposited in the entire ejecta can contribute to the luminosity of the kilonova. On the other hand, the dynamical ejecta can hardly become completely optically thin within $10.5\,$days after the merger. This is especially true for Case II, in which the evolution rate of the photosphere is so slow  that it stays at the front edge of the dynamical ejecta at early times after the merger. For Case I, the photosphere can quickly penetrate into the ejecta but the photosphere never completely penetrate the ejecta within $10.5\,$days. For both cases, the low velocity region of the dynamical ejecta has higher density and temperature, indicating that this region contributes more to the observed luminosity.

Some caveats are worth mentioning. For simplicity, we model the temperature profile evolution with a constant gray opacity approximation. The opacities of a mixture of $r$-process elements are strongly temperature- and wavelength-dependent \citep[e.g.,][]{Kasen2013,Tanaka2013,Tanaka2017}. \cite{Tanaka2019} found that the gray opacities are nearly constant for temperatures $T = 5-10\times 10^3\,{\rm K}$ and  decease steeply at lower temperatures. {However, a steep decrease of the gray opacities significantly occurs at the temperature $T \lesssim 1500\,{\rm K}$. The fitting temperature of the emergent spectrum, shown in the following sections (see Figure \ref{fig:ViewingAngleSpectrum}), cool to $\sim1500 \,{\rm K}$ at $t\sim7-10\,{\rm day}$ after the merger.} This means that the photosphere can penetrate deeper into the ejecta at late times and the ejecta may become completely optically thin at late times. We also use a simple blackbody approximation for our model. For more sophisticated simulations, one should also consider temperature ``floor'' behavior. For example, \cite{Barnes2013} found that the the {effective} temperature remains unchanged as the photosphere recedes when lanthanides-rich ejecta cool to the first ionisation temperature of the lanthanides ($T_{\rm La} \approx 2500\, {\rm K}$).

\subsection{Photosphere Evolution in the Observer Frame}

\begin{figure}[htbp]
\centering
\includegraphics[width = 0.49\linewidth]{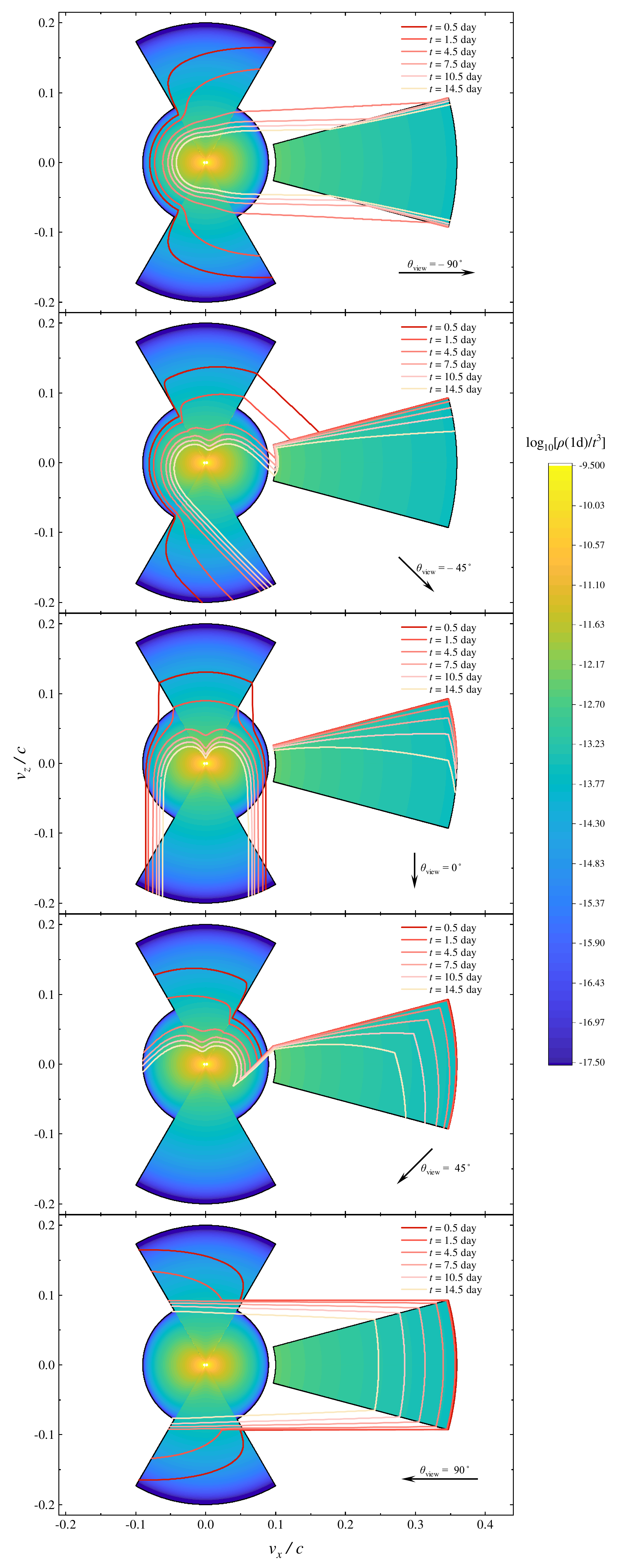}\quad\includegraphics[width = 0.49\linewidth]{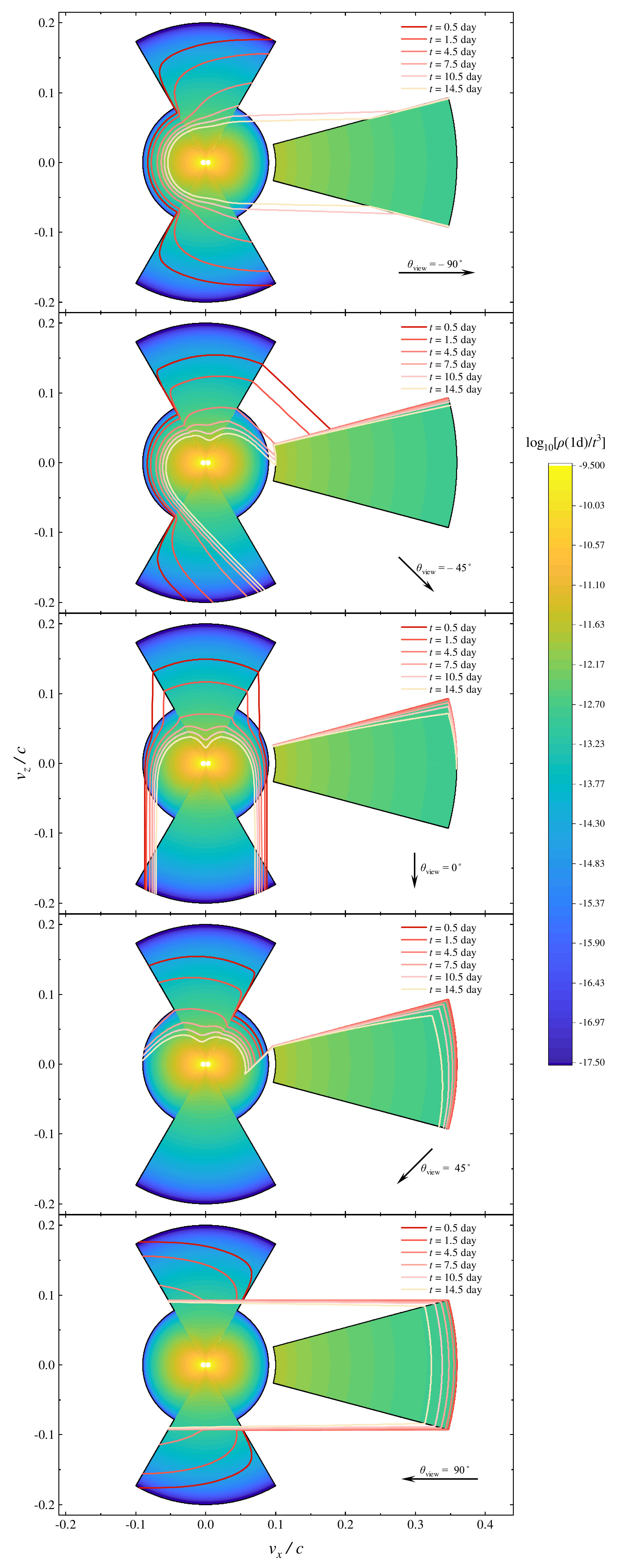}
\caption{Sectional drawings of the density profile and the photosphere evolution for observers in different viewing directions. Case I and II are presented in the left and right panels, respectively. From the top to bottom, the panels show the shape of the photosphere for viewing angle $\theta_{\rm view} = -90^\circ$, $-45^\circ$, $0^\circ$, $45^\circ$ and $90^\circ$, respectively. The color lines from dark to light represent the photosphere contour at $0.5$, $1.5$, $4.5$, $7.5$, $10.5$ and $14.5\,$days, respectively. \label{fig:PhotosphereEvolution}}
\end{figure}

The calculations presented in Section \ref{sec:4.1} are presented in the rest frame of the source, also called the laboratory frame. The photosphere defined there does not reflect the photosphere of a certain observer. The photosphere in the frame of an observer is defined by the optical depth in a particular direction, so that it is viewing angle dependent. One should also consider the light propagation effect, such that the photosphere is defined at ``retarded'' times for the same observational time, i.e. emission comes from the so-called equal-arrival-time surface. These effects are fully considered in our calculations (see Appendix \ref{app:A}).

Figure \ref{fig:PhotosphereEvolution} shows the 2D sectional drawings of the photosphere evolution in the observer frame in velocity space, for the two cases (Case I left and Case II right) with different viewing angles. We take $\varphi_{\rm view}$=0 and adopt five values of the viewing angle $\theta_{\rm view}$, i.e., $\theta_{\rm view} = -90^\circ$, $-45^\circ$, $0^\circ$, $45^\circ$, $90^\circ$.

The photosphere in the observer frame also evolves rapidly in the beginning and slows down later, becoming very slow after $\sim 10$ days of merger. For all the directions, the photosphere as seen by the observer cannot pass through the central matter of BH disk wind ejecta due to its high density. The matter behind the core is also blocked. Therefore, an observer cannot see emission from all the ejecta during the entire evolution period of kilonova emission. 

Within $14.5$ days {(which is the  timescale of our calculation)}, the photosphere in the frame of any observer also cannot pass through the entire dynamical ejecta. However, compared with the photosphere in the source frame, the photosphere in the observer frame continuously evolve, with the evolution rate not slowing down with time. For a long enough time, one may finally see the entire dynamical ejecta matter. A special case is for $\theta_{\rm view} = 90^\circ$. The wind ejecta will be hidden from view by the dynamical ejecta. 
The contribution of the wind ejecta emission can be essentially ignored after the neutrino-driven ejecta becomes optically thin. The observer can only see the emission from the dynamical ejecta.

Since the half opening angle in the latitudinal direction of the dynamical ejecta is really thin, there is a significant change of  the projected photosphere area for observers with different lines of sight. Obviously, the face-on projected photosphere area would be larger than those at large latitudinal viewing angles $\theta_{\rm view}$. The projected photosphere area variation with latitudinal viewing angle is roughly a factor of $\sim (1.5 - 3)$. The change of projected  photosphere area would affect the kilonova luminosity. Also, the projected photosphere shape is highly asymmetric. Both change of the area and the change of the projected photosphere shape would have effects on the polarization of the kilonova \citep[e.g.,][]{Li2019,Shapiro1982}, which we do not study in this work.

Our calculation results of the photosphere evolution in the observer frame are again based on the constant gray opacity approximation. As we mention before, the photosphere properties can be also affected by the steeply decreasing of gray opacity when the temperature drops below $\sim5000\ {\rm K}$ \citep{Tanaka2019}. The actual evolution of the observer-frame photophere emission at late times could be more complicated.

\subsection{Evolution of the Emergent Spectra \label{sec:4.3}}

\begin{figure}[htbp]
\centering
\includegraphics[width = 0.43\linewidth , trim = 55 70 80 55, clip]{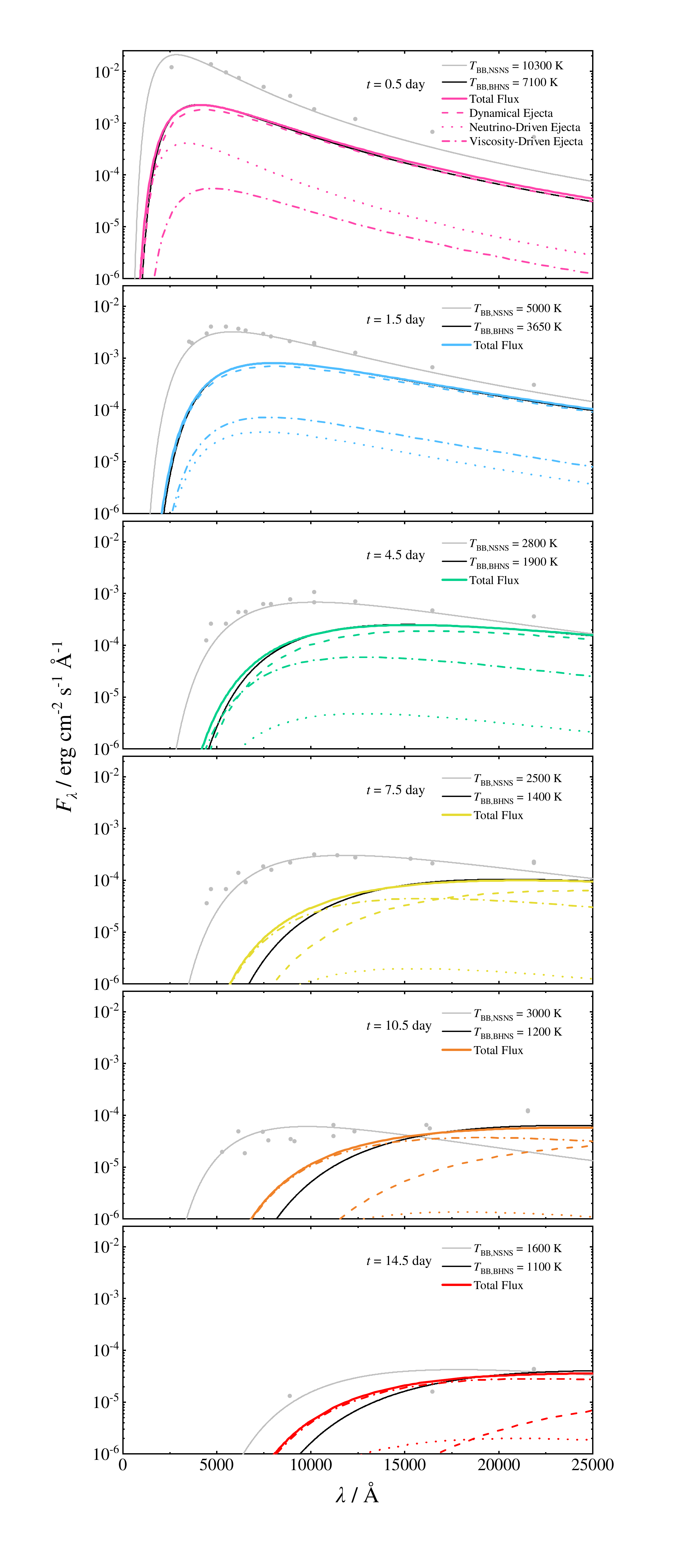}\quad\includegraphics[width = 0.43\linewidth , trim = 55 70 80 55, clip]{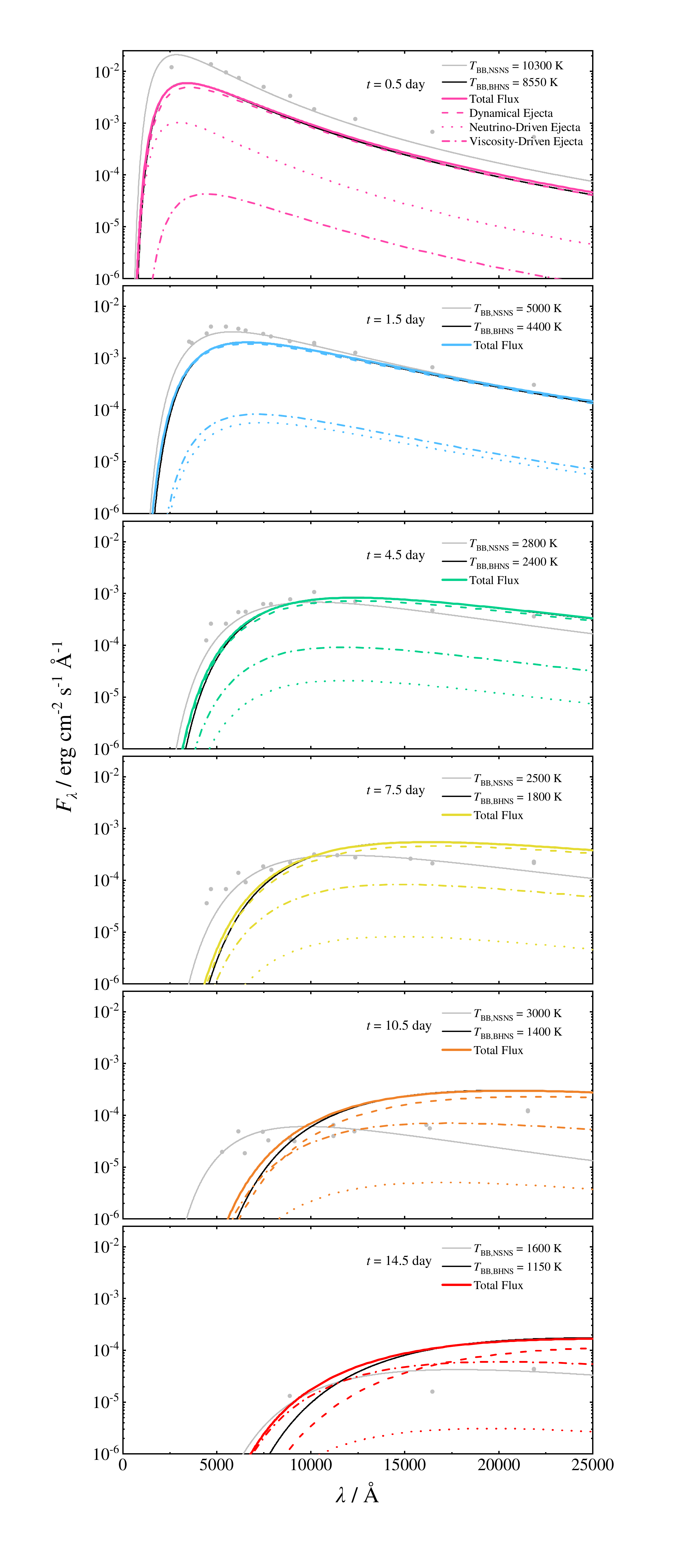}
\caption{Examples of face-on ($\theta_{\rm view} = 0^
\circ$, $\varphi_{\rm view} = 0^\circ$) emergent spectra for  Case I (left panels) and Case II (right panels), where $D_{\rm L} = 10\ {\rm pc}$ is adopted. The panels from top to bottom show six epochs after the merger: $0.5$ day, $1.5$ days, $4.5$ days, $7.5$ days, $10.5$ days and $14.5$ days, respectively. The colored solid lines, dashed lines, dotted lines and dash-dotted lines denote the total observed spectrum, spectrum contributed from the dynamical ejecta, spectrum contributed from the neutrino-driven wind ejecta, and spectrum contributed from the viscosity-driven wind ejecta, respectively. The solid black lines represent the single temperature blackbody temperature fits to total observed BH-NS kilonova spectrum . The gray solid lines, obtained from Waxman et al. (2017), are the blackbody fits to the photometric data of GW170817/AT2017gfo taken from Villar et al. (2017). \label{fig:ViewingAngleSpectrum}} 
\end{figure}

With the projected surface and thermal temperature of the photosphere in the observer's frame solved, the total observed flux density can be expressed as (see Appendix \ref{app:A} for detail)
\begin{equation}
\label{Eq. FluxDensity}
    F_\lambda(\lambda , t_{\rm obs}) \approx \frac{2}{h^4c^3D_{\rm L}^2}\iint_S\frac{\mathcal{D}^3(hc/\mathcal{D}\lambda)^5}{\exp{(hc/\mathcal{D}k_{\rm B}\lambda T^{ij}_{\rm mesh}}) - 1}{\rm d}\sigma'^{ij},
\end{equation}
where $D_{\rm L}$ is the luminosity distance, $h$ is the Planck constant, $k_{\rm B}$ is the Boltzmann constant, $\lambda$ is wavelength, $T^{ij}_{\rm mesh}$ is the temperature at the mesh grid of our spatial discretization model, and ${\rm d}\sigma'^{\rm ij}$ is the infinitesimal projected photosphere area, which reads
\begin{equation}
    {\rm d}\sigma'^{ij} = \left(\frac{t_{\rm obs}}{1 - p_{\rm phot} ^ {ij} / c} \right)^2{\rm d}v'^{ij}_{{\rm mesh},x}{\rm d}v'^{ij}_{{\rm mesh},y}.
\end{equation}
{Here $t_{\rm obs}$ is the observational time, $p_{\rm phot}^{ij}$ is the velocity space distances between the photophere points and the mesh grid plane}, and $v^{ij}_{{\rm mesh},x}$,$v^{ij}_{{\rm mesh},y}$ are velocity components in the mesh grid. {The factor of $t_{\rm obs} / (1 - p_{\rm phot}^{ij} / c)$ is introduced to account for the light propagation effect.} We set the luminosity distance as $D_{\rm L} = 10\,{\rm pc}$ hereafter, so that magnitude stands for the absolute magnitude. 

We show examples of face-on ($\theta_{\rm view} = 0^\circ$, $\varphi_{\rm view} = 0 ^ \circ$) emergent spectra and different component contributions for both  cases in Figure \ref{fig:ViewingAngleSpectrum}. One can see that throughout the evolution, the emission is mainly contributed from the radiation of the dynamical ejecta. 
For both cases, the peak flux density of the neutrino-driven ejecta is only about half of that of the dynamical ejecta, even though the neutrino-driven ejecta has a higher temperature. The emission from the viscosity-driven ejecta is over-shone by the neutrino-driven ejecta at early time of emission, so that one can only see a little radiation from the viscosity-driven ejecta for the face-on geometry. Since the neutrino-driven ejecta has a lower opacity and a smaller mass than other components, its radiation fades out rapidly with time. After $\sim 1.5\,{\rm days}$ post-merger, the wavelengths of the peak emission move from optical to infrared. As the neutrino-driven ejecta becomes optically thin, the contribution of the viscosity-driven ejecta increases significantly. At late time of  emission, for Case I, the viscosity-driven ejecta contributes to relatively short wavelengths due to its low opacity compared with the dynamical ejecta. The dynamical ejecta, on the other hand, contributes more to the long-wavelength band. As for Case II, the late time emission is always contributed from the dynamical ejecta. We fit the total emergent spectra with a single blackbody model. For both cases, the spectra can be approximately fitted by the single temperature blackbody model. Consistent with intuition, the temperatures of Case II are higher than Case I, because more mass is outside the BH remnant. Case II has a total ejecta mass that is only about four times of that of Case I. The blackbody fit temperature is only higher by a factor of $\sim (1.1-1.2)$.

\begin{figure}[tbp]
\centering
\includegraphics[width = 0.49\linewidth , trim = 60 185 85 60, clip]{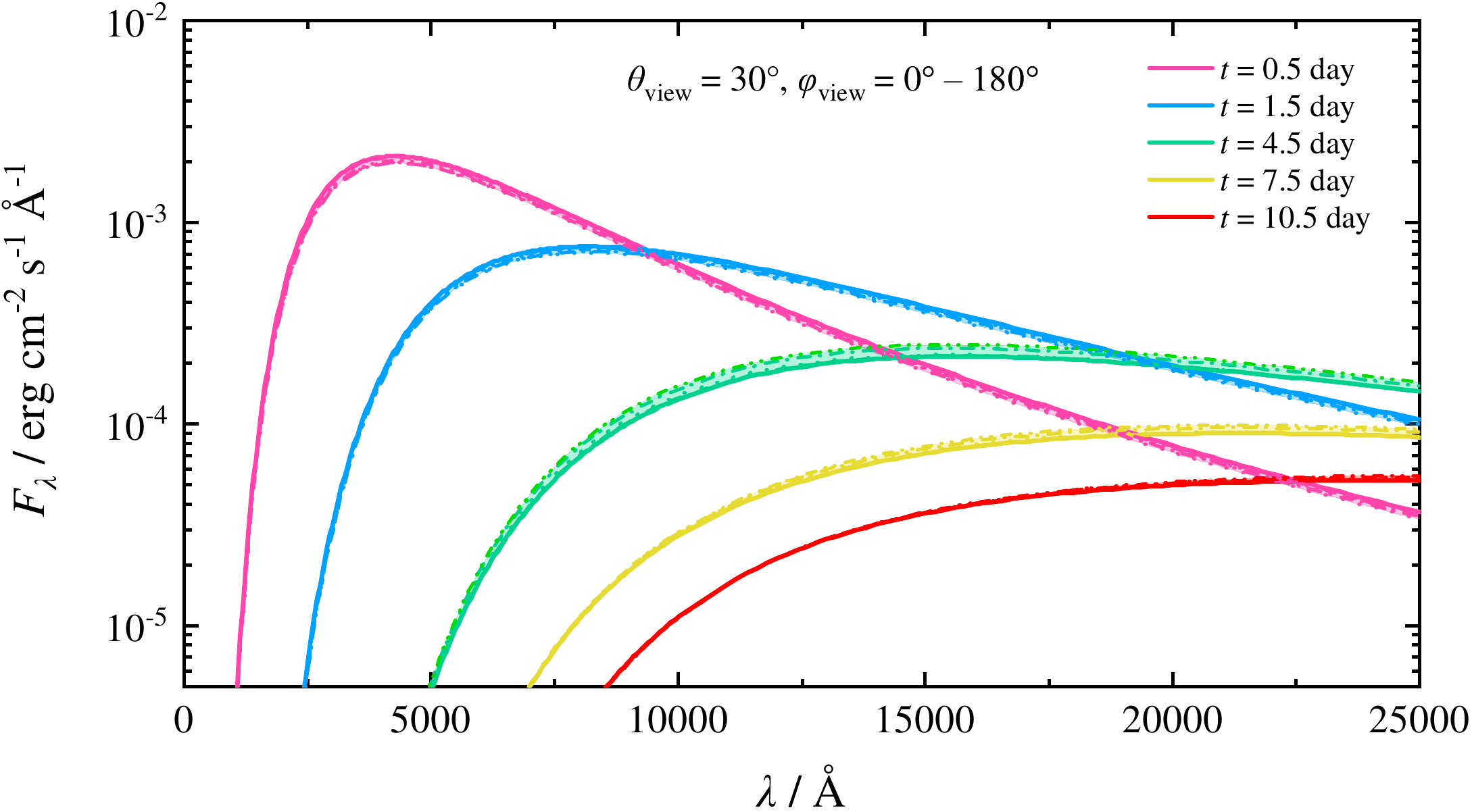}\quad\includegraphics[width = 0.49\linewidth , trim = 60 185 85 60, clip]{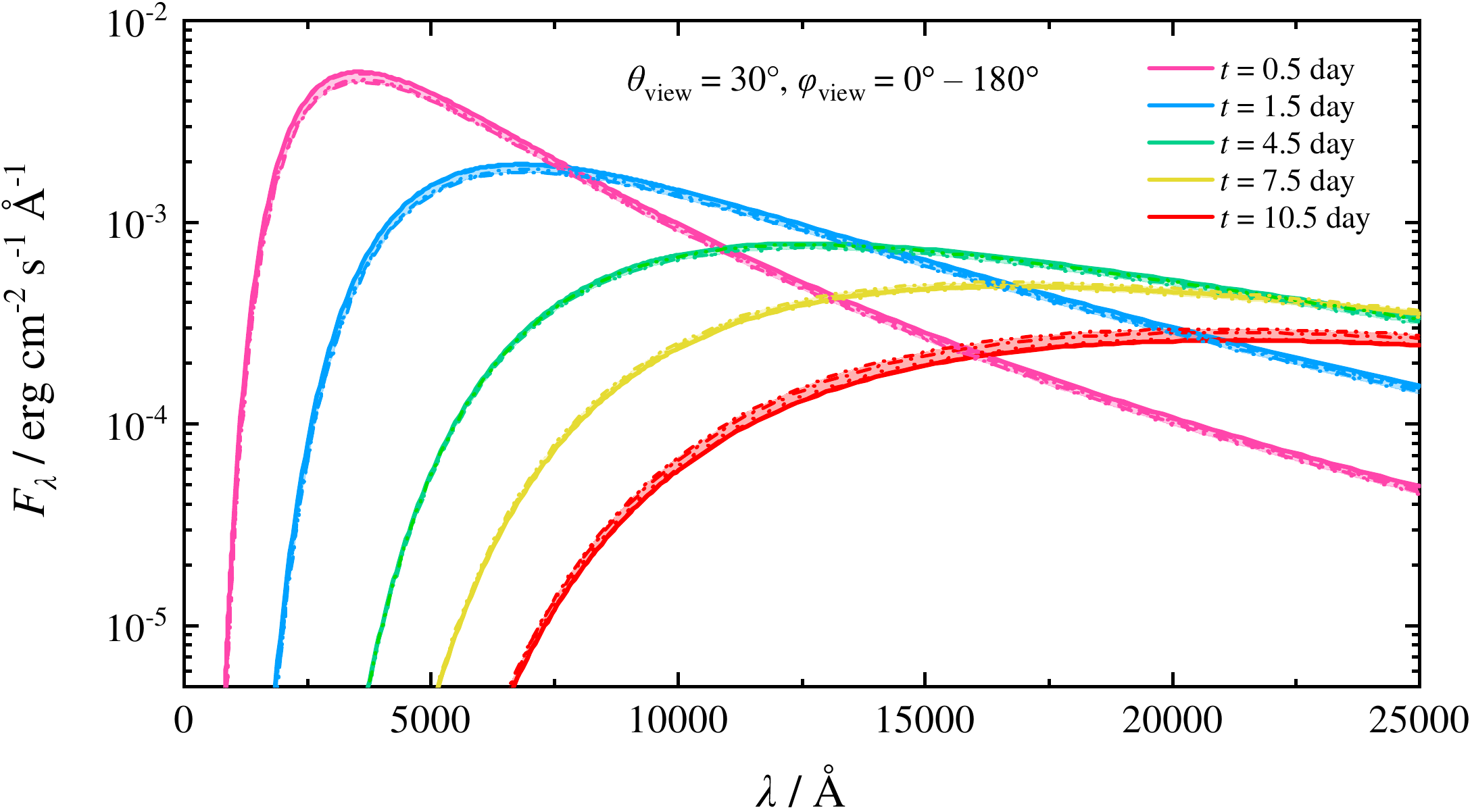}
\includegraphics[width = 0.49\linewidth , trim = 60 185 85 60, clip]{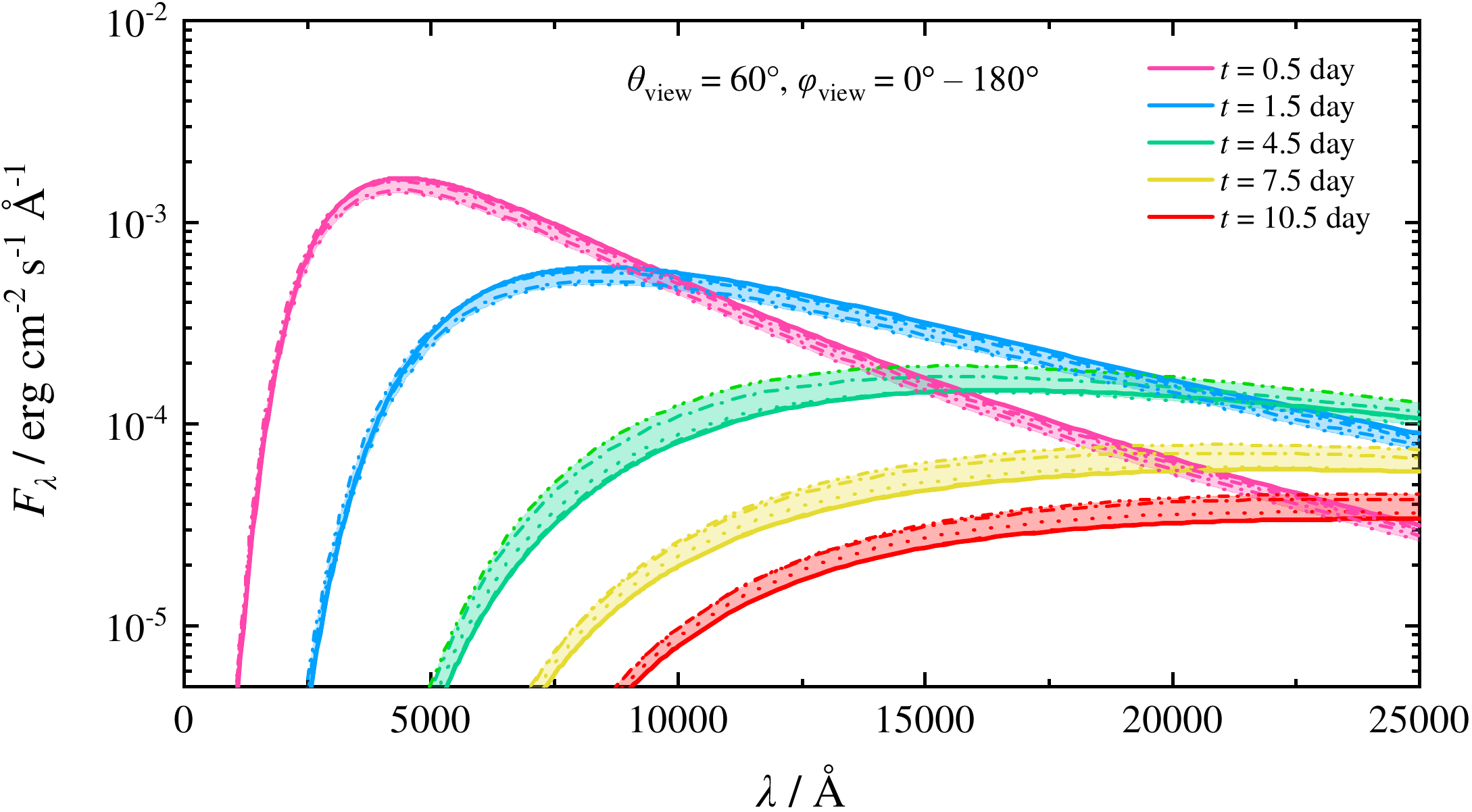}\quad\includegraphics[width = 0.49\linewidth , trim = 60 185 85 60, clip]{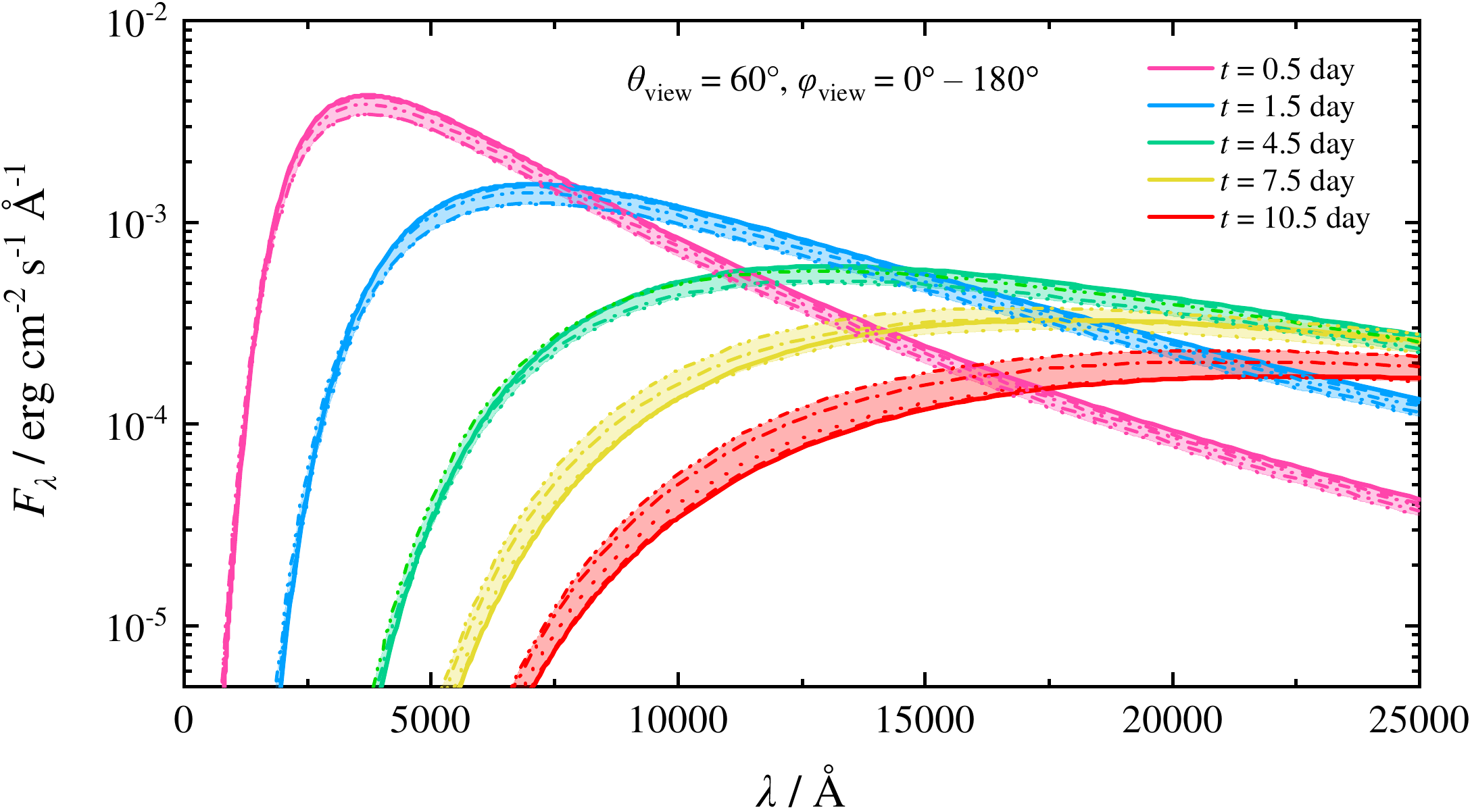}
\includegraphics[width = 0.49\linewidth , trim = 60 185 85 60, clip]{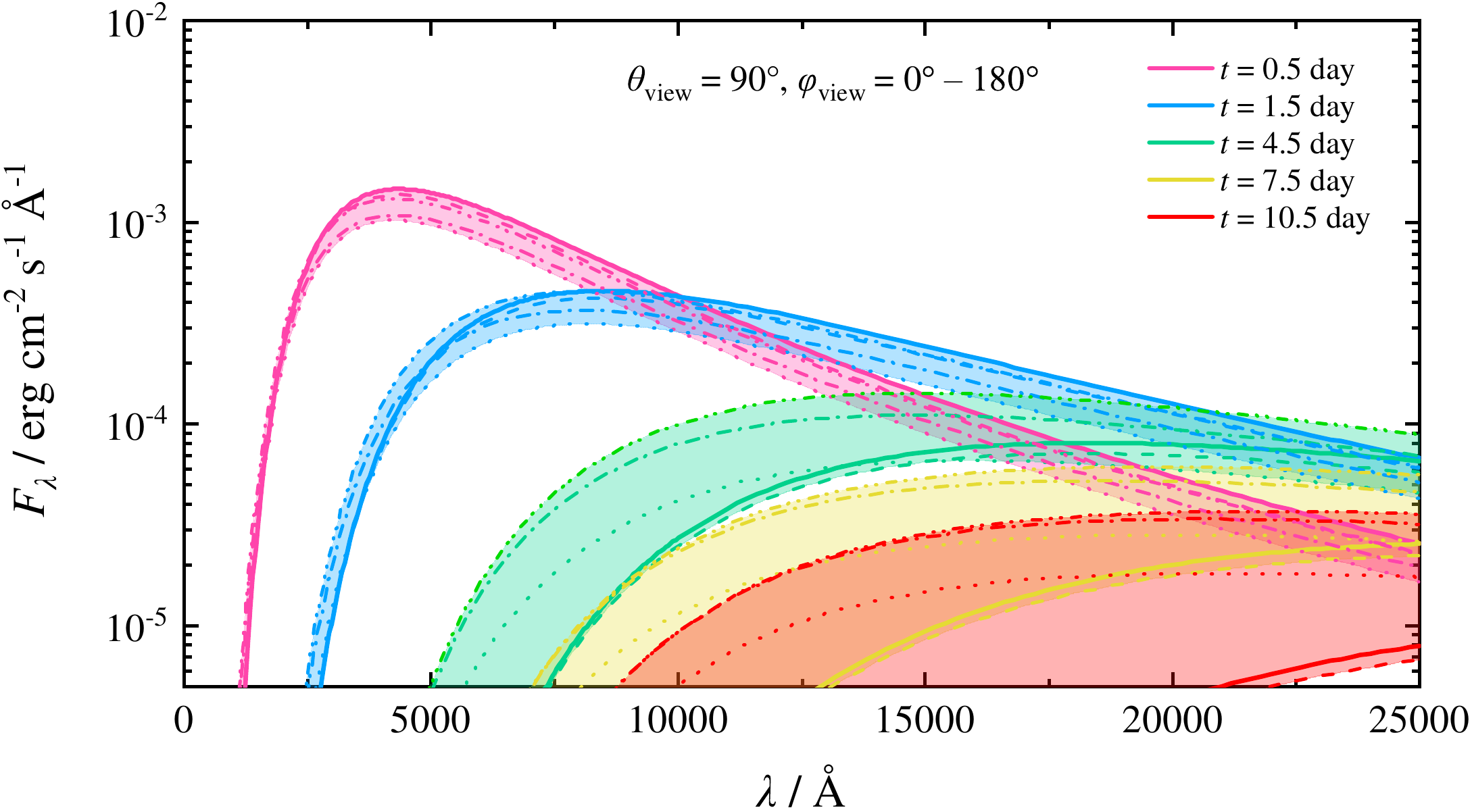}\quad\includegraphics[width = 0.49\linewidth , trim = 60 185 85 60, clip]{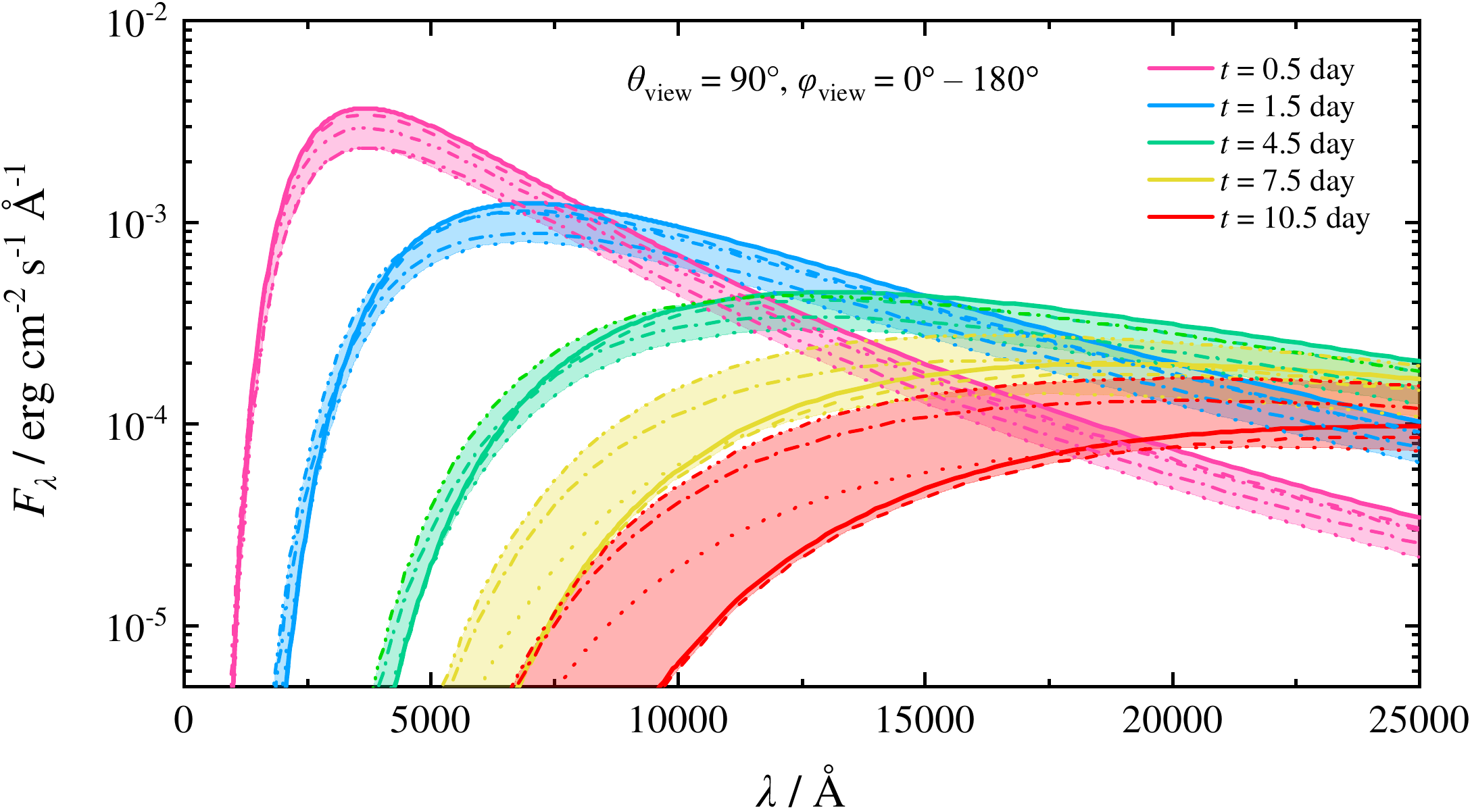}
\caption{$\varphi_{\rm view}-$dependent emergent spectra for Case I (left panels) and Case II (right panels). From top to bottom, the three panels are for  $\theta_{\rm view} = 30^\circ$, $60^\circ$, and $90^\circ$, respectively. The range for each spectra spans five possible $\varphi_{\rm view}$ values with respect to the observer: $0^\circ$ (solid), $45^\circ$ (dashed), $90^\circ$ (dotted), $135^\circ$ (dashed-dotted), and $180^\circ$ (dashed-dotted-dotted), respectively. \label{fig:ViewingSpectrumTheta}}
\end{figure}

\begin{figure}[tbp]
\centering
\includegraphics[width = 0.49\linewidth , trim = 60 185 85 60, clip]{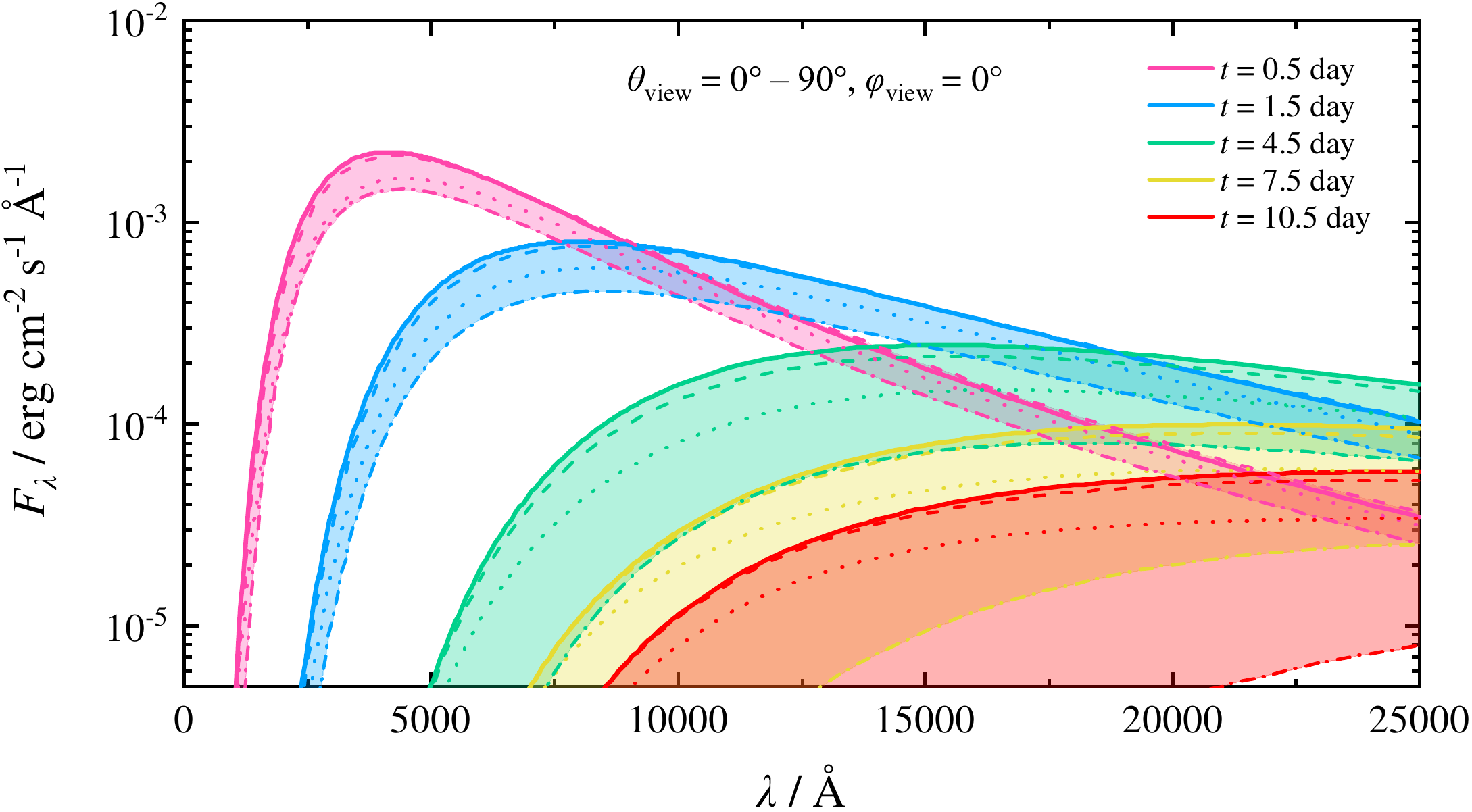}\quad\includegraphics[width = 0.49\linewidth , trim = 60 185 85 60, clip]{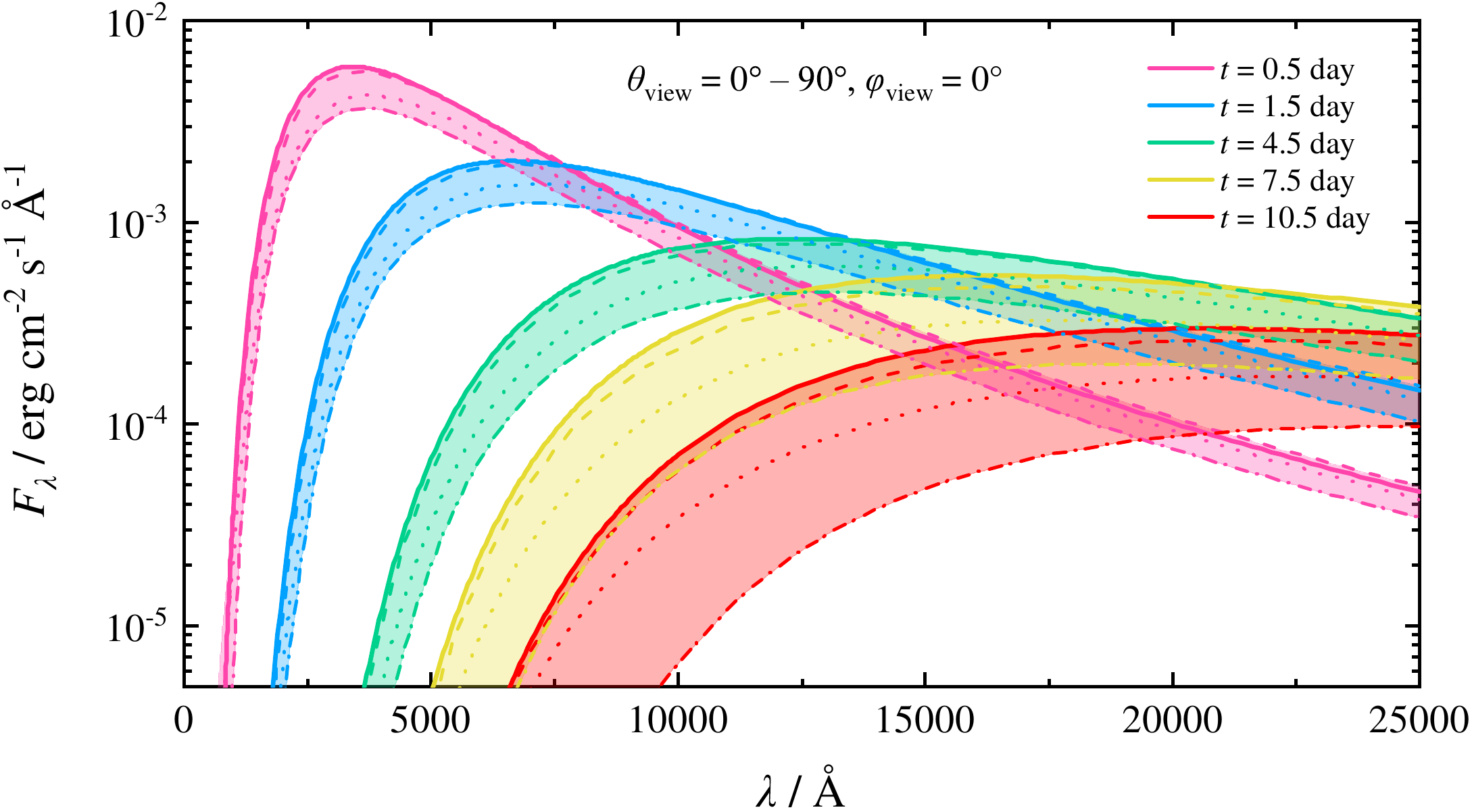}
\caption{$\theta_{\rm view}-$dependent emergent spectra for Case I (left panel) and Case II (right panel), where we set $\varphi_{\rm view} = 0 ^ \circ$ as a constant. The range for each spectra spans four possible $\theta_{\rm view}$ values with respect to the observer: $0^\circ$ (\textbf{thick} solid), $30^\circ$ (dashed), $60^\circ$ (dotted), and $90^\circ$ (dashed-dotted), respectively. \label{fig:ViewingSpectrumPhi=0}}
\end{figure}

In Figure \ref{fig:ViewingSpectrumTheta} and Figure \ref{fig:ViewingSpectrumPhi=0}, we show the $\varphi_{\rm view}-$ and $\theta_{\rm view}-$dependent emergent spectra at the same {observational} epochs after the merger. We first discuss the $\varphi_{\rm view}-$dependent emergent spectra. {One can find that for increasing $\varphi_{\rm view}$, the early-stage observed luminosity would decrease while the late-stage observed luminosity would increase.} For $\theta_{\rm view} = 30^\circ$ and $\theta_{\rm view} = 60^\circ$, the observer can simultaneously observe all three components. Changing $\varphi_{\rm view}$ has little effect on the observed wavelength at peak flux density. Since the observed temperature is not significantly modified by the change of the viewing angle, the difference of the peak flux density can approximately represent that of the observed luminosity. Changing $\varphi_{\rm view}$ results in a difference {of observed luminosity} by a factor of {$\sim (1.05 - 1.1)$ for both cases at early times ($t \lesssim 1.5\,{\rm day}$) and a factor $\sim(0.75-0.85)$ at late times ($t \gtrsim 7.5\,{\rm day}$).} For $\theta_{\rm view} = 90^\circ$, the dynamical ejecta can outshine the other two components completely when $\varphi_{\rm view} \approx 0^\circ - 60^\circ$. Therefore, under this condition, the variation of $\varphi_{\rm view}$ can have a relatively larger effect on the fitting temperature of the mergent spectrum and the observed luminosity. {The variation can be particularly significant for Case I at late times because its late-stage luminosity mainly contributes from the viscosity-driven ejecta.} {The observed luminosity of $\varphi_{\rm view}-$dependent emergent spectra varies by a factor from $\sim1.1$ for Case I and $\sim1.3$ for Case II at early times ($t \lesssim 1.5\,{\rm day}$) to $\sim0.2$ for Case I and $\sim0.5$ for Case II at late times ($t \gtrsim 7.5\,{\rm day}$).}

We next discuss the $\theta_{\rm view}-$dependence. Figure \ref{fig:ViewingSpectrumPhi=0} shows the $\theta_{\rm view}-$dependent emergent spectra, where we set $\varphi_{\rm view} = 0^\circ$, which always has the largest observed luminosity for each $\theta_{\rm view}$ {at the early stage}. The observed luminosity is the largest at the same epoch if $\theta_{\rm view} = 0 ^ \circ$. This is because {the projected photosphere area is the largest along the line of sight}. The observed luminosity would decrease with increasing $\theta_{\rm view}$. At each epoch for Case I, the maximum observed luminosity is $\sim1.5 - 7$ times of the minimum observed luminosity. For Case II, the factor vary to $\sim1.5 - 3$ times of the minimum observed luminosity.

One can conclude that the dynamical ejecta is the main contributor to the kilonova emission of BH-NS mergers. This is mainly because the projected photosphere area of the dynamical ejecta is far larger than those of the other two components along any line of sight. Another reason is that BH-NS mergers lack lanthanide-free ejecta. Due to the large projected photosphere area of the dynamical ejecta, there is only a small part of the observed photosphere that can be covered by the other two ejecta components if we only change $\varphi_{\rm view}$. This is why changing $\varphi_{\rm view}$ has little effect on the variation of the observed spectra. {The light propagation effect would mainly affect the late-stage observed luminoisty. For a large $\theta_{\rm view}$ condition, the observer would see the photons emitted from the dynamical ejecta $\sim(v_{\rm max,d}/c)t_{\rm obs}$ days earlier if the dynamical ejecta move away from the observer (i.e., $\varphi_{\rm view} \sim 180^\circ$). The earlier high-temperature dynamical ejecta can enhance the observed luminosity. This is the reason why the late-stage observed luminosity would enhance by increasing $\varphi_{\rm view}$. However, the light propagation effect has little effect on the early-stage observed luminosity due to the rapid change of the projected area.} In contrast, the projected photosphere area would significantly decrease with the increasing $\theta_{\rm view}$. However, this decreasing trend cannot cause a significant change in luminosity. With viewing angle change, the variations of the observed luminosity are mainly derived from the change of the projected photosphere area of the dynamical ejecta which causes the difference of the observed luminosity by only a factor of $\sim (2 - 3)$ for different epochs.

\subsection{Viewing-Angle-Dependent Lightcurve}\label{sec:4.4}

\begin{figure}[tbp]
\centering
\includegraphics[width = 0.49\linewidth , trim = 65 35 90 60, clip]{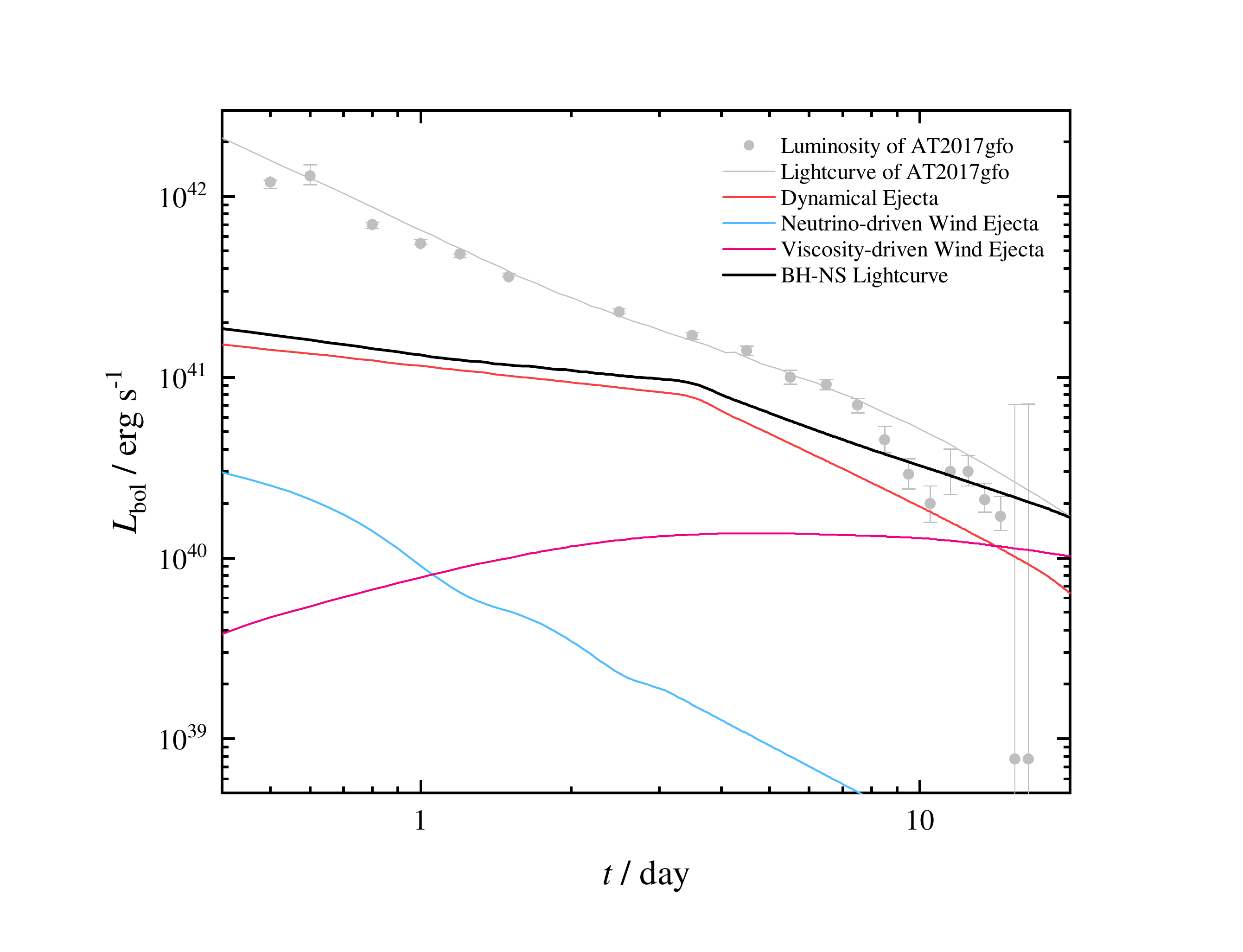}\quad\includegraphics[width = 0.49\linewidth , trim = 65 35 90 60, clip]{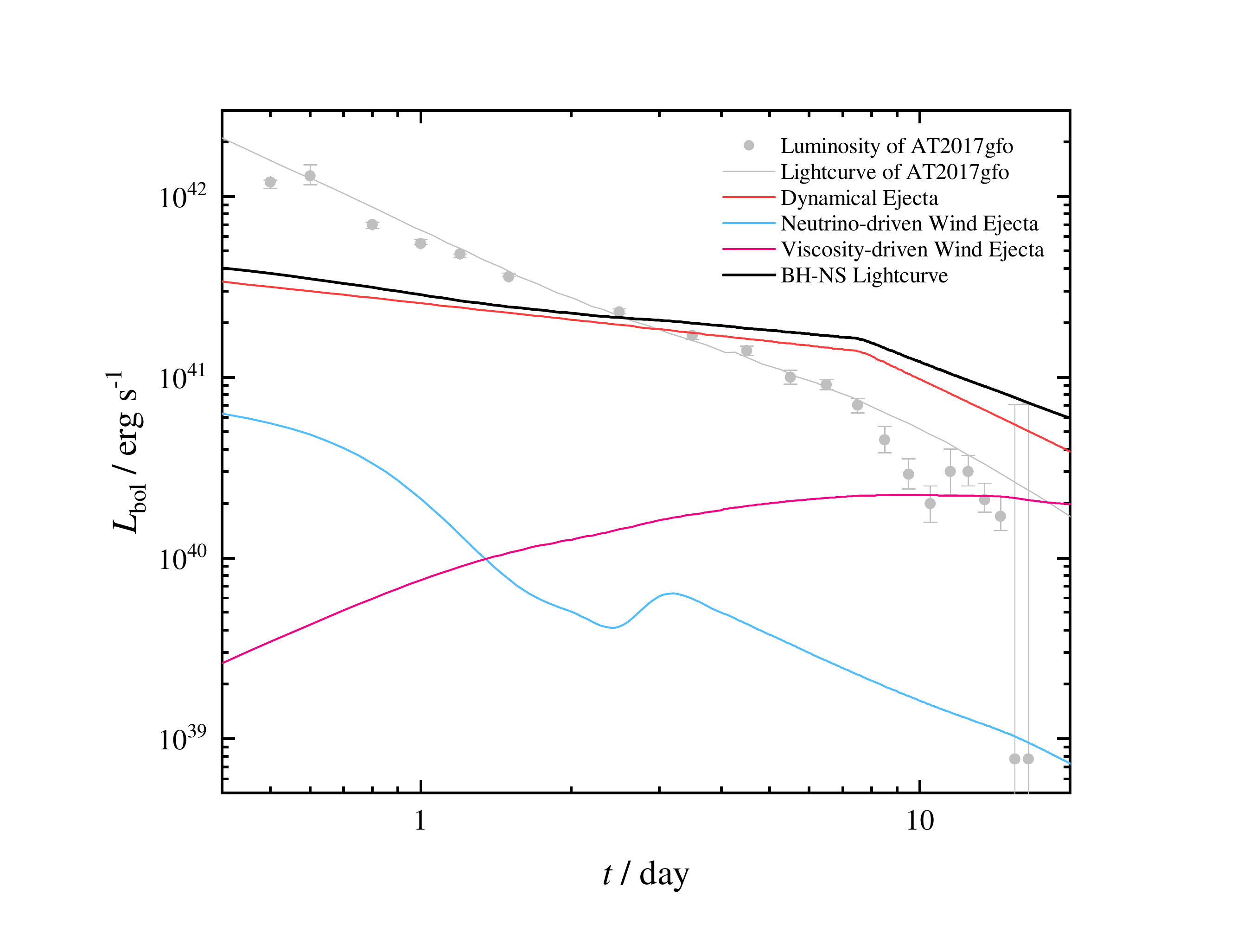}
\includegraphics[width = 0.49\linewidth , trim = 65 35 90 60, clip]{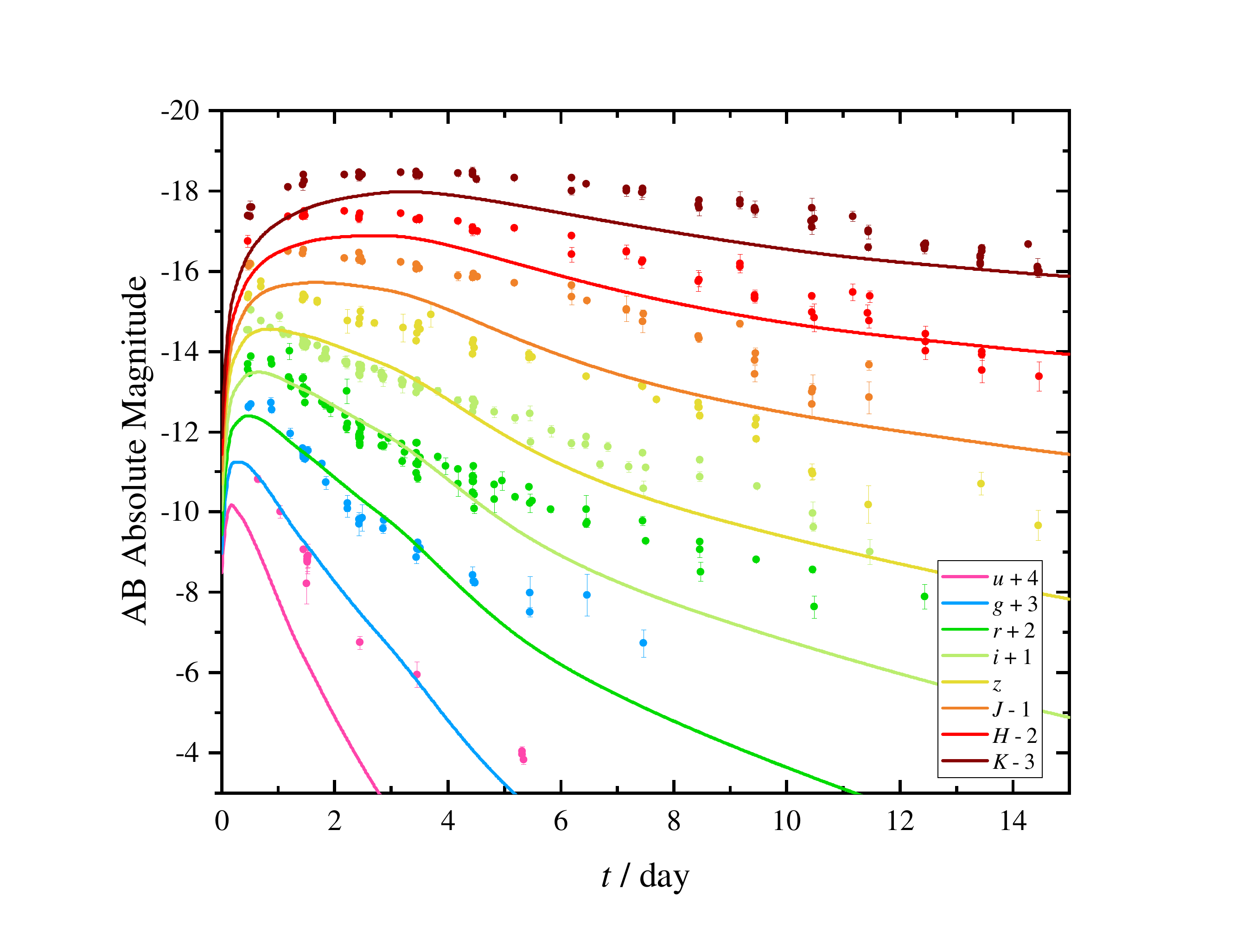}\quad\includegraphics[width = 0.49\linewidth , trim = 65 35 90 60, clip]{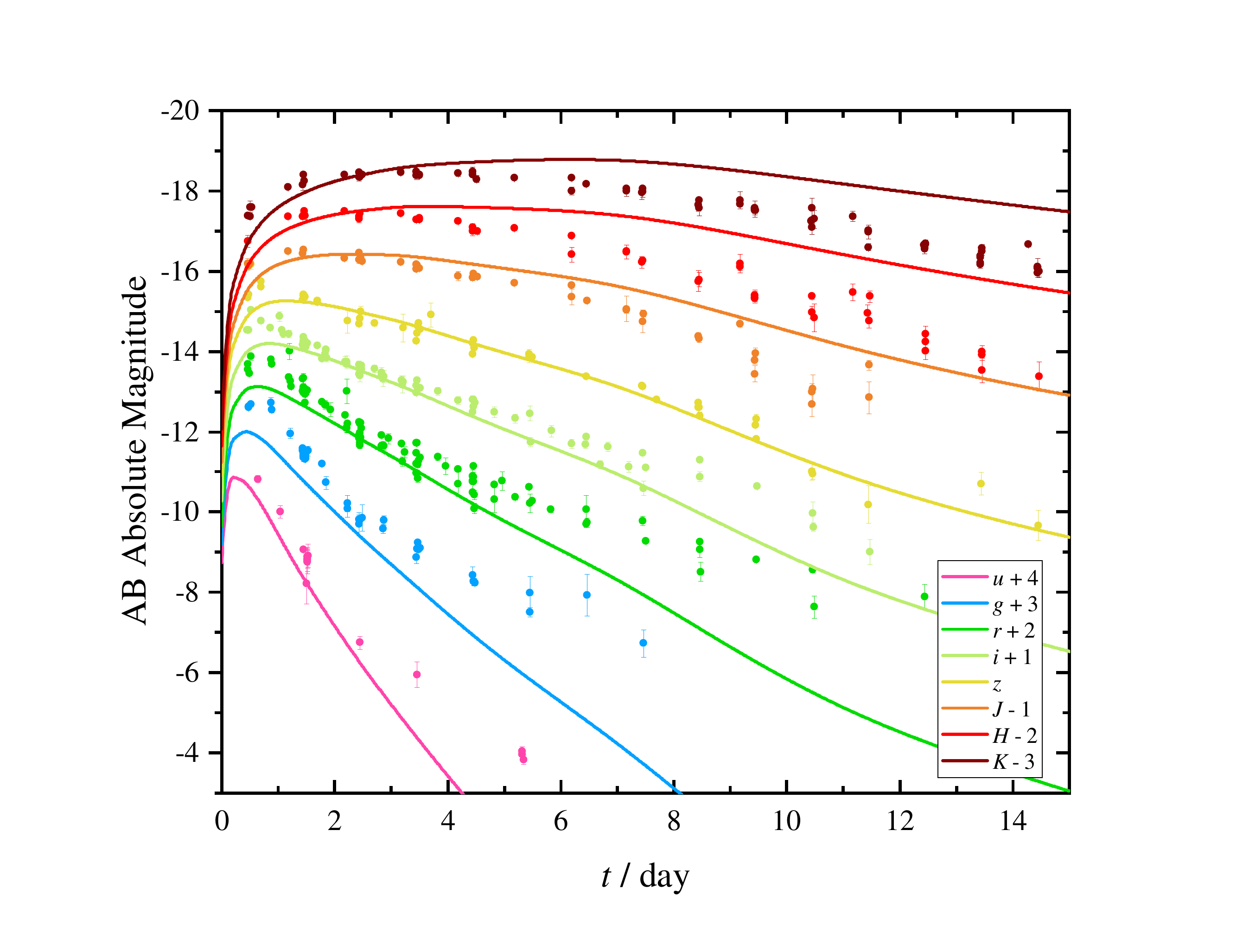}
\caption{Top panels: face-on($\theta_{\rm view} = 0^
\circ$, $\varphi_{\rm view} = 0^\circ$) observed bolometric lightcurves for Case I (left panel) and Case II (right panel). The black, red, blue, purple solid lines represent the total bolometric lightcurves, contributions from the dynamical ejecta, the neutrino-driven wind ejecta, and the viscosity-driven wind ejecta, respectively. The gray points denote the bolometric luminosity points of AT2017gfo which are taken from Waxman et al. (2017). The gray lines denote the fitting bolometric lightcurve of AT2017gfo which is taken from Wu et al. (2018). Bottom panels: face-on $ugrizJHK$-band observed lightcurves for Case I (left panels) and Case II (right panels). The photometric data points of AT 2017gfo are taken from Villar et al. (2017). \label{fig:ViewingAngleLightCurve}}
\end{figure}

With the time-dependent kilonova emergent spectra calculated in Section \ref{sec:4.3}, one can calculate viewing-angle-dependent bolometric and color lightcurves. The total bolometric luminosity is given by $L_{\rm bol}(t) = 4\pi D^2_{\rm L}\int_0^{\infty}F_\lambda{\rm d}\lambda$, where the flux density at the photon wavelength $\lambda$ is given by Equation (\ref{Eq. FluxDensity}). Based on the conversion between frequency and wavelength, i.e., $F_\nu{\rm d}\nu = F_\lambda{\rm d}\lambda$, the flux density at photon frequency $\nu$ can be expressed as $F_\nu = \lambda^2F_\lambda/c$. One can also obtain multi-frequency monochromatic AB magnitude defined by $M_\nu = -2.5\log_{10}(F_\nu/3631{\rm Jy})$. Since we set $D_{\rm L} = 10\,{\rm pc}$, $M_\nu$ actually is the AB absolute magnitude. In Figure \ref{fig:ViewingAngleLightCurve}, we show the examples of face-on bolometric lightcurves and $ugrizJHK$-band lightcurves for both cases. We also show the contributions of different components in the face-on bolometric lightcurves. Similar to the discussion on the emergent spectra in Section \ref{sec:4.3}, one can clearly see that most of the radiation energy is contributed from the dynamical ejecta. The face-on bolometric lightcurves can be approximately  described by a power-law with two breaks. In the broken-power-law description, the features of the face-on bolometric lightcurves are as follows. (i) Before $t \sim 1\,{\rm day}$, the luminosity of a BH-NS merger kilonova is about a few times $10^{41}\,{\rm erg\,s^{-1}}$. During this time range, the radiation of the neutrino-driven ejecta is significant, even though only $\sim 1 / 5$ of the radiation energy is contributed from the neutrino-driven ejecta at $t \sim 0.5\,{\rm day}$. The bolometric lightcurves are well represented by a power-law decay with a power-law index of $-0.37$ for {both cases}. (ii) With the neutrino-driven ejecta becoming optically thin, almost all of the kilonova emission is contributed by the dynamical ejecta. Between $t \gtrsim 1\,{\rm day}$ and $t = t_{\rm c}$ {($t_{\rm c}\sim4\,{\rm day}$ for Case I and $t_{\rm c}\sim7\,{\rm day}$ for Case II)}, the temporal index for both cases becomes $\sim-0.26$. (iii) There is another break at $t = t_{\rm c}$ for the dynamical ejecta emission, which is followed by a steeper decay with index $\sim-0.93$ {for Case I and $\sim-1$ for Case II}. The contribution from the viscosity-driven ejecta becomes progressively important at late times. 

The resulting lightcurves on a timescale $\sim0.5\,{\rm day}$ have a bolometric luminosity $\sim 1.7\times 10 ^ {41}\,{\rm erg}\,{\rm s}^{-1}$ for Case I and $\sim 3.7\times 10 ^ {41}\,{\rm erg}\,{\rm s}^{-1}$ for Case II. Case II also have a longer evolution time compared with Case I. One can see that kilonovae from the BH-NS mergers are typically fainter than $\sim -14.5\,{\rm mag}$ in optical and $\sim -15\,{\rm mag}$ in infrared for Case I, and fainter than $\sim -15\,{\rm mag}$ in optical and $\sim -16\,{\rm mag}$ in infrared for Case II. As for multi-band lightcurves, different bands peak at different times but for the same band there is no significant difference in the peak time for the two cases except the $K$-band, which shows a coincidence of the peak time with $t_{\rm c}$. Since $t_{\rm c}$ carries the information of ejecta mass, one may use the $K$-band peak time to estimate the mass in the ejecta. Compare with the peak AB absolute magnitudes of each filter for the two cases, the differences are in the range of $0.6\, {\rm mag}$ to $0.8\,{\rm mag}$.

\begin{figure}[htbp]
\centering
\includegraphics[width = 0.49\linewidth , trim = 60 30 100 60, clip]{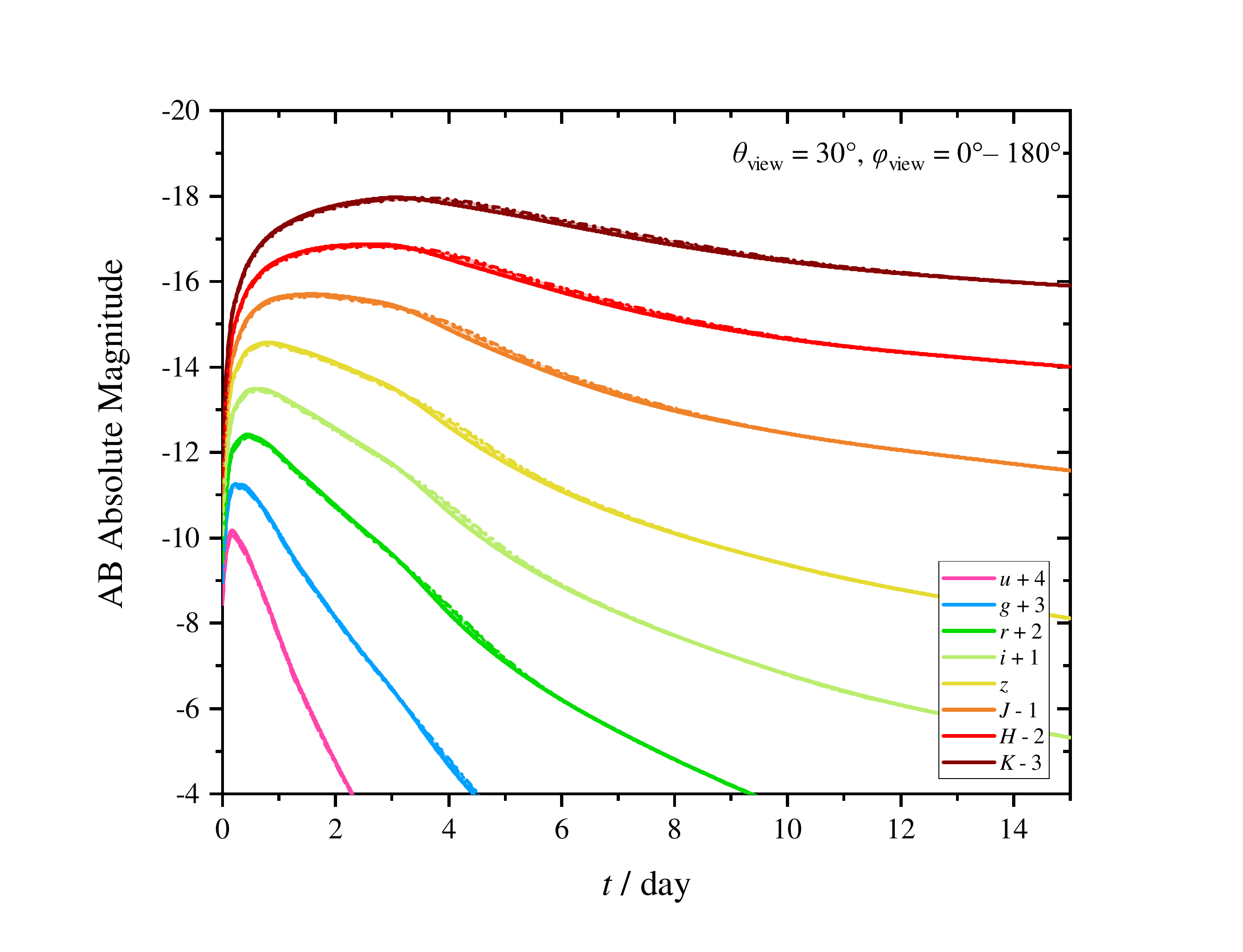}\quad\includegraphics[width = 0.49\linewidth , trim = 60 30 100 60, clip]{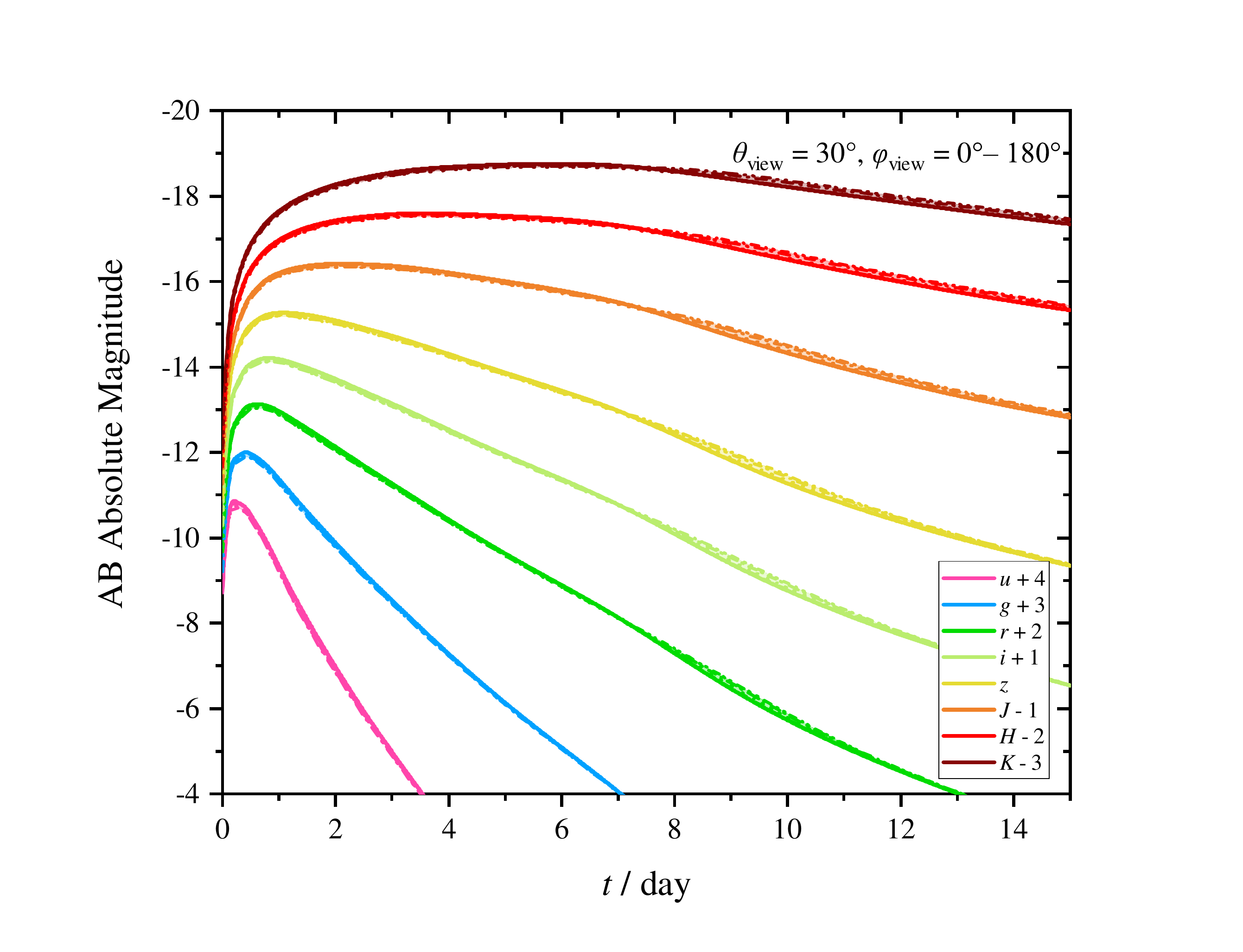}
\includegraphics[width = 0.49\linewidth , trim = 60 30 100 60, clip]{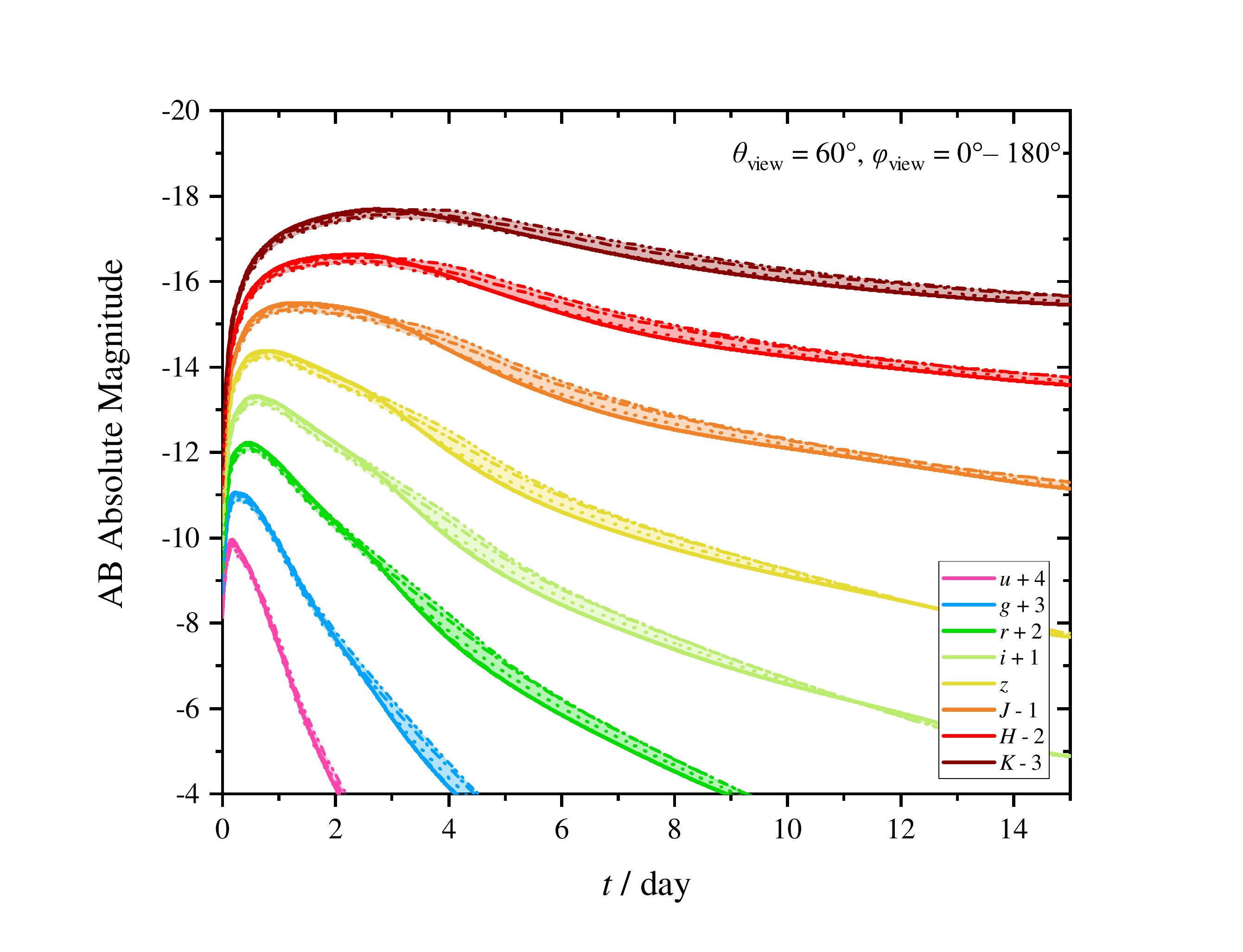}\quad\includegraphics[width = 0.49\linewidth , trim = 60 30 100 60, clip]{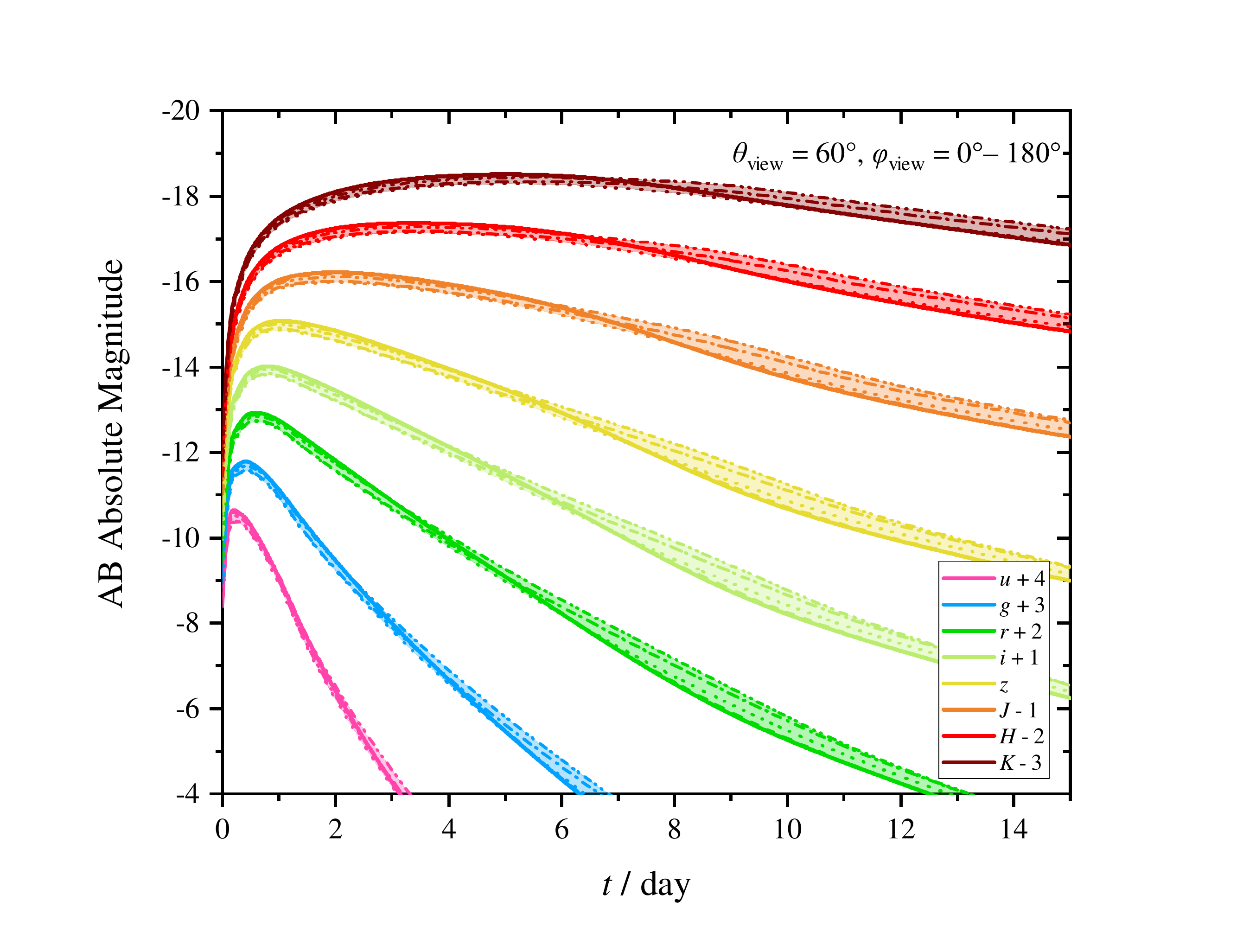}
\includegraphics[width = 0.49\linewidth , trim = 60 30 100 60, clip]{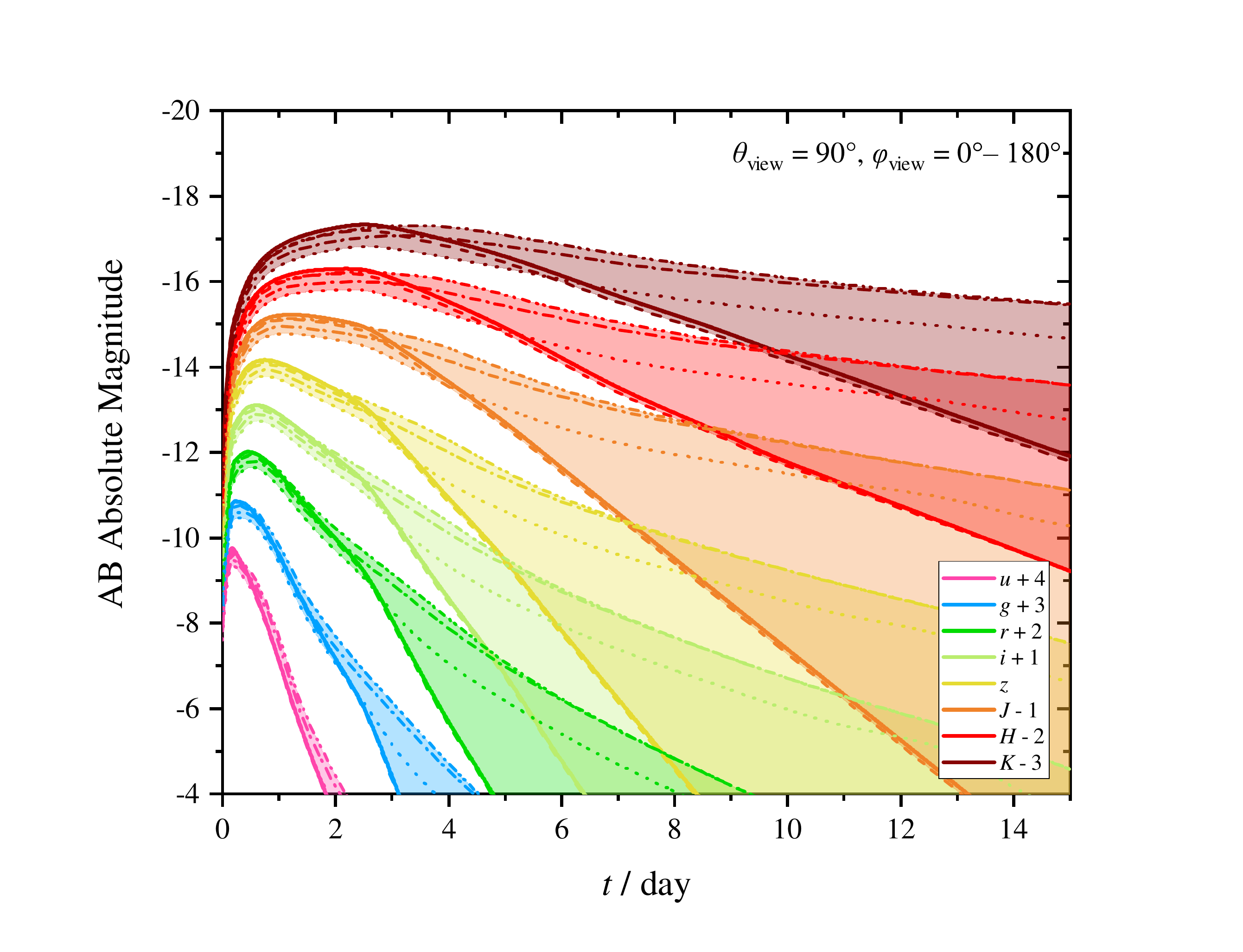}\quad\includegraphics[width = 0.49\linewidth , trim = 60 30 100 60, clip]{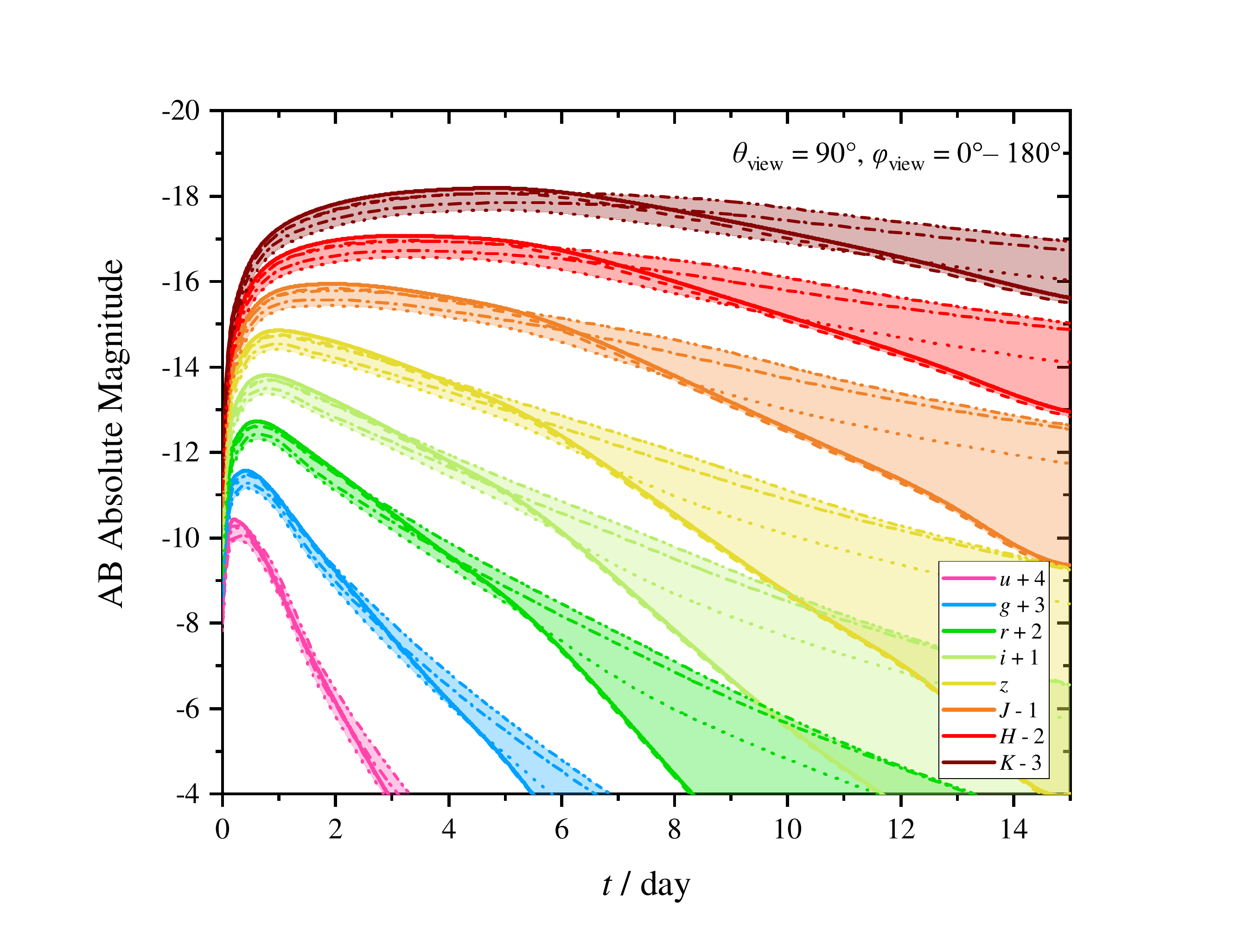}
\caption{Predicted $\varphi_{\rm view}-$dependent lightcurves for Case I (left panels) and Case II (right panels). From top to bottom, the three panels show $\theta_{\rm view} = 30^\circ$, $60^\circ$, and $90^\circ$, respectively. The range for each light curve spans five possible $\varphi_{\rm view}$ values: $0^\circ$ (\textbf{thick} solid), $45^\circ$ (dashed), $90^\circ$ (dotted), $135^\circ$ (dashed-dotted), and $180^\circ$ (dashed-dotted-dotted), respectively. \label{fig:ViewngLightCurveTheta}}
\end{figure}

\begin{figure}[htbp]
\centering
\includegraphics[width = 0.49\linewidth , trim = 60 30 100 60, clip]{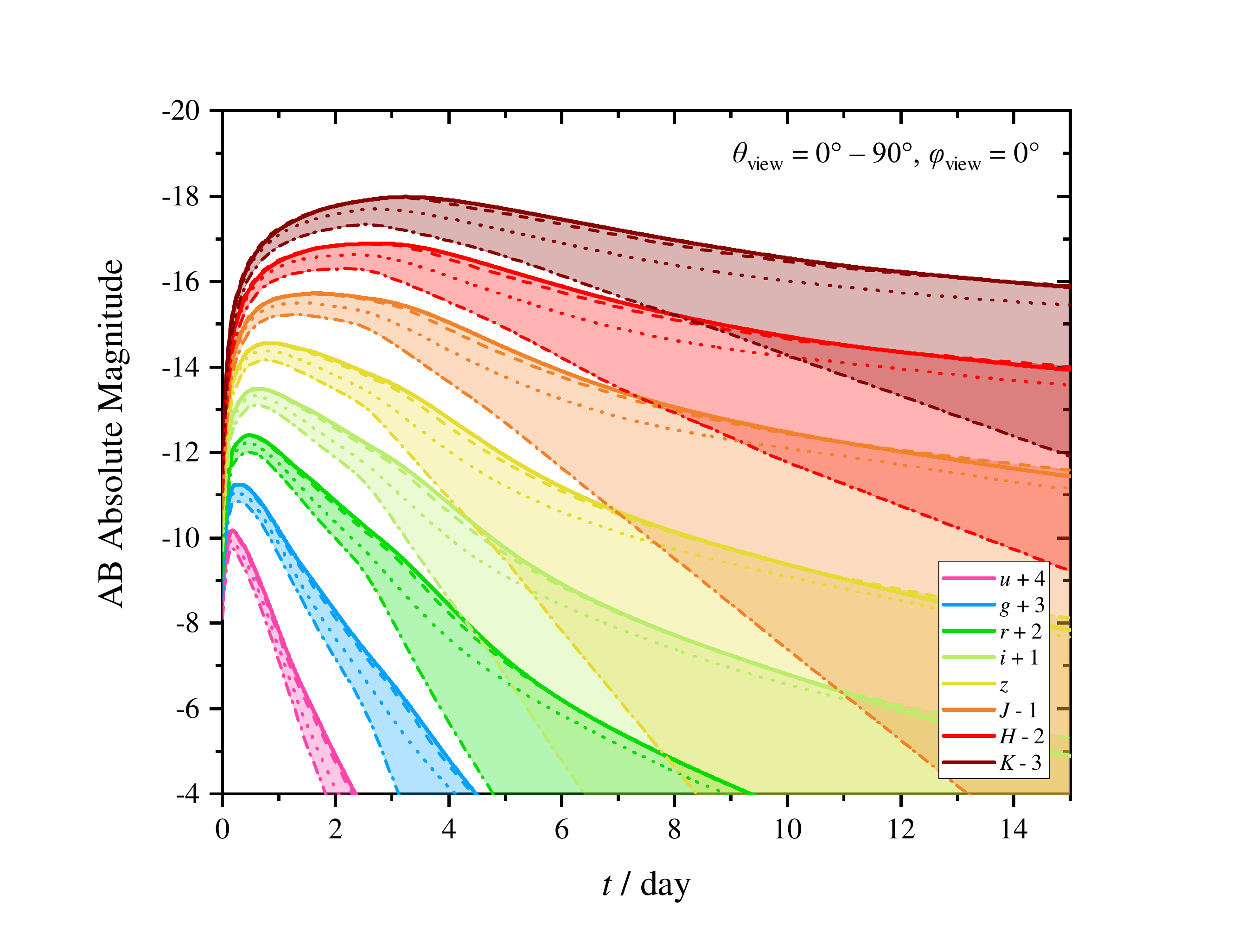}\quad\includegraphics[width = 0.49\linewidth , trim = 60 30 100 60, clip]{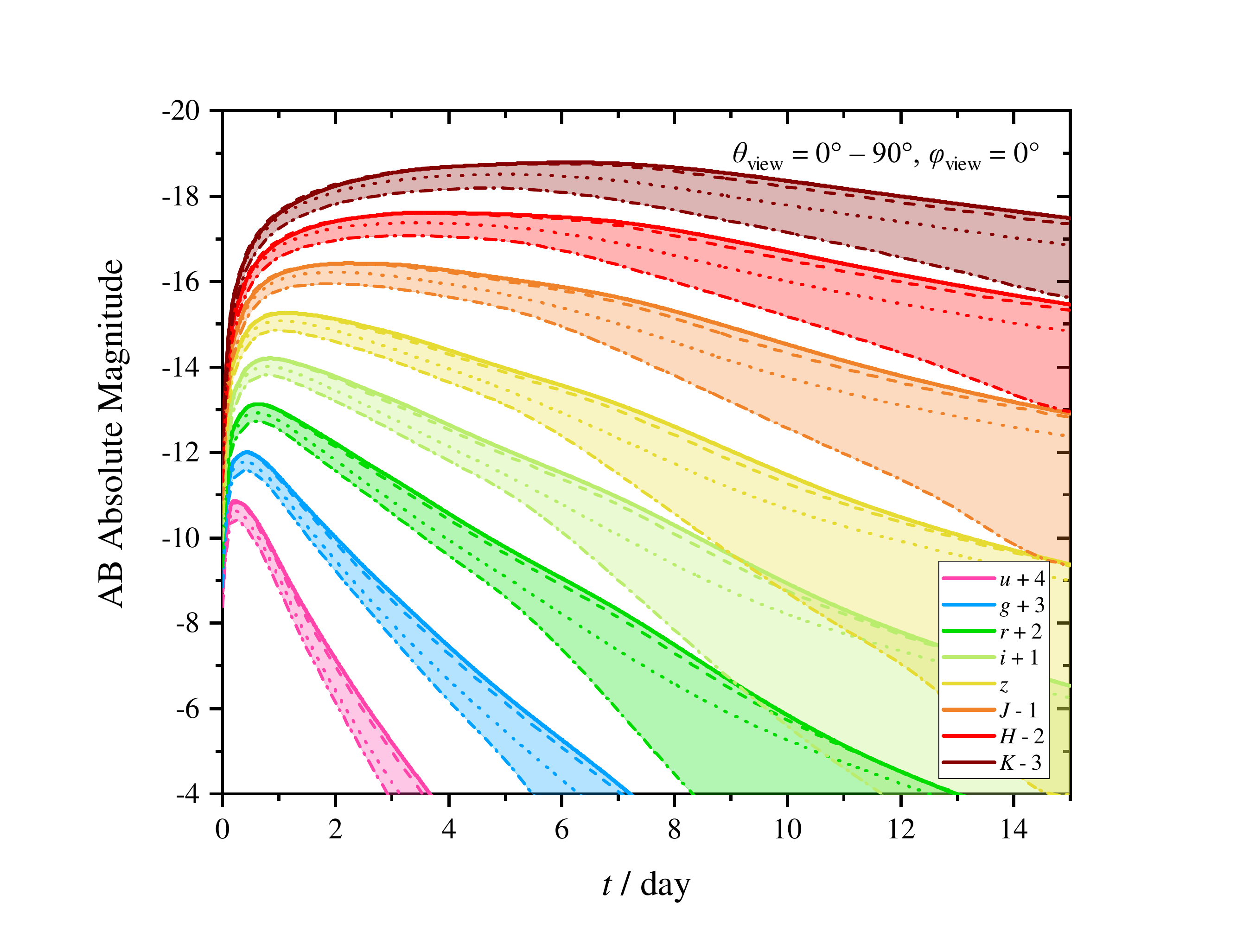}
\caption{Predicted $\theta_{\rm view}-$dependent lightcurves for Case I (left panel) and Case II (right panel), where we set $\varphi_{\rm view} = 0 ^ \circ$ as a constant. The range for each light curve spans four possible $\theta_{\rm view}$ values: $0^\circ$ (\textbf{thick} solid), $30^\circ$ (dashed), $60^\circ$ (dotted), and $90^\circ$ (dashed-dotted), respectively. \label{fig:ViewingLightCurvePhi=0}}
\end{figure}

Figure \ref{fig:ViewngLightCurveTheta} and Figure \ref{fig:ViewingLightCurvePhi=0}, show the predicted $\varphi_{\rm view}-$ and $\theta_{\rm view}-$dependent lightcurves. {Both $\varphi_{\rm view}$ and $\theta_{\rm view}$ can change the peak time in each band which is caused by the light propagation effect. The variation of the peak time in each band can be approximately estimated by $\sim -(v_{\rm max,d} / c)\sin\theta_{\rm view}\cos\varphi_{\rm view}t_{\rm peak}$, where $t_{\rm peak}$ is  roughly the peak timescale of each band. Therefore, the variation of the peak time depends on the relative motion direction of the dynamical ejecta: the peak time would increase if the dynamical ejecta moves away from the observer (i.e., $\varphi_{\rm view} = 90^\circ-180^\circ$) and decrease if it moves toward the observer (i.e., $\varphi_{\rm view} = 0^\circ-90^\circ$).} As $\varphi_{\rm view}$ varies when we set $\theta_{\rm view} = 30 ^ \circ$ and $\theta_{\rm view} = 60^\circ$, the change on the shape and the magnitude of multi-band lightcurves is tiny. For both cases, the differences of magnitude in each band between the maximum to the minimum are $\sim0.1\,{\rm mag}$ and $\sim0.2\,{\rm mag}$ at $\theta_{\rm view} = 30^\circ$ and $\theta_{\rm view} = 60^\circ$, respectively. At $\theta_{\rm view} = 90^\circ$, varying $\varphi_{\rm view}$ can give a relatively obvious effect on the shape of late time multi-band lightcurves. The differences of maximum magnitude of each band can be larger, i.e in the range of $0.4 - 0.6\,{\rm mag}$. At $\varphi_{\rm view} \sim 0 -45^\circ$, the decay index of each band can be steeper. This is because along the line of sight in this range, the dynamical ejecta would block the emission of the other two components. However, at late times, the viscosity-driven ejecta would mainly contribute to near infrared band while the dynamical ejecta contributes to longer wavelengths. The lack of contribution from the viscosity-driven ejecta causes this steeper decay of each multi-band lightcurves. 

We show $\theta_{\rm view}-$dependent lightcurves in Figure \ref{fig:ViewingLightCurvePhi=0}, where we set $\varphi_{\rm view} = 0^\circ$. One can see that the multi-band lightcurves at $\theta_{\rm view} = 0^\circ$ are almost the same as those at $\theta_{\rm view} = 30^\circ$. The differences of the maximum magnitude of each filter are in the range of $0.5 - 0.6\,{\rm mag}$. Altogether, the magnitude differences caused by the viewing angle changes are approximately $\sim1\,{\rm mag}$.

\cite{Kyutoku2015} indicated that the dynamical ejecta is always smaller than $0.1\,M_\odot$. This means that BH-NS merger kilonovae are always less luminous than $\sim4.5 \times10^{41}\,{\rm erg\,s^{-1}}$. Corresponding to AB absolute magnitudes, they are fainter than $\sim -15\,{\rm mag}$ in optical and $\sim -16\,{\rm mag}$ in infrared. Both $\varphi_{\rm view}$ and $\theta_{\rm view}$ {can} change the peak time of each filter
{due to the light propagation effect}. They have little effect on the shape of the multi-band lightcurves, expect for $\theta_{\rm view} = 90^\circ$ and $\varphi_{\rm view} \sim 0 - 45^\circ$, in which case the decay index of the late time multi-band lightcurves can be steeper. The observed luminosity differences caused by viewing angle changes are in the range of a factor of $\sim2 - 3$, which corresponds to approximately $\sim1\,{\rm mag}$.

\subsection{Viewing-Angle-Dependence for Different Mass Ratio}

\begin{figure}[tbp]
\centering
\includegraphics[width = 0.32\linewidth , trim = 40 35 120 60, clip]{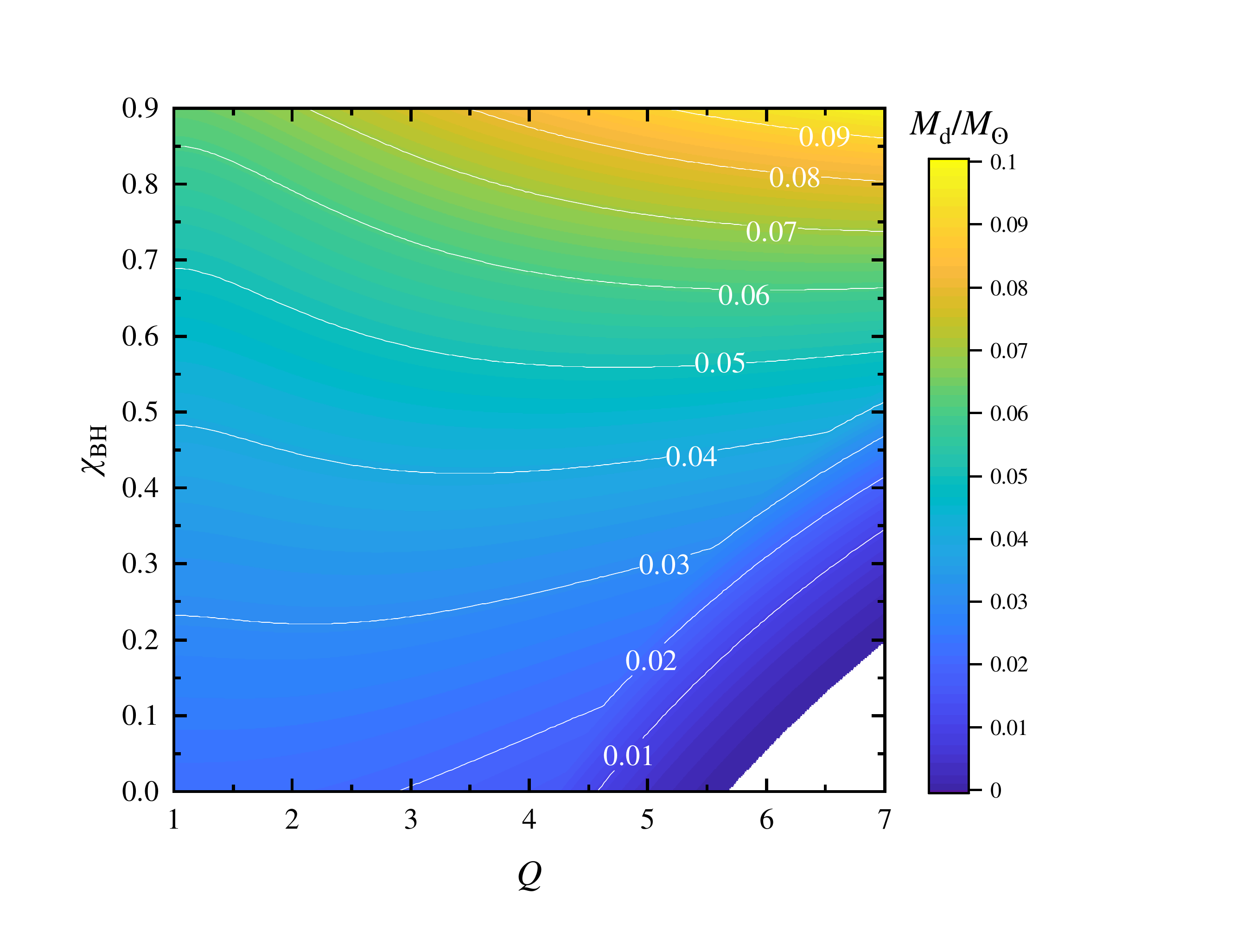}\quad\includegraphics[width = 0.32\linewidth , trim = 40 35 120 60, clip]{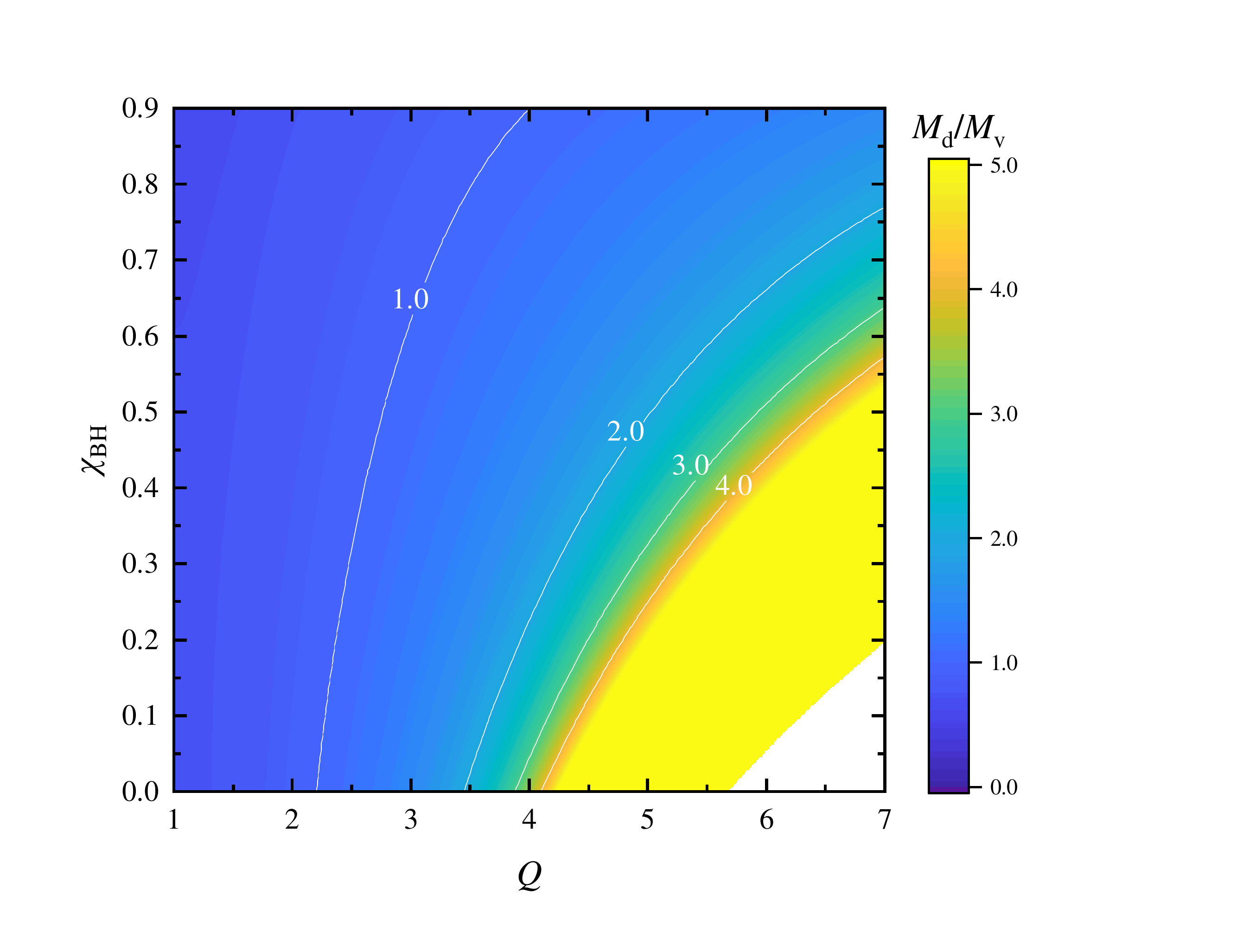}\quad\includegraphics[width = 0.32\linewidth , trim = 65 35 90 60, clip]{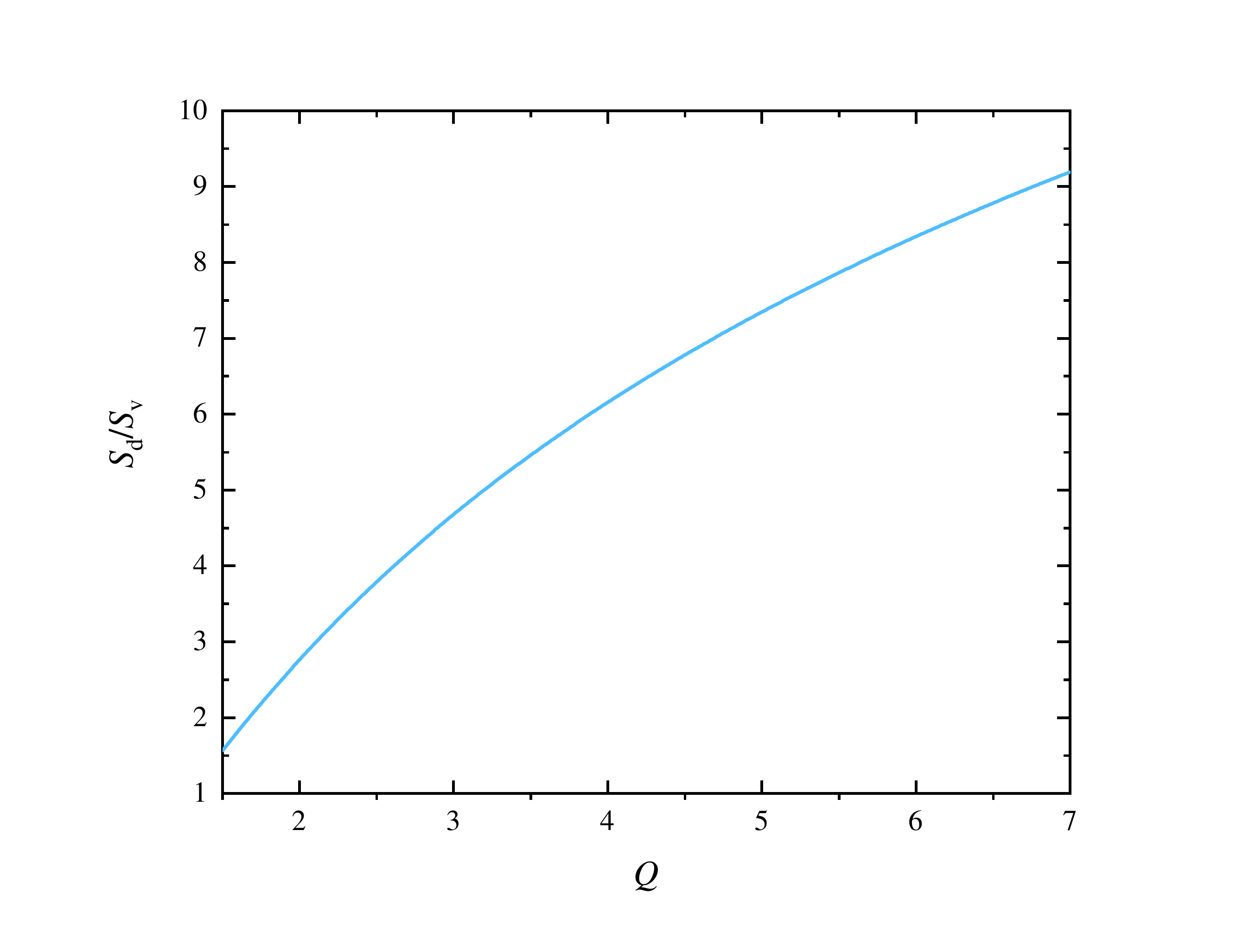}
\caption{Left panel: the parameter space in the $Q-\chi_{\rm BH}$ plane with color indicating the dynamical ejecta mass $M_{\rm d}$. The blank area represents no disruption happening when BH-NS mergers. Here, we show an example by assuming the NS mass is $M_{\rm BH} = 1.35\,M_\odot$ and the compactness of the NS is $C_{\rm NS} = 0.130$. Middle panel: similar with the left panel while color representing the ratio between dynamical ejecta mass $M_{\rm d}$ and viscosity-driven ejecta mass $M_{\rm v}$. Right panel: the face-on projected surface area ratio between dynamical ejecta and viscosity-driven ejecta depending on the mass ratio.  \label{fig:Ratio}}
\end{figure}

In the above subsections, we discussed the viewing angle effect on the emergent spectra and lightcurves by setting two cases with certain parameters. The masses of the dynamical ejecta and the viscosity-driven ejecta for both cases have a similar order of magnitude. The BH-NS binary system with a high mass ratio, e.g., $Q = 5$, can produce a relativistic tidal dynamical ejecta whose projected surface area is much larger than those of other components. Under these premises, we can thus draw the conclusions that the dynamcial ejecta is the main contributor to the kilonova emission of BH-NS mergers, and that the variation of the observed luminosity is only a factor of $\sim(2-3)$ as the viewing angle varies. However, for the near-equal-mass BH-NS mergers, simulation results from \cite{Foucart2019} showed that only a small amount of matter can become unbound dynamical ejecta. Moreover, \cite{Kyutoku2015} indicated that the lower energy material remaining outside the apparent horizon for a smaller value of $Q$ tend to form a slower dynamical ejecta. The typical velocity for a near-equal-mass BH-NS merger lies in the range of $\sim (0.1 - 0.15)\,c$ \citep{Foucart2019}, which can reduce the area of the dynamical ejecta. Therefore, it is necessary to discuss the properties of kilonovae produced from the near-equal-mass BH-NS mergers, and the applicability of our conclusions with different mass ratios $Q$.

In the left and the middle panels of Figure \ref{fig:Ratio}, by using our new fitting formula presented in Section \ref{sec:2}, we plot the dynamical ejecta mass and the mass ratio between the dynamical ejecta and the viscosity-driven ejecta associated with $Q$ and $\chi_{\rm BH}$ where we set a certain NS mass ($M_{\rm NS} = 1.35\,M_\odot$) and a stiff compactness of the NS ($C_{\rm NS} = 0.130$). Here, since the mass of the neutrino-driven ejecta is only $\sim 1/20$ of the mass of the viscosity-driven ejecta, we mainly compare the difference between the dynamical ejecta and the viscosity-driven ejecta. The right panel in Figure \ref{fig:Ratio} also describes the relation of $Q$ and the face-on projected surface area ratio between the dynamical ejecta and the viscosity-driven ejecta. As shown in Figure \ref{fig:Ratio}, for a mass range with $Q \gtrsim 3$, the projected area of the dynamical ejecta would be significantly larger than that of the viscosity-driven ejecta. Also, the lanthanide-rich dynamical ejecta would occupy a considerable portion of the material outside the remnant BH. Therefore, the emission from the dynamical ejecta would mainly contribute to the BH-NS merger kilonovae, while the observed luminosity only varies by a factor $\sim(2-3)$ due to the variation of the projected photosphere area of the dynamical ejecta with respect to the viewing angle.

For the near-equal-mass regime, as shown in Figure \ref{fig:Ratio}, most of the materials outside the remnant would form a bound disk to produce the viscosity-driven ejecta. Besides, the velocities of the unbound dynamical ejecta decrease, which can significantly reduce the area ratio between the dynamical ejecta and the viscosity-driven ejecta. It can be predicted that the kilonovae from near-equal-mass BH-NS mergers are much dimmer due to the lack of a fast moving dynamical ejecta. The dynamical ejecta also cannot block the emission of the other two components at any viewing angle, so that observers would simultaneously observe the emission from all components along the line of sight. As a result, the viewing-angle variation of the observed luminosity and the variation of the lightcurve peak time due to the photon propagation effect are both minor.

\subsection{Comparison With GW170817/AT2017gfo}

Even though AT2017gfo was powered by a BNS merger rather than a BH-NS merger, it is still interesting to compare the two since AT2017gfo is the most carefully studied kilonova so far. In order to better see the differences between the emergent spectra of BH-NS mergers and that of AT2017gfo, we compare the face-on ($\theta_{\rm view} = 0^\circ$, $\varphi_{\rm view} = 0 ^ \circ$)\footnote{The fitting results for the viewing angle of AT2017gfo is in the range of $\theta_{\rm view}\approx20-40^\circ$ (e.g., \citealt{Alexander2017,Lyman2018,Margutti2017}). As discussed in Section \ref{sec:4.4}, the variation of BH-NS merger lightcurves is insignificant if $\theta_{\rm view}\leq30^\circ$. Therefore, we can consider that the comparison between AT2017gfo and BH-NS merger kilonovae is along the same line of sight. }  emergent spectra which are presented in Figure \ref{fig:ViewingAngleSpectrum} with the blackbody fits to the photometric data of AT2017gfo \citep{Waxman2018} at the same epoch after the merger. The photometric data are taken from \citep{Villar2017} who collected data from \cite{Arcavi2017,Coulter2017,Cowperthwaite2017,Diaz2017,Drout2017,Evans2017,Hu2017,Kasliwal2017,Lippuner2017,Pian2017,Pozanenko2018,Shappee2017,Smartt2017,Tanvir2017,Troja2017,Utsumi2017,Valenti2017}. In Figure \ref{fig:ViewingAngleLightCurve}, we compare the predicted bolometric and multi-band lightcurves of BH-NS mergers with the AT2017gfo lightcurves. In the top panels of Figure \ref{fig:ViewingAngleLightCurve}, the bolometric luminosity points (gray points) of AT2017gfo are directly taken from \cite{Waxman2018} while the fitting lightcurve (gray line) corresponds to the DZ31 model presented in \cite{Wu2019}. The multi-band photometric points of the bottom panels are also taken from \cite{Villar2017}.

As shown in Figure \ref{fig:ViewingAngleSpectrum} and Figure \ref{fig:ViewingAngleLightCurve}, the predicted kilonova emission from the BH-NS mergers is dimmer than AT2017gfo at early times but are possible to be more luminous at late times if the remnant mass is large enough. This is because different from the BH-NS merger case that favors mainly one emission component by the dynamical ejecta, the BNS case may have comparable contributions from at least two (blue and red) components. The early-stage of AT2017gfo can be explained by a lanthanide-free blue component which may be generated by shocked-heating \citep[e.g.,][]{Oechslin2006,Radice2016,Sekiguchi2016,Wanajo2014} or neutrino irradiation from the remnant massive NS \citep[e.g.,][]{Metzger2014,Perego2014,Yu2018}. The late-stage emission is mainly contributed from a lanthanide-rich red component, likely from the tidal dynamical ejecta \citep[e.g.,][]{Bauswein2013,Goriely2011,Korobkin2012,Radice2016}. Without considering energy injection \citep[cf.][]{Yu2018,Li2018}, the total mass of the blue component to explain the high luminosity of AT2017gfo lies in the range of $\sim 0.01-0.025\,M_\odot$ \citep[e.g.,][]{Cowperthwaite2017,Murguia-Berthier2017,Perego2017,Kasen2017,Kasliwal2017,Tanaka2017,Villar2017}. NR simulations indicated that the bound disk mass is always $\lesssim 0.3\,M_\odot$ \cite{Kyutoku2015}, which corresponds to the neutrino-driven ejecta mass of $\lesssim 3\times 10^{-3}\,M_\odot$ in BH-NS mergers. This is much less than that required to explain  AT2017gfo. Therefore, the lack of a large-mass blue component ejecta in BH-NS mergers makes their kilonovae much dimmer than AT2017gfo at early times. More specifically, the BH-NS merger kilonovae are always less luminous than $4.5\times10^{41}\,{\rm erg\, s^{-1}}$, which is $\sim 1 / 4$ of the bolometric luminosity of AT2017gfo. On the other hand, since the dynamical ejecta mass can be up to $\sim0.1\,M_\odot$ \citep{Kyutoku2015} which is much larger than that invoked to interpret AT2017gfo, the kilonovae of BH-NS mergers can be more luminous than AT2017gfo at the late times. Figure \ref{fig:ViewingAngleSpectrum} also shows another significant difference between BH-NS kilonovae and AT2017gfo in terms of the blackbody fit temperatures. The temperatures of BH-NS merger kilonovae are only $\lesssim4/5$ of that of AT2017gfo at the same epoch after the merger.

One can conclude that the kilonovae from BH-NS mergers are optically dim, but possibly infrared bright compared with GW170817/AT2017gfo. The differences in the temperature evolution and lightcurves may be used to differentiate BNS mergers from BH-NS mergers.

\section{Gamma-Ray Burst Afterglow \label{sec:5}}

Another EM counterpart for BH-NS mergers is a sGRB and its broadband afterglows. The sGRB afterglow lightcurve sensitively depends on the viewing. It is therefore interesting to simultaneously model the predicted lightcurves for both kilonova and sGRB afterglow from the BH-NS mergers. In this section, we discuss the viewing-angle-dependent lightcurves for kilonovae and sGRB afterglows, and how the different parameter values affect the detectability of the kilonova for the on-axis configuration.

In BH-NS mergers with tidal disruption, the remnant BH would accrete from the remnant disk and launch a relativistic jet via the Blandford-Znajek mechanism \citep{Blandford1977}. The kinetic energy of the jet \citep{Barbieri2019} may be estimated as
\begin{equation}
    E_{\rm K,jet} = \epsilon(1 - \xi_{\rm w} - \xi_{\rm s})M_{\rm disk}c^2\Omega_{\rm H}^2f(\Omega_{\rm H}),
\end{equation}
where $\epsilon = 0.015$, $\Omega_{\rm H}$ is the dimensionless angular frequency at the horizon which is determined by the final spin of the BH,
\begin{equation}
    \Omega_{\rm H} = \frac{\chi_{\rm BH,f}}{2(1 + \sqrt{1 - \chi_{\rm BH,f}^2})},
\end{equation}
and $f(\Omega_{\rm H}) = 1 + 1.38\Omega_{\rm H}^2 - 9.2\Omega_{\rm H}^4$ is a correction factor for high-spin values \citep{Tchekhovskoy2010}. We use Equation (11) from \cite{Pannarale2013} to calculate the final spin of the BH.

We apply a power-law structured jet model \citep{Zhang2002,Rossi2002}. The angular distributions of the kinetic energy and Lorentz Factor $\Gamma$ are adopted as \citep{Ghirlanda2019,Salafia2019}
\begin{equation}
\begin{split}
    \frac{{\rm d}E}{{\rm d}\Omega}(\theta) &= \frac{E_{\rm c}/4\pi}{1 + (\theta/\theta_{\rm c})^{s_1}},\\
    \Gamma(0 , \theta) &= 1 + \frac{\Gamma_c - 1}{1 + (\theta/\theta_{\rm c})^{s_2}},
\end{split}
\end{equation}
where we set $\Gamma_{\rm c} = 250$, $\theta_{\rm c} = 5^\circ$, $s_1 = 5.5$ and $s_2 = 3.5$, and $E_{\rm c} = E_{\rm K,jet}/\pi\theta_{\rm c}^2$. These parameters have been used to model GW170187/GRB 170817A. The standard GRB afterglow model is briefly introduced in Appendix \ref{app:B}. 

\subsection{Viewing Angle Dependence}

\begin{figure}[htbp]
\centering
\includegraphics[width = 0.49\linewidth , trim = 50 30 100 60, clip]{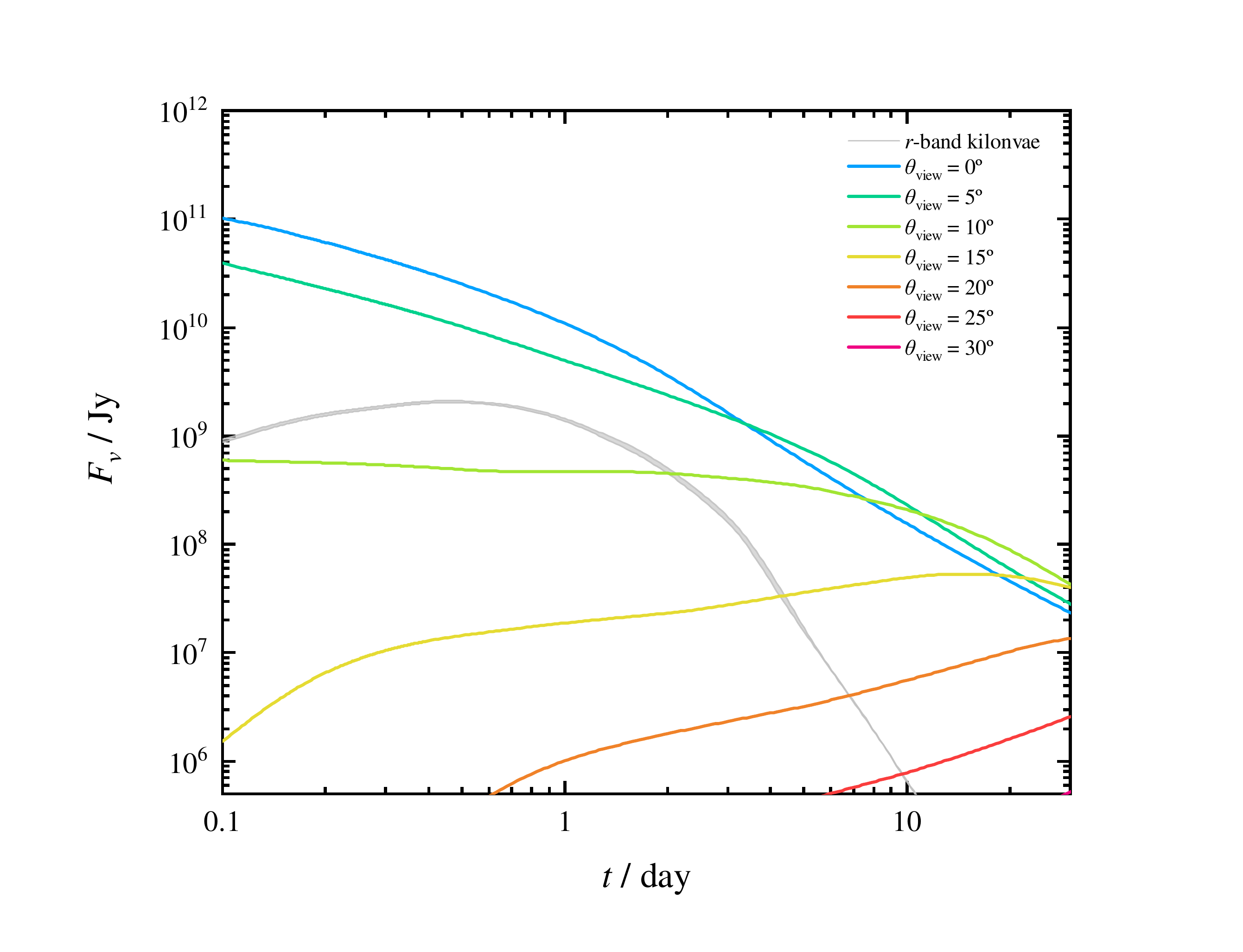}\quad\includegraphics[width = 0.49\linewidth , trim = 50 30 100 60, clip]{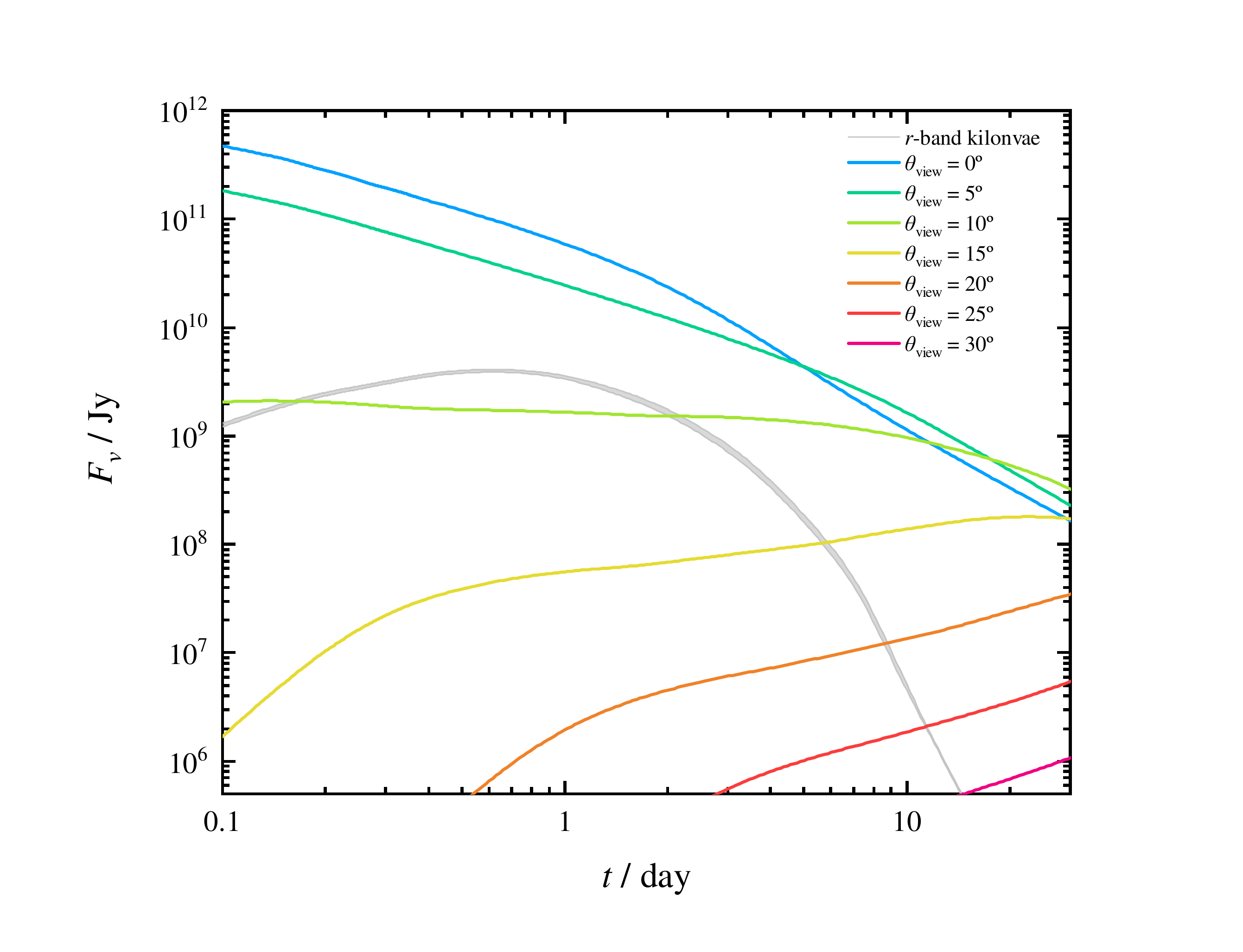}
\includegraphics[width = 0.49\linewidth , trim = 50 30 100 60, clip]{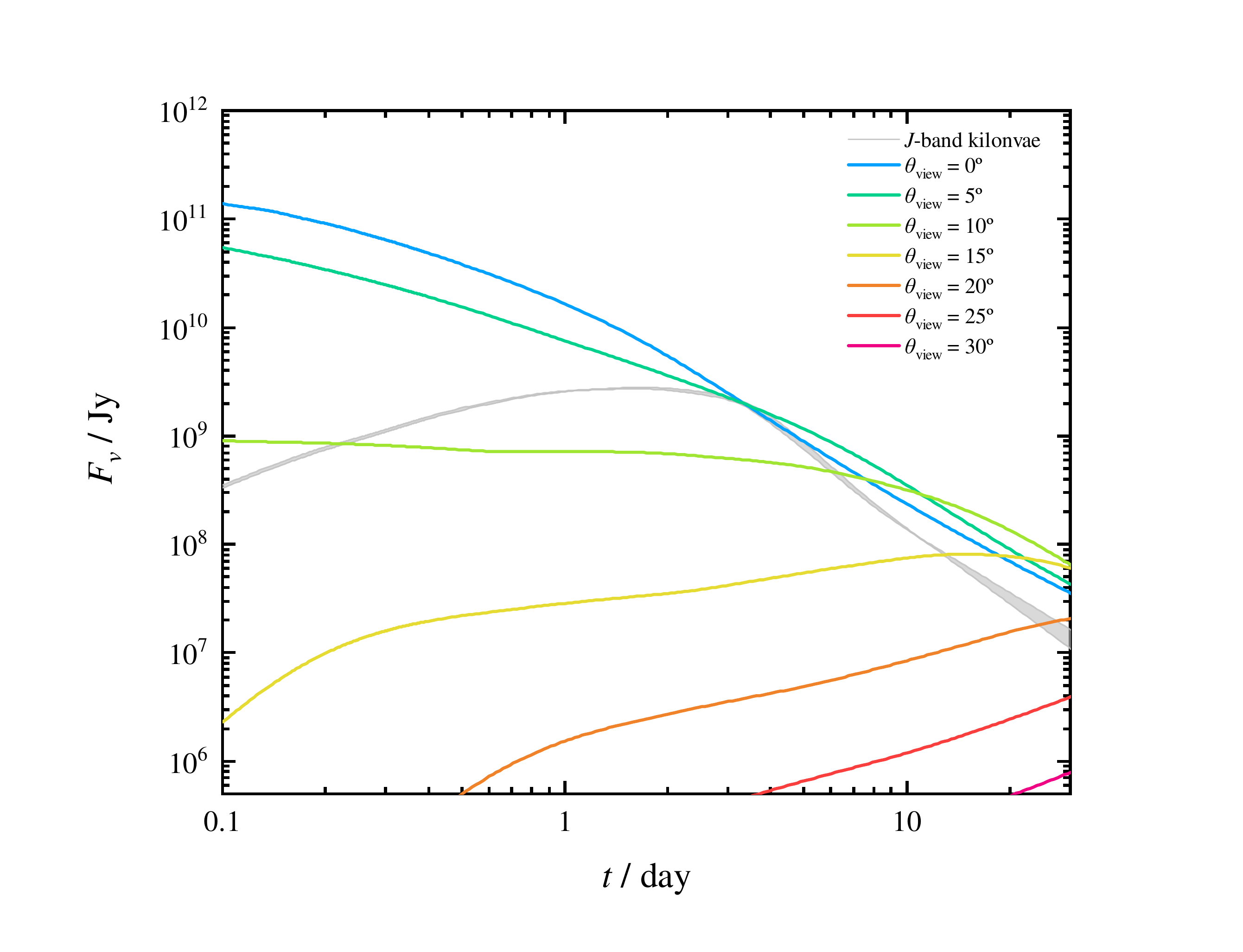}\quad\includegraphics[width = 0.49\linewidth , trim = 50 30 100 60, clip]{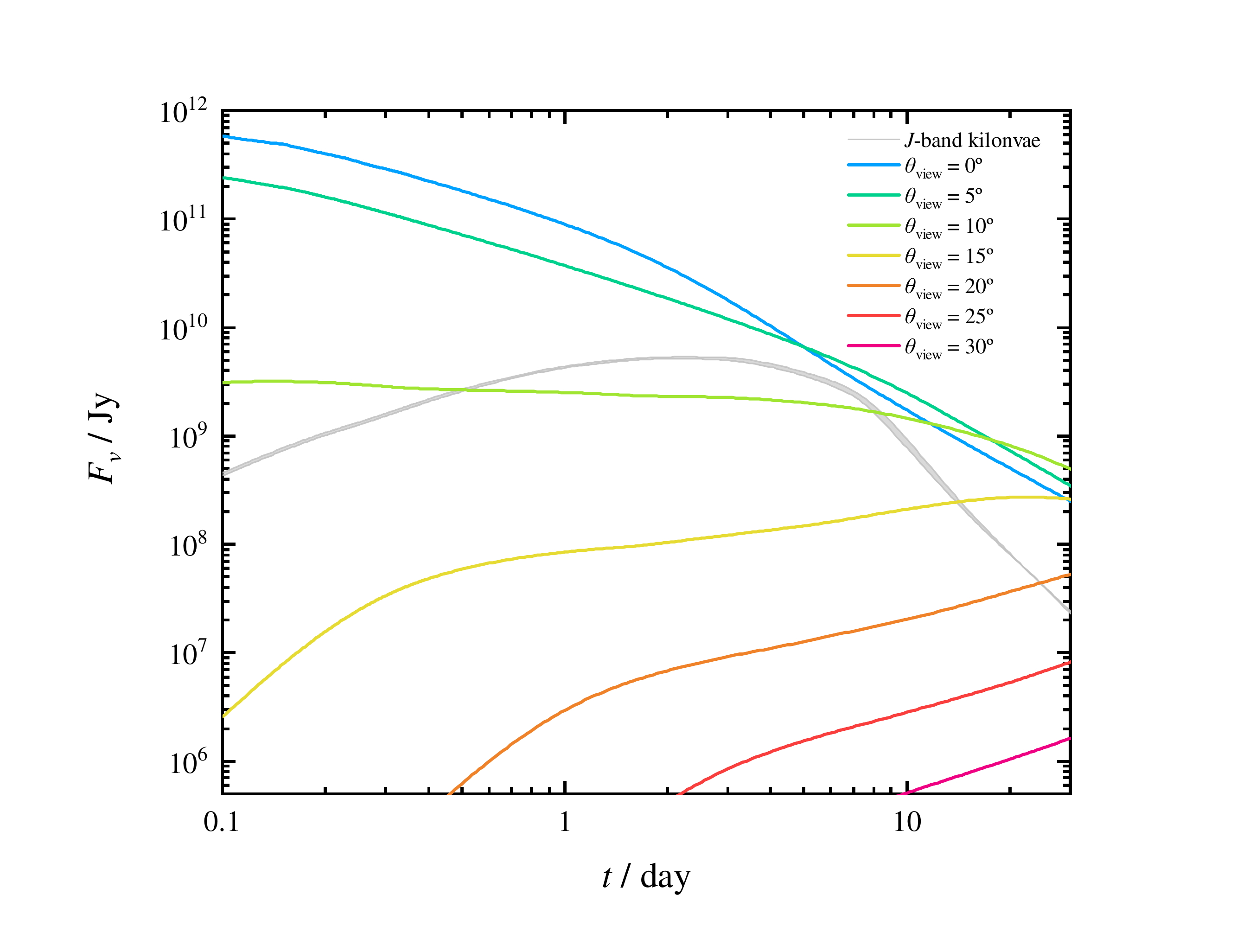}
\includegraphics[width = 0.49\linewidth , trim = 50 30 100 60, clip]{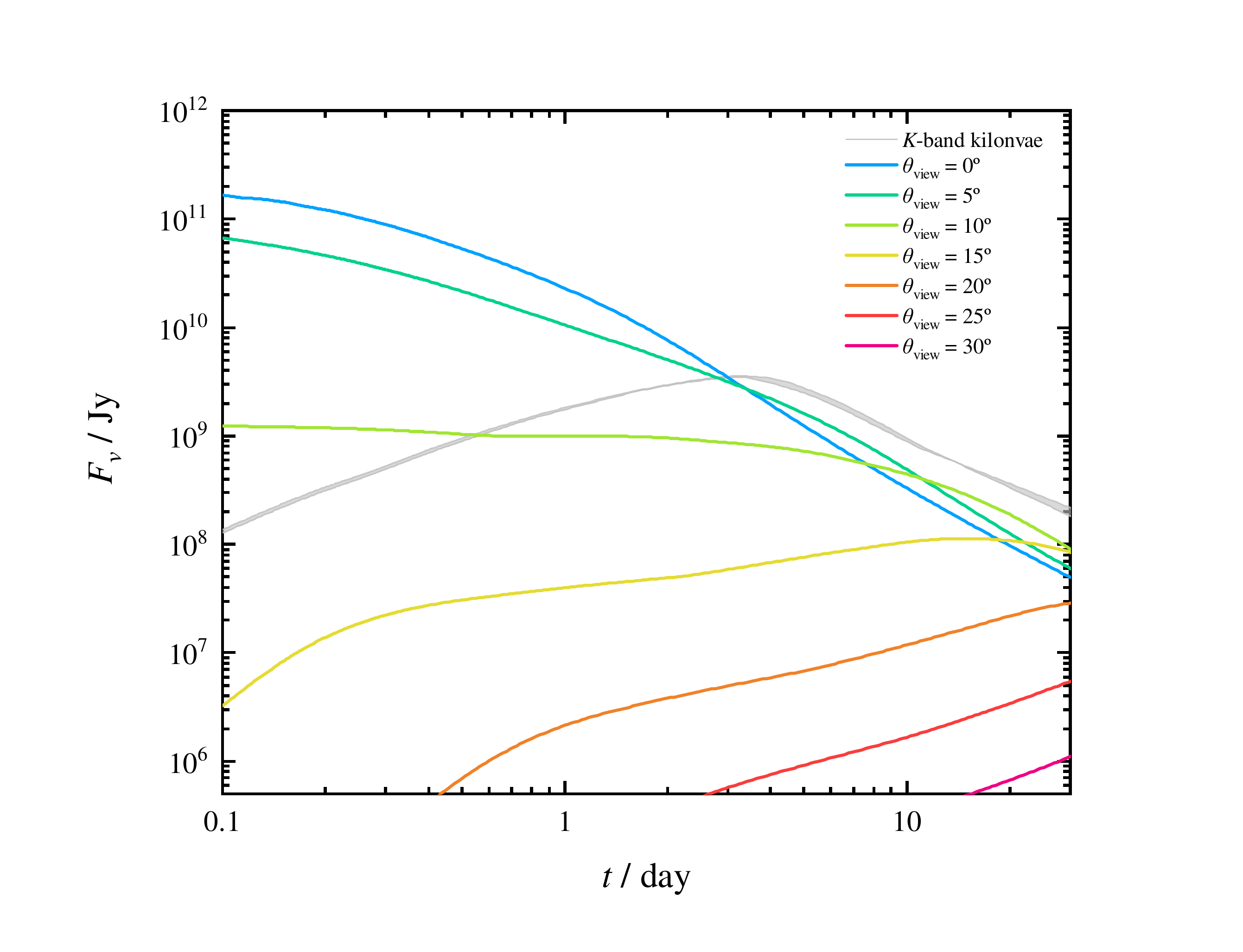}\quad\includegraphics[width = 0.49\linewidth , trim = 50 30 100 60, clip]{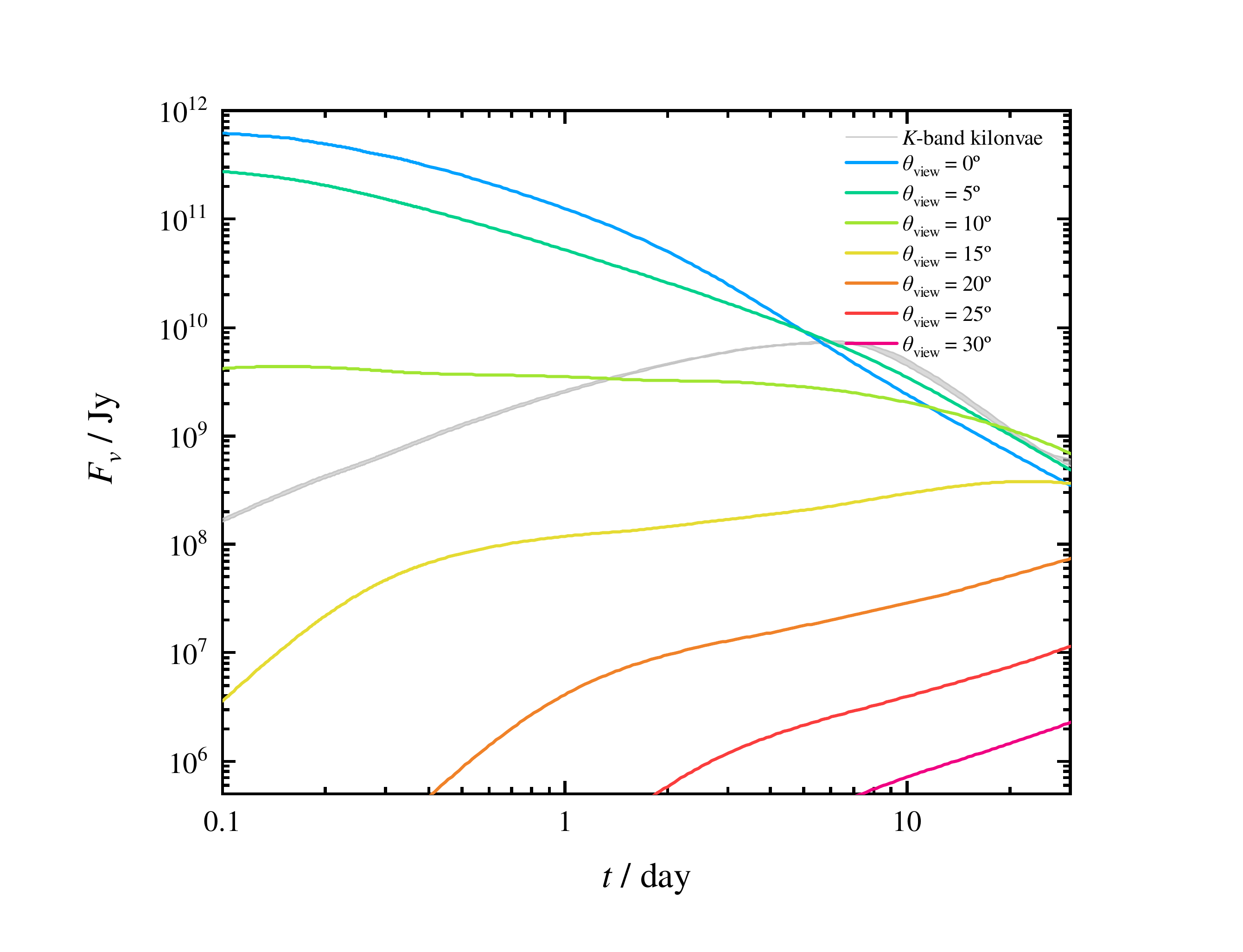}
\caption{Viewing-angle-dependent $rJK$-band kilonovae lightcurves compared with the sGRB afterglow lightcurves for Case I (left panels) and Case II (right panels). From top to bottom, we show r-band, J-band and K-band lightcurves. Seven $\theta_{\rm view}$ values are calculated: $\theta_{\rm view} = 0^\circ$, $5^\circ$, $10^\circ$, $15^\circ$, $20^\circ$, $25^\circ$ and $30^\circ$. The range for kilonova lightcurves (gray regions) span from $\theta_{\rm view}= 0^\circ$ to $\theta_{\rm view}= 30^\circ$. 
\label{fig:AfterGlow}}
\end{figure}

In our calculations to discuss the viewing-angle-dependent lightcurves of sGRB afterglows, the following typical afterglow model parameters are adopted: the fraction of shock energy carried by magnetic field $\epsilon_B = 0.001$, the fraction of shock energy carried by electrons $\epsilon_e = 0.1$, and the ISM number density $n = 5\times10^{-3}\,{\rm cm}^{-3}$.

In Figure \ref{fig:AfterGlow}, we present the lightcurves in three representative filters, i.e., the $r$-, $J$-, and $K$-bands. Both the kilonova lightcurves and afterglow lightcurves are shown for both Case I (left panel) and Case II (right panel). We adopt several possible viewing angles from $\theta_{\rm view}= 0^\circ$ to $30^\circ$. One can see that the change of kilonova lightcurves is essentially negligible.

For an on-axis view, i.e., $\theta_{\rm view}=0^\circ-5^\circ$ ($\theta_{\rm view} = 0-\theta_{\rm c}$), the $r$-band flux density from the kilonovae is much less than the $r$-band flux density from the afterglow. For $J$-band, the kilonovae are less luminous than afterglow most of the time, but may show up in $\sim(2-7)\,$days when the $J$-band flux densities between the two are comparable. The $K$-band emission after $\sim2\,$days becomes dominated by the kilonova emission. Therefore, along the on-axis line of sight, the best filter to observe kilonovae from BH-NS mergers is $K$-band, even though the emission shows up at a relatively late epoch. 

For an off-axis view, the early-stage kilonova emission is always much luminous than the sGRB afterglow emission. The effect becomes more prominant as the viewing angle increases. However, as the sGRB blastwave decelerates (i.s. Lorentz factor $\Gamma$ of the external shock decreases), the more energetic jet core becomes visible, so that the afterglow lightcurves continuously rise and eventually outshines the kilonova emission in $r$- and $J$- bands. The $K$-band is likely dominated by the kilonova emission for an extended period of time. 

\subsection{Parameter Dependence}

\begin{figure}[tbp]
\centering
\includegraphics[width = 0.32\linewidth , trim = 50 30 100 60, clip]{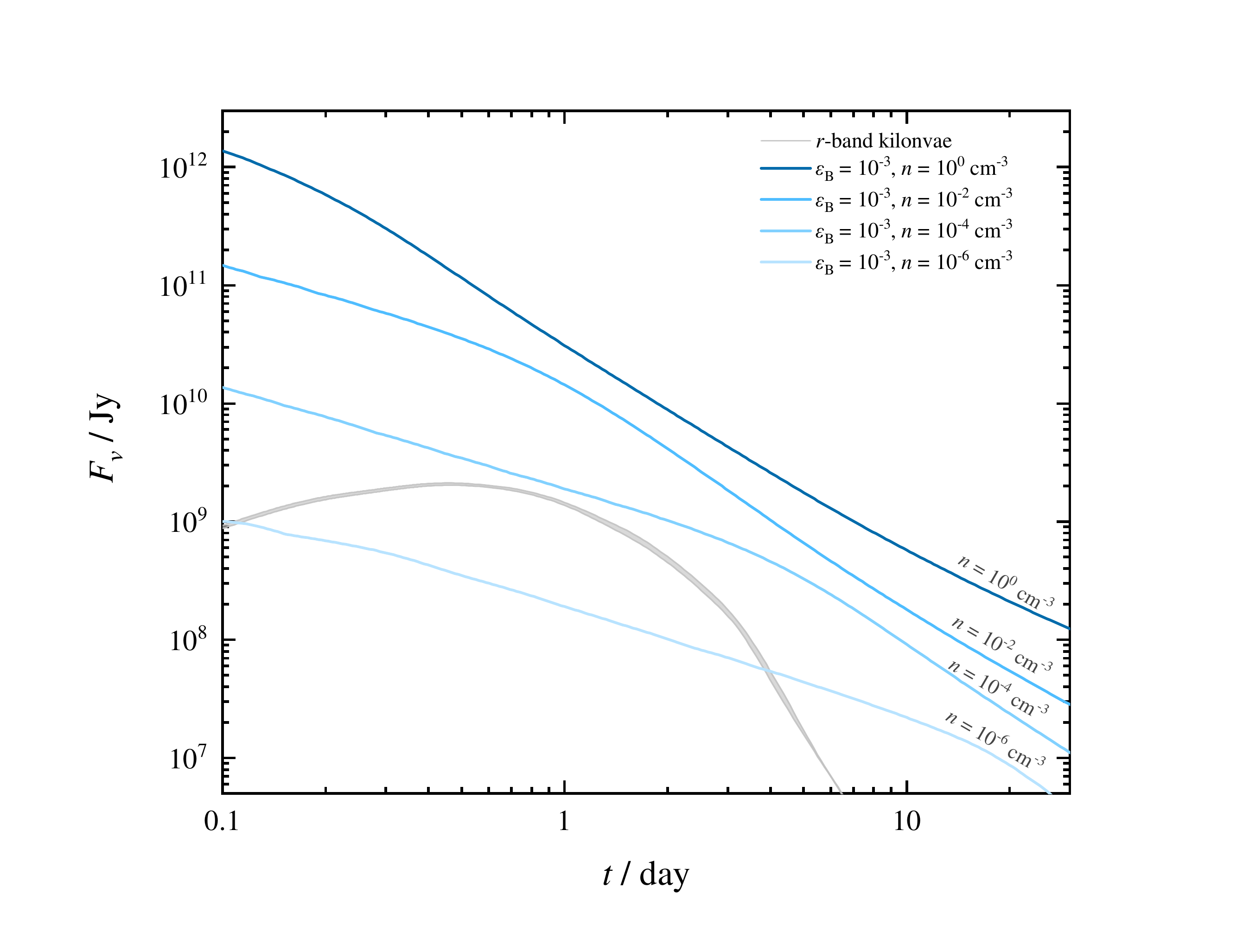}\quad\includegraphics[width = 0.32\linewidth , trim = 50 30 100 60, clip]{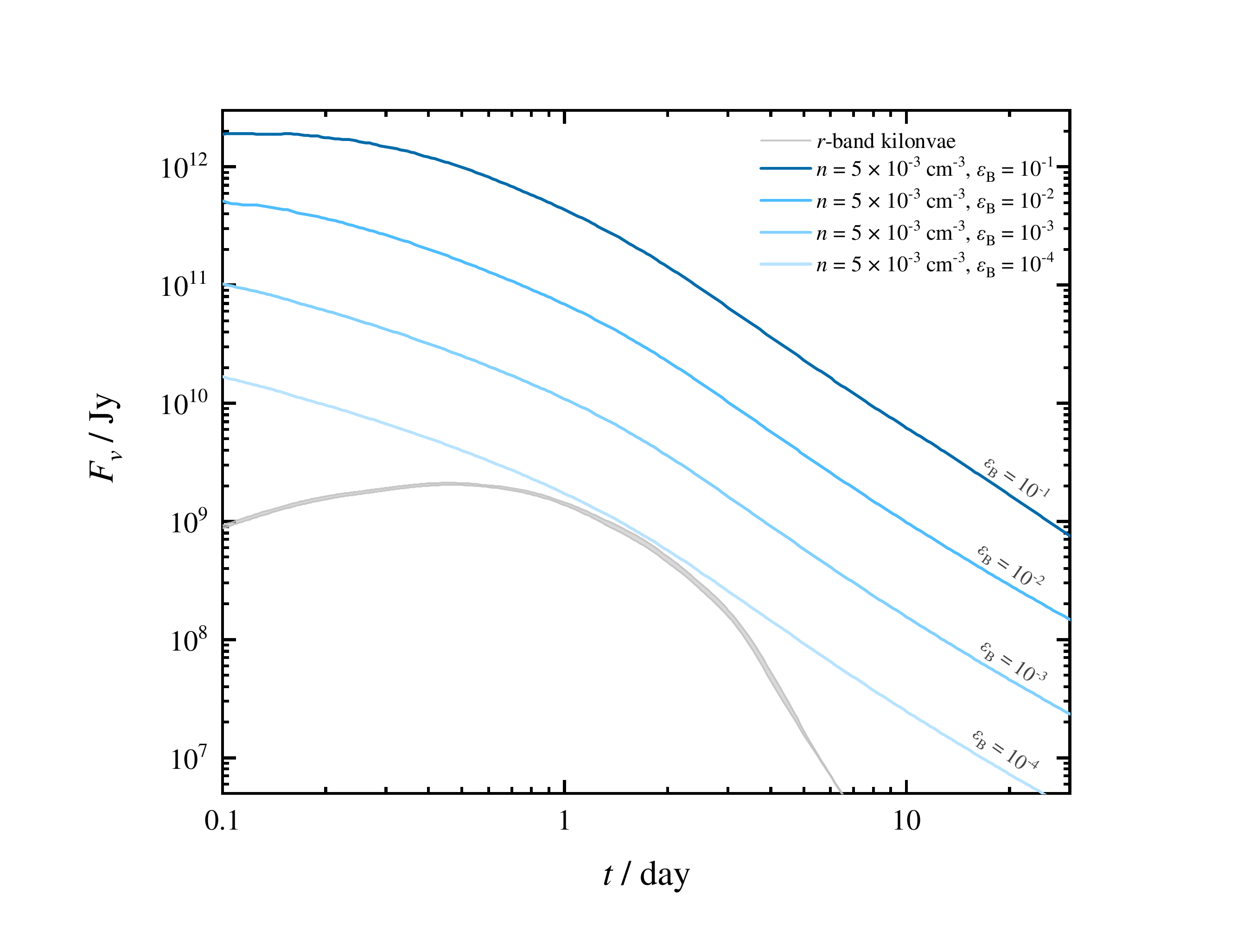}\quad\includegraphics[width = 0.32\linewidth , trim = 50 30 100 60, clip]{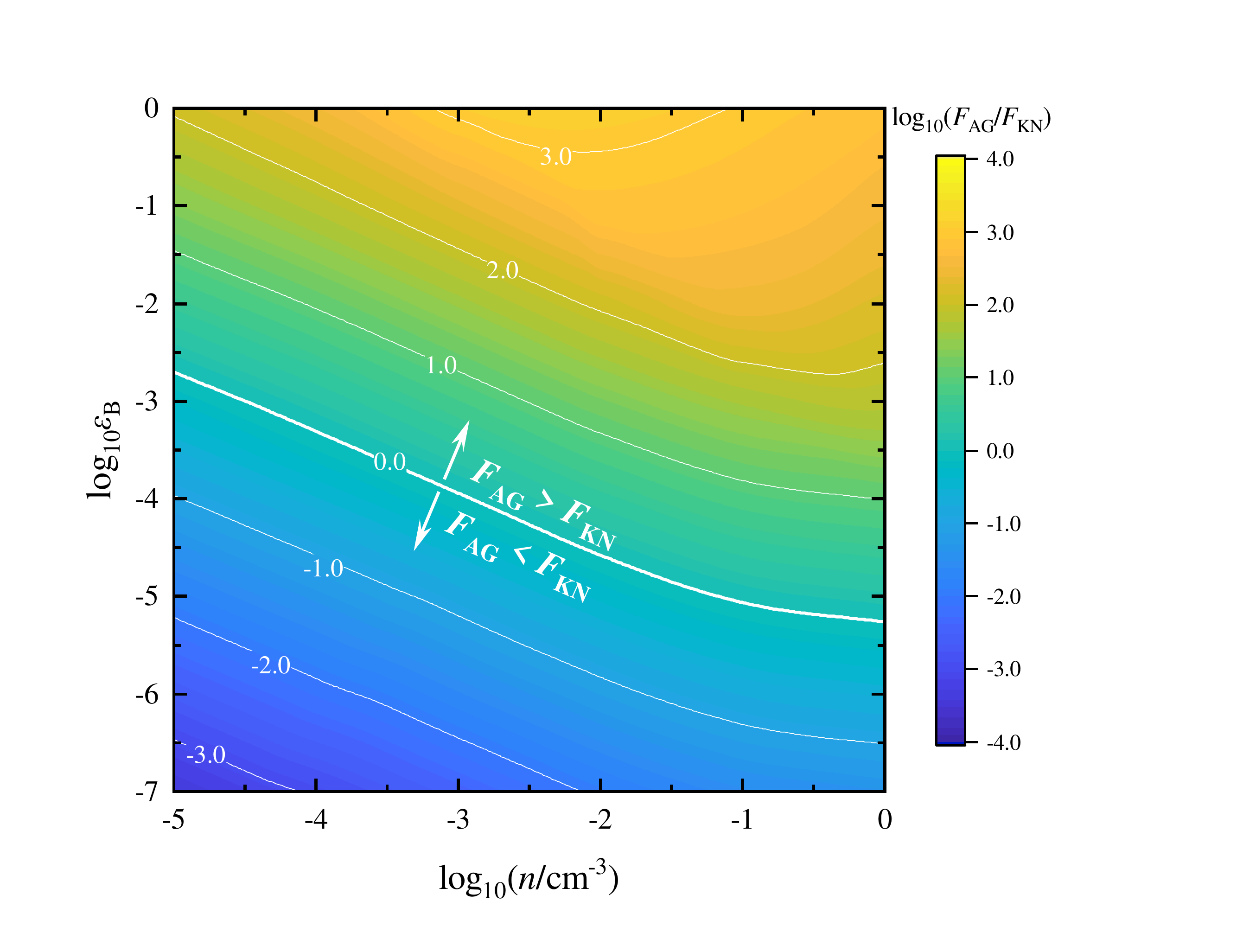}
\includegraphics[width = 0.32\linewidth , trim = 50 30 100 60, clip]{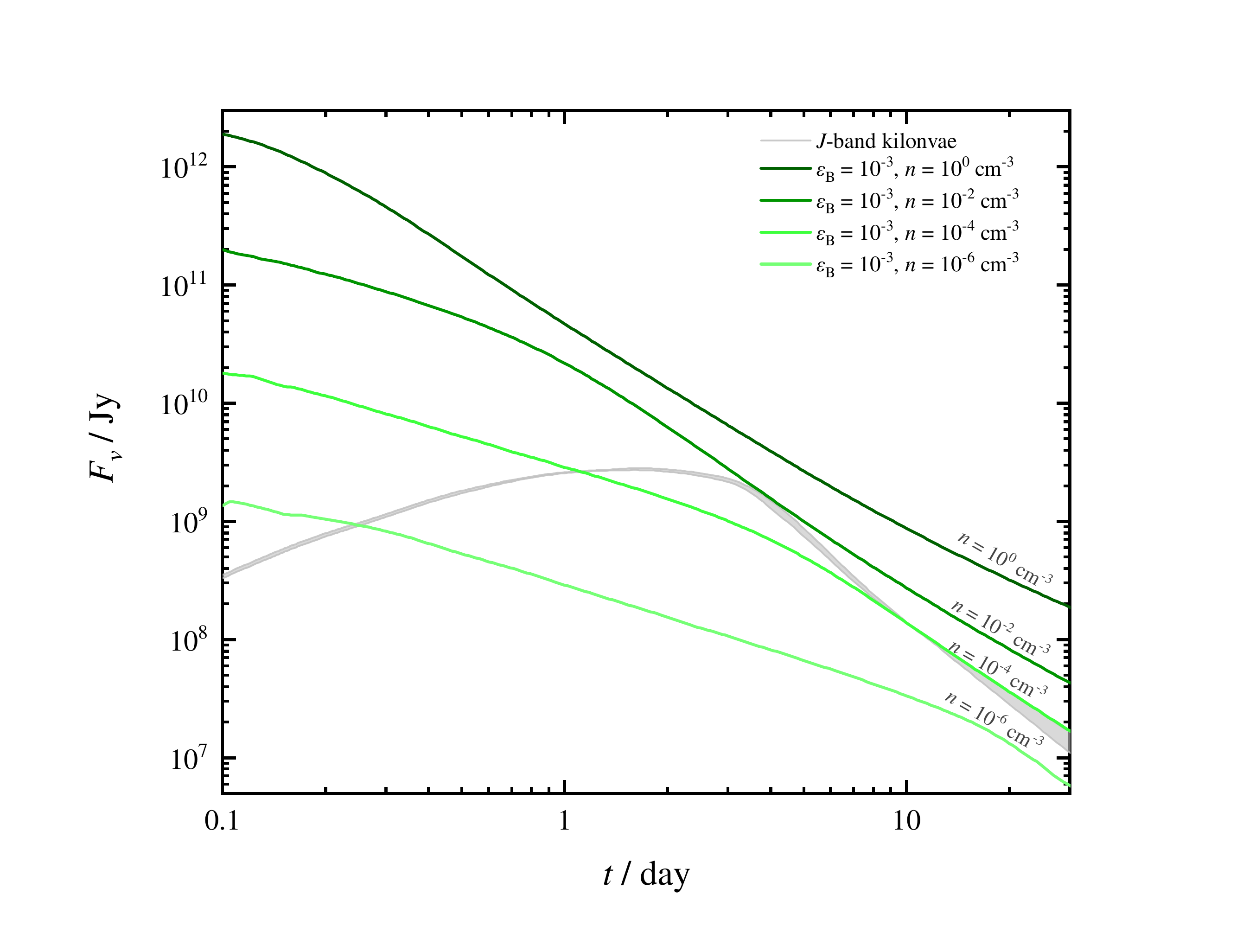}\quad\includegraphics[width = 0.32\linewidth , trim = 50 30 100 60, clip]{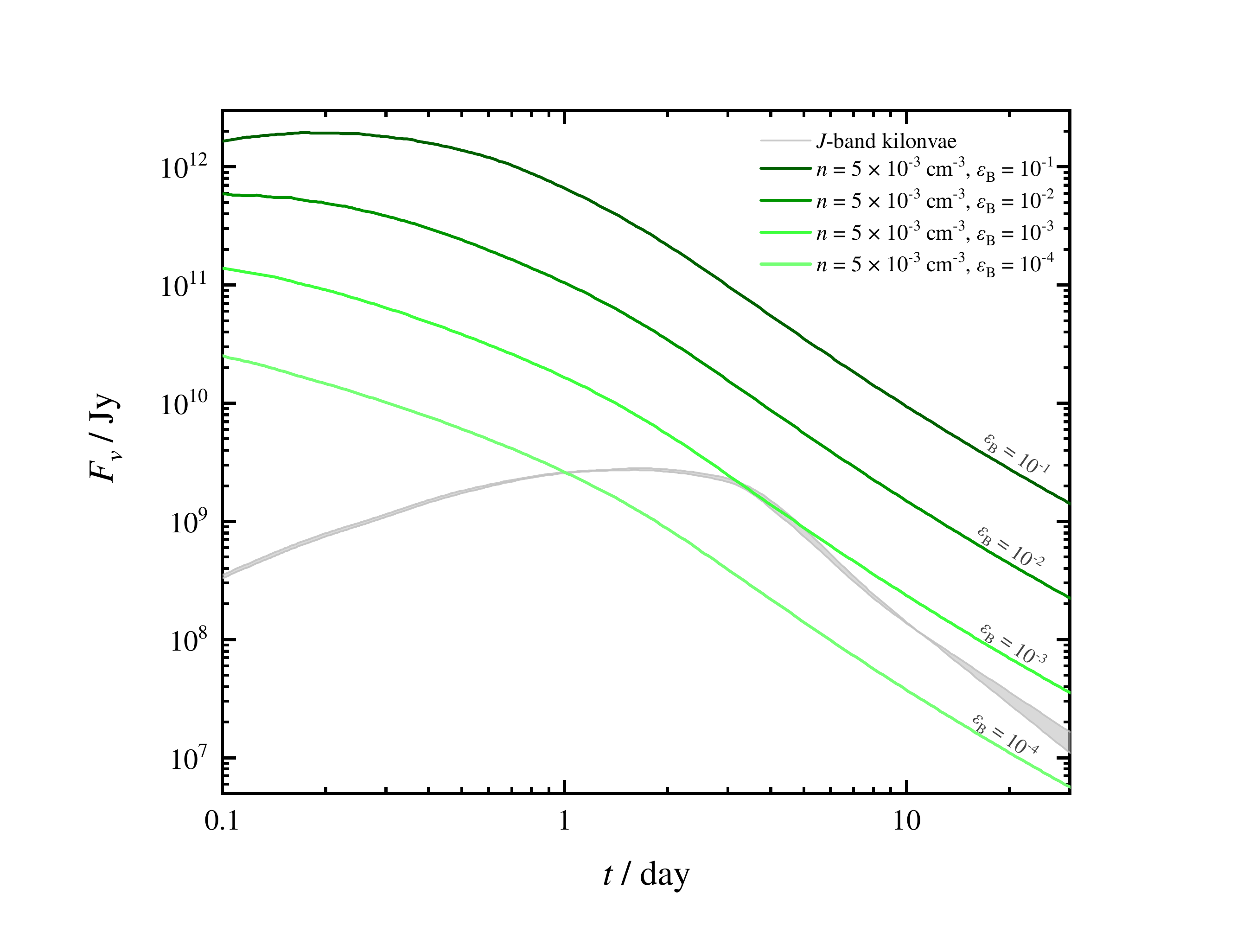}\quad\includegraphics[width = 0.32\linewidth , trim = 50 30 100 60, clip]{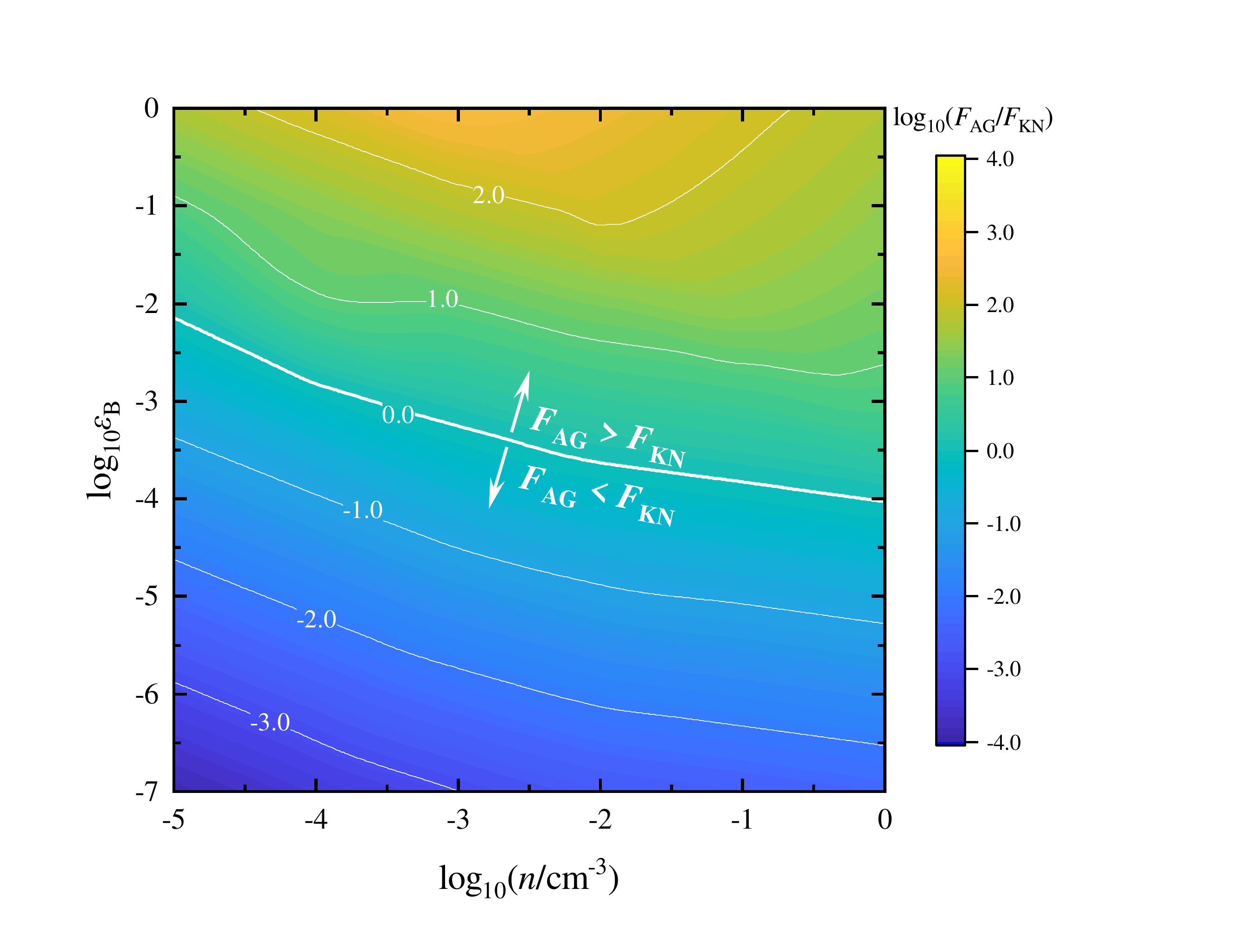}
\includegraphics[width = 0.32\linewidth , trim = 50 30 100 60, clip]{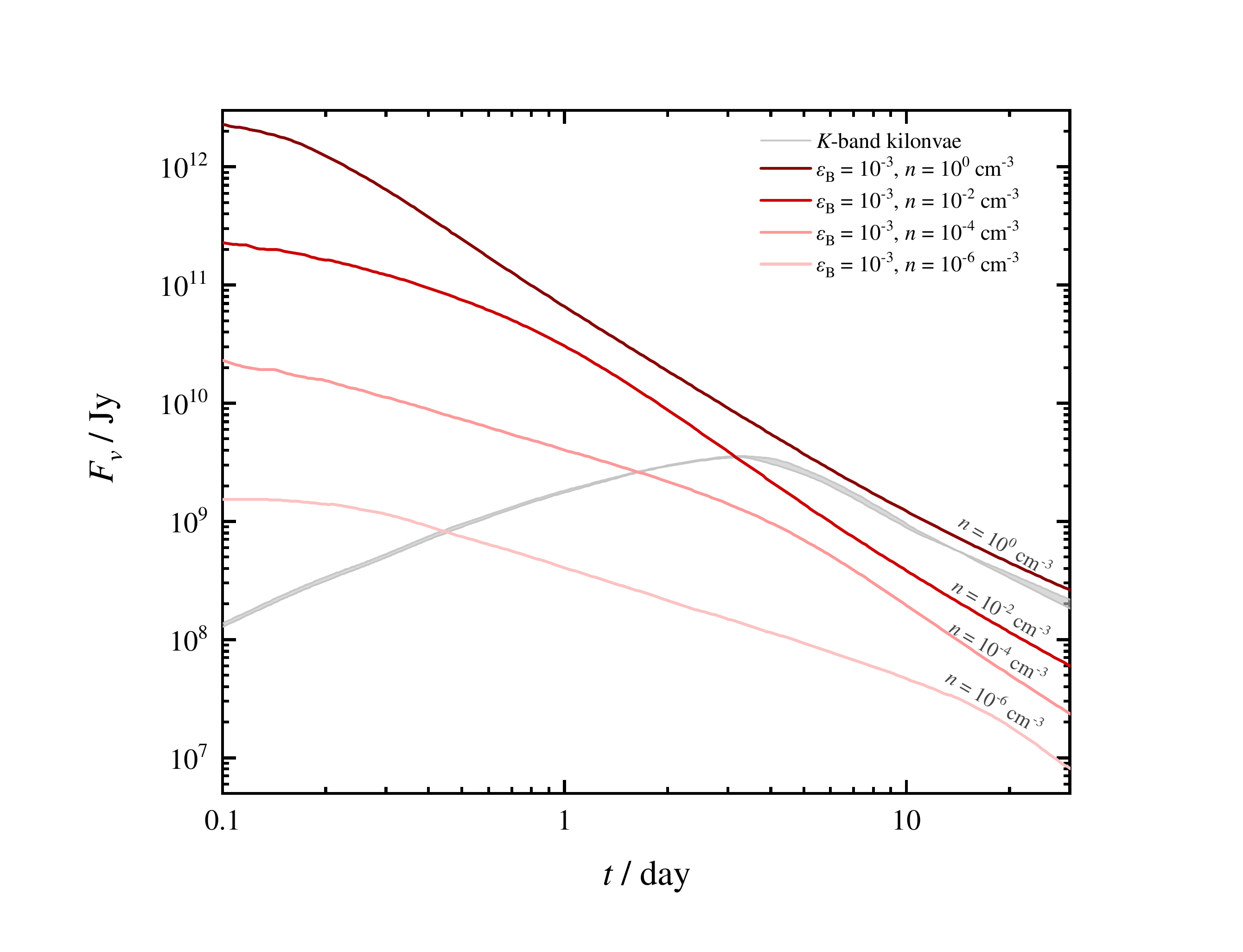}\quad\includegraphics[width = 0.32\linewidth , trim = 50 30 100 60, clip]{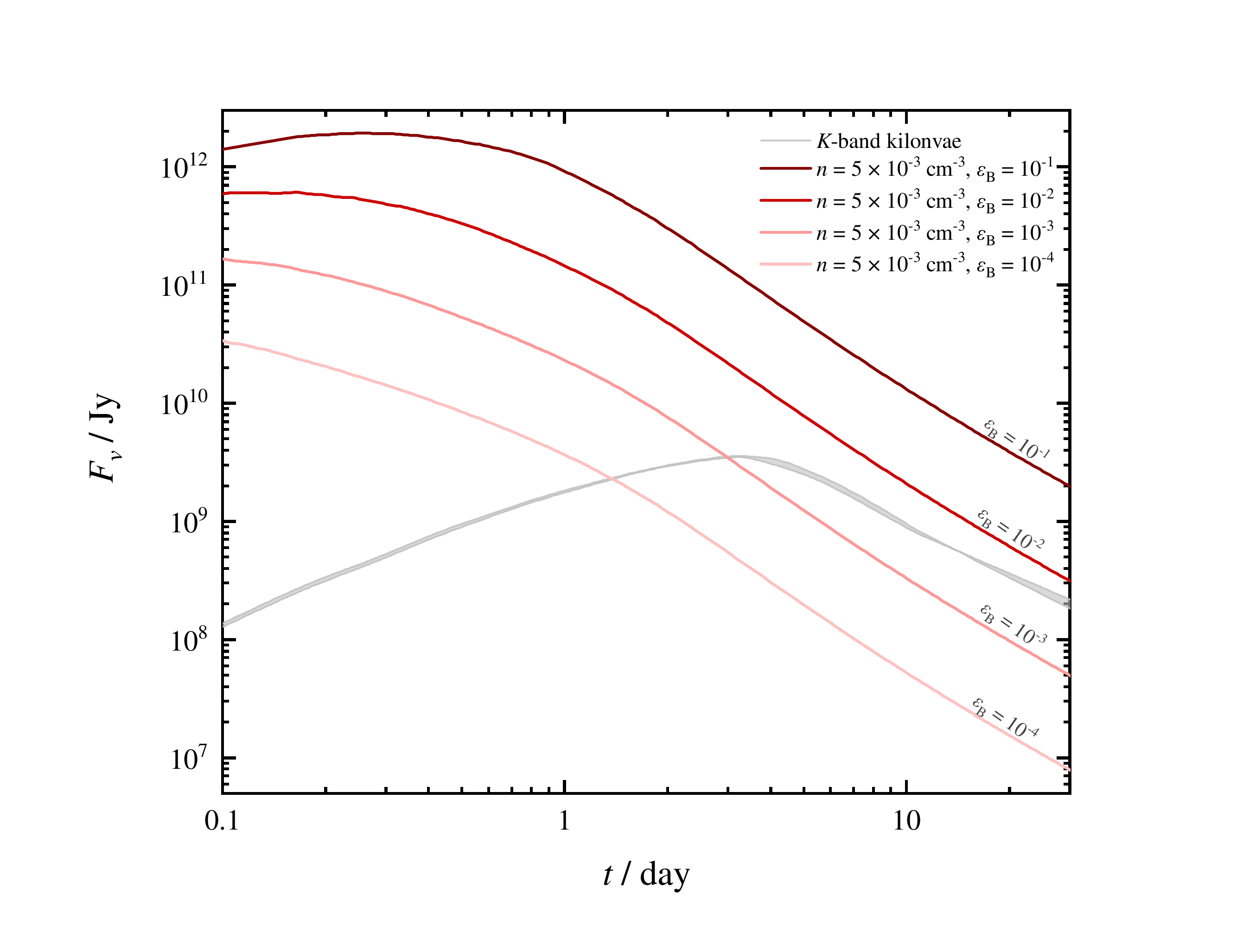}\quad\includegraphics[width = 0.32\linewidth , trim = 50 30 100 60, clip]{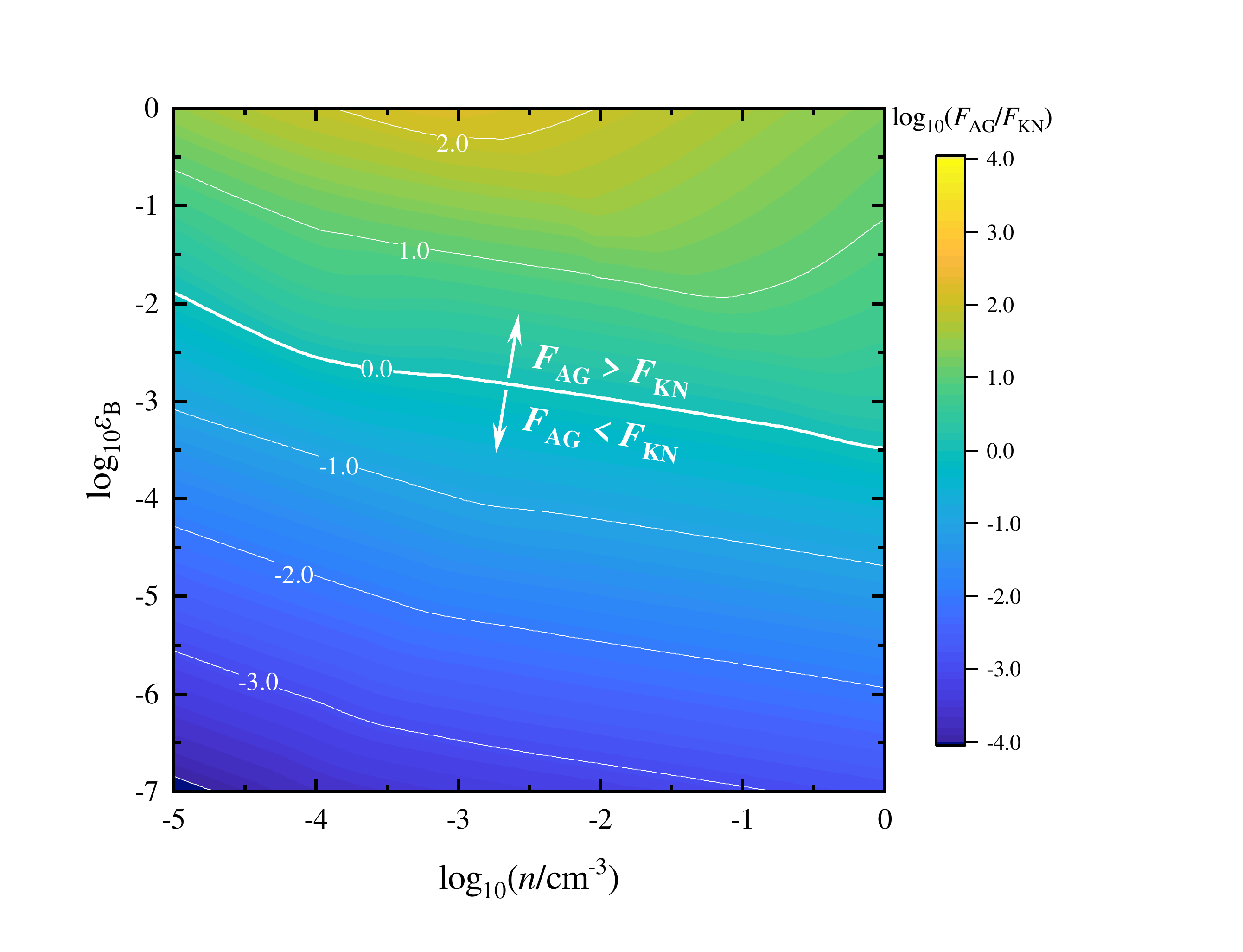}
\caption{Left panels: face-on $rJK$-band kilonovae lightcurves compared with the sGRB afterglow lightcurves, where we set $\epsilon_B = 10^{-3}$ as constant. The color lines from dark to light represent four possible ISM density $n$ values: $1$, $10^{-2}$, $10^{-4}$, and $10^{-6}\,{\rm cm}^{-3}$. Middle panels: face-on $rJK$-band kilonovae lightcurves compared with the sGRB afterglow lightcurves, where we set $n = 5\times10^{-3}\,{\rm cm}^{-3}$ as constant. The color lines from dark to light represent four possible $\epsilon_B$ values: $10^{-1}$, $10^{-2}$, $10^{-3}$, and $10^{-4}$. Right panels: the parameter space in the $\log_{10}n-\log_{10}\epsilon_B$ plane with the color contour representing the logarithmic flux density ratio between the afterglow and kilonova emissions at the peak time of the kilonova lightcurve for each band. The lines in bold represent that the flux density of afterglow is equal to that of the kilonova emission. 
\label{fig:AfterGlow2}}
\end{figure}

In order to discuss the parameter space that the kilonova can be observed in the on-axis view, we only show the comparison between the kilonova lightcurves and the afterglow lightcurves for Case I, since similar results can be obtained for Case II. 

The left and middle panels of Figure \ref{fig:AfterGlow2} respectively show the effects of the ISM number density $n$ and the fraction of the shock energy carried by magnetic field $\epsilon_B$ on the afterglow lightcurve. Here, we set the fraction of the shock energy carried by electrons is $\epsilon_e = 0.1$, since the variation of this parameter constrained by observations have a relatively small change (e.g., \citealt{Santana2014}). The right panels of Figure \ref{fig:AfterGlow2} compare the difference of the flux density between the afterglow component and the kilonova component at the peak time of the kilonova lightcurve for each band. One can still conclude that the best filter to observe the kilonova from a BH-NS merger is $K$-band. Compared with the $J$-band and $K$-band in which the kilonova emission is observable for a large parameter space, the optical emission of the kilonova can only be observed in a very low-density environment with low $\epsilon_B$.

Observationally, \cite{Fong2015} (see also \citealt{O'Connor2020}) found that most of sGRBs occur in low-density environments with a median density  $3-15\times 10^{-3}\,{\rm cm}^{-3}$. The systematic studies on magnetic fields in external forward shocks \citep{Santana2014,Wang2015} showed that $\epsilon_B$ has a wide range of $\sim 10^{-8}-10^{-3}$ ($10^{-6}-10^{-1}$) and is mainly centered at $\sim{\rm few}\times 10^{-5}$ ($\sim{\rm few}\times 10^{-3}$) by taking $n = 1\,{\rm cm}^{-3}$ ($n = 10^{-2}\,{\rm cm}^{-3}$). Therefore, as shown in the right panels of Figure \ref{fig:AfterGlow2},  the optical emission of kilonovae is often outshone by the afterglow and not detectable, while the emission in $J$-band and $K$-band can be easily detected. Only a small amount of BH-NS mergers kilonovae can have their optical emission observable along the jet axis.

\section{Conclusions and Discussion  \label{sec:6}}

By considering three radioactivity-powered components, i.e., lanthanide-rich tidal dynamical ejecta, intermediate opacity viscosity-driven wind ejecta, and lanthanide-free neutrino-driven wind ejecta, we modeled the dynamics and temperature profile evolution of the BH-NS merger kilonovae in great detail. Since numerical simulations show that these three components are highly anisotropic, we have paid our special attention to the viewing angle effect on BH-NS merger kilonovae. We presented a numerical method to model the evolution of the photosphere in the observer frame and studied the emergent spectra and lightcurves for different lines of sight in two degrees of freedom for the viewing angles, i.e., the latitudinal viewing angle $\theta_{\rm view}$, and the longitudinal viewing angle $\varphi_{\rm view}$. The ejecta models presented here are simplified with the assumption of homologous expansion, constant gray opacity, and simple radiative transfer. More realistic models should consider complex dynamical evolution, thermodynamical evolution, and temperature- and wavelength-dependent opacities. Nevertheless, this simplified model provides valuable information for us to comprehend the characteristics of the BH-NS merger kilonovae and the viewing angle effect on multi-band kilonova lightcurves. 

We find that the dynamical ejecta would contribute to the majority of the kilonova emission from the BH-NS mergers due to its largest projected photosphere area. Since NR simulations of BH-NS mergers revealed that the mass of the dynamical ejecta is always $\lesssim 0.1\,M_\odot$ \citep{Kyutoku2015}\footnote{{This value applies to the initial BH whose spin is $\chi_{\rm BH}\leq0.9$. Those BHs whose spin is close to extremal (see \citealt{Lovelace2013}.) can generate more massive dynamical ejecta}}, the peak luminosity of BH-NS merger kilonovae would be always less luminous than $\lesssim 4.5 \times 10^{41}\,{\rm erg\,s^{-1}}$. Corresponding to the AB absolute magnitudes commonly used by observers, its maximum absolute magnitude is $\sim-15\,{\rm mag}$ in optical and $\sim-16\,{\rm mag}$ in infrared. 

Long-lived energy injection from the remnant BH may produce an additional source of ejecta heating in excess of the contribution from $r$-process radioactivity. A fraction of gravitationally bound materiel would fall back onto the BH and enter the disk during a range of timescales from seconds to days \citep{Rosswog2007,Kyutoku2015}. The energy release from fallback accretion may enhance the peak brightness of the kilonova. For example, \cite{Ma2018} proposed an energy injection mechanism invoking a wind driven by the Blandford-Payne mechanism from an accretion disk \citep{Blandford1982}. According to our prediction, a  radioactivity-powered BH-NS merger kilonova would be always fainter than the critical luminosity, i.e., $\sim4.5\times 10^{41}\,{\rm erg\,s^{-1}}$, at the peak time. Future observations of a BH-NS merger kilonova brighter than this critical luminosity would suggest additional energy injection from the central engine.

We compare our theoretical results of BH-NS merger kilonovae with the observational properties of AT2017gfo. At each epoch after the merger, we find that the blackbody fit temperature of BH-NS mergers kilonovae is lower than that of AT2017gfo. Due to lack of abundant lanthanide-free ejecta like AT2017gfo, the BH-NS mergers kilonovae are optically dim, but possibly infrared bright. 

We showed that the observed luminosity of BH-NS merger kilonovae varies with the projected photosphere area determined by the viewing angles. The variation of the longitudinal viewing angle $\varphi_{\rm view}$ has little effect on the variation of the observed luminosity, while the variation of the latitudinal viewing angle $\theta_{\rm view}$ can significantly change the projected photosphere area, and hence, affect on the observed luminosity. In total, the difference of the observed luminosity caused by the variation of viewing angles is only a factor of $\sim (2 - 3)$, corresponding to the change of the multi-band magnitude by which are $\sim1\,{\rm mag}$. This is similar to the result of \cite{Roberts2011}, who adopted three dimensional radiation simulations to study the viewing angle effect on the emission of tidal tails. Furthermore, the blackbody fitting temperature and the shape of the observed multi-band lightcurves are not significantly dependent on line of sight. However, for the case that the dynamical ejecta blocks most of emission from the disk wind outflows along the line of sight, the decay index of multi-band lightcurves would become steeper at late times. {In addition, both $\varphi_{\rm view}$ and $\theta_{\rm view}$ can affect the peak time of the multi-band lightcurves. The variation of the peak time, caused by the light propagation effect, depends on the relative motion direction of the dynamical ejecta. More specifically, the peak time would increase if the dynamical ejecta moves away from the observer and decrease if it moves toward the observer.} Recently, \cite{Darbha2020} provided a simple analytic estimate of the viewing-angle-dependent lightcurves as a function of the projected surface area along the line of sight by considering three specific geometries, i.e., an ellipsoid, a ring torus, and a conical section embedded in a sphere. Their conclusions are qualitatively similar to ours. However, they derived that the viewing angle effect can cause a factor of $\sim 5-10$ difference in the BH-NS merger kilonova luminosity, which is larger than ours. This may be introduced by their simplified treatment of the geometry. In particular, they modeled the BH-NS merger tidal tail as an oblate ellipsoid with an axial ratio $R = 5$ or a torus with a radius ratio $K = 5$. However, according to simulation results \citep[e.g.,][]{Roberts2011,Kyutoku2015,Brege2018}, the dynamical ejecta is concentrated near the equator with the opening angle in the longitudinal direction filling an arc of about $\pi$, which is shaped like a crescent. Therefore, one may simply assume that the dynamical ejecta is like a moving ellipsoid with a relatively broad axial ratio $R\sim2 - 3$. This would reduce the the peak luminosity variation due to viewing-angle variation to be consistent with our results. 

The sGRB afterglows are very senstive to the viewing angle and can significantly affect the detectability of BH-NS merger kilonovae. For an on-axis observer, optical filter is not recommended to be used to observe the emission from the kilonova since it is completely outshone by the afterglow emission. The optical emission can only be observed if the sGRB occurs in a very low-density environment with a low $\epsilon_B$. Redder filters are more preferred to detect the kilonova emission. In $J$-band, the kilonova emission may be barely detected several days after the merger. The $K$-band is most ideal band since the kilonova emission becomes dominant after $\sim2\,{\rm days}$ and last for an extended period of time. 

For an off-axis geometry, the early-stage kilonova emission is always much luminous compared with the afterglow emission. The effect is more prominent for larger viewing angles. In relatively blue bands, the afterglow emission will outshine the kilonova emission at late times as the bright jet core becomes visible.

{\cite{Fujibayashi2020} recently found that the viscosity-driven ejecta can be lanthanide-free, whose electron fraction mainly lies in the range of $Y_e\sim0.3-0.4$, if the viscous coefficient is not extremely high. A substantial mass of the blue component ejecta may be formed after a BH-NS merger, which may be similar to or even more than the mass of the blue component invoked to explain AT2017gfo. As a result, a AT2017gfo-like kilonova may be possible to generate for a BH-NS merger as well. We may simply estimate the viewing-angle-dependent observed luminosity using the fitting results of AT2017gfo. For the face-on configuration, the peak observed luminosity may reach $\sim1\times10^{42}\,{\rm erg\ s^{-1}}$. The observed luminosity drops in view angles where the dynamical ejecta blocks the emission from the blue component ejecta, with the minimum at e.g., $\theta_{\rm view} \sim 90 ^\circ$ and $\varphi_{\rm view}\sim0^\circ-45^\circ$. Since the mass of the dynamical ejecta mainly lies in range of $(0.01-0.05)\, M_\odot$ \citep[e.g.,][]{Cowperthwaite2017,Murguia-Berthier2017,Perego2017,Kasen2017,Kasliwal2017,Tanaka2017,Villar2017}, the peak observed luminosity would be in the range of $\sim (1 - 2)\times 10^{41}\,{\rm erg\ s^{-1}}$ for the case when the blue component is blocked. Overall, if BH-NS mergers indeed have a blue component, the peak observed luminosity can vary by a factor of $\sim 5 - 10$ for different viewing angles.} 

We point out that our results of the viewing angle effect only apply to the mass ratio range of $Q \gtrsim 3$. The population-synthesis simulation results of the final mass distribution of BH-NS mergers \citep{Giacobbo2018,Mapelli2018} showed that the mass ratio of almost all merging systems is larger than $Q > 2.5$, and the most likely value is $Q = 5$. We can conclude that our results are always relevant for  BH-NS merger kilonovae. Moreover, the existence of low-mass BHs cannot be theoretically ruled out. We also discuss the properties of BH-NS merger kilonovae in the near-equal-mass regime. We extend the fitting formulae for the mass and velocity of the dynamical ejecta across a wider range of mass ratio from $Q = 1$ to $Q = 7$ validated with 66 simulations. For near-equal-mass regime, our results show that the dynamical ejecta is slow for ejecta produced during the tidal disruption of a neutron star (see also \citealt{Foucart2019}). The projected photosphere of the dynamical ejecta would decrease remarkably so that the viewing angle effects on the peak luminosity of the lightcurves would change. Besides, a less massive dynamical ejecta can be ejected after the merger. One can predict that the kilonovae of near-equal-mass BH-NS mergers are much dimmer than those of large mass ratios. More detailed studies on the parameter dependence and viewing angle dependence of the properties of BH-NS merger kilonovae, detection rate, and polarization are subject to further studies in the following articles of this series. Apparently, the characteristics of the BH-NS merger kilonovae are complex. As more BH-NS mergers are detected by LIGO/Virgo, one expects that their associated kilonovae will be eventually detected \citep{Bhattacharya2019}, c.f. \cite{Zappa2019,Tsujimoto2020}.
Future GW-lead multi-messenger observations of BH-NS mergers can help us to better understand the characteristics these kilonovae. 

\acknowledgments
We thank an anonymous referee for constructive suggestions, Koutarou Kyutoku, Masaomi Tanaka for valuable discussion and comments, Kyohei Kawaguchi for providing helpful data and  information. The work of J.P.Z is partially supported by the National Science Foundation of China under Grant No. 11721303 and the National Basic Research Program of China under grant No. 2014CB845800. L.D.L. is supported by the National Postdoctoral Program for Innovative Talents (grant No. BX20190044), China Postdoctoral Science Foundation (grant No. 2019M660515), and “LiYun” postdoctoral fellow of Beijing Normal University. Z.L is supported by the National Natural Science Foundation of China under Grant No. 11773003, U1931201. Y.W.Y is supported by the National Natural Science Foundation of China under Grant No. 11822302, 1183303. H.G. is supported by the National Natural Science Foundation of China under Grant No. 11722324, 11690024, 11603003, 11633001, the Strategic Priority Research Program of the Chinese Academy of Sciences, Grant No. XDB23040100 and the Fundamental Research Funds for the Central Universities.
\software{Matlab,  \url{https://www.mathworks.com}}

\appendix
\section{Definitions of frequently used variables\label{app:A0}}

\begin{deluxetable*}{cll}[h!]
\tablecaption{Definitions of frequently used variables}
\tablecolumns{3}
\tablenum{3}
\tablewidth{0pt}
\tablehead{
\colhead{Case} &
\colhead{Variable} &
\colhead{Definition} 
}
\startdata
\multirow{6}{*}{BH-NS merger} & $M_{\rm BH}$ & Mass of the BH \\
{} & $M_{\rm NS}$ & Mass of the NS \\
{} & $M_{\rm NS}^{\rm b}$ & Baryon mass of the NS \\
{} & $\chi_{\rm BH}$ & Dimensionless spin of the BH \\
{} & $C_{\rm NS}$ & Compactness of the NS \\
{} & $Q$ & Mass ratio between BH mass and NS mass \\
\hline
\multirow{17}{*}{Ejecta} & $Y_e$ & Electron fraction \\
{} & $M_{\rm disk}$ & Mass of the remnant disk \\
{} & $M_{\rm d}$ & Mass of the tidal dynamical ejecta \\
{} & $\kappa_{\rm d}$ & Opacity of the dynamical ejecta; the value is $\kappa_{\rm d} = 20\, {\rm cm^2\,g^{-1}}$ \\
{} & $\theta_{\rm d}$ & Half opening angle in the latitudinal direction of the dynamical ejecta; the value we set is $\theta_{\rm d} \approx 15 ^ \circ$ \\
{} & $\varphi_{\rm d}$ & Opening angle in the longitudinal direction of the dynamical ejecta; the value is $\varphi_{\rm d} \approx \pi$ \\
{} & $v_{\rm min,d}$ & Minimum velocity of the dynamical ejecta; the value we set is $v_{\rm min,d} = 0.1\, c$ \\
{} & $v_{\rm rms,d}$ & rms velocity of the dynamical ejecta \\
{} & $M_{\rm n}$ & Mass of the neutrino-driven wind ejecta \\
{} & $\kappa_{\rm n}$ & Opacity of the neutrino-driven ejecta; the value we set is $\kappa_{\rm n} = 1\, {\rm cm^2\,g^{-1}}$ \\
{} & $\theta_{\rm n}$ & Opening angle in the latitudinal direction of the neutrino-driven ejecta; the value is $\theta_{\rm n} \approx 30 ^ \circ$ \\
{} & $v_{\rm max,n}$ & Maximum velocity of the neutrino-driven ejecta; the value we set is $v_{\rm max,n} = 0.2\, c$ \\
{} & $v_{\rm rms,n}$ & rms velocity of the neutrino-driven ejecta \\
{} & $M_{\rm v}$ & Mass of the viscosity-driven wind ejecta \\ 
{} & $\kappa_{\rm v}$ & Opacity of the viscosity-driven ejecta; the value we set is $\kappa_{\rm v} = 5\,{\rm cm^2\,g^{-1}}$ \\
{} & $v_{\rm max,v}$ & Maximum velocity of the viscosity-driven ejecta; the value we set is $v_{\rm max,v} = 0.09\, c$ \\
{} & $v_{\rm rms,v}$ & rms velocity of the viscosity-driven ejecta \\
\hline
\multirow{2}{*}{Viewing angle} & $\theta_{\rm view}$ & Latitudinal viewing angle \\
{} & $\varphi_{\rm view}$ & Longitudinal viewing angle \\
\hline
\multirow{4}{*}{Numerical method} & $v'^{ij}_{\rm mesh,x}$ & $v'_x$-component of mesh grid points\\
{} & $v'^{ij}_{\rm mesh,y}$ & $v'_y$-component of mesh grid points \\
{} & $p^{ij}_{\rm phot}$ & {Velocity space distance between photosphere and the mesh grid plane} \\
{} & $T^{ij}_{\rm mesh}$ & Thermal temperature of mesh grid points \\
\hline
\multirow{5}{*}{Observational parameter} & $D_{\rm L}$ & Luminosity distance; the value we set is $D_{\rm L} = 10\, {\rm pc}$\\
{} & $F_\lambda$ & Flux density at photon wavelength $\lambda$ \\
{} & $F_\nu$ & Flux density at photon frequency $\nu$ \\
{} & $M_\nu$ & AB absolute magnitude \\
{} & $t_{\rm obs}$ & Observational time \\
\hline
\multirow{6}{*}{Afterglow} & {$\nu'_{\rm m}$} & Synchrotron frequency of the accelerated electrons with the minimum Lorentz factor \\
{} & $\nu'_{\rm c}$ & Cooling frequency \\
{} & $\nu'_{\rm a}$ & Synchrotron self-absorption frequency \\
{} & $n$ & Interstellar medium number density \\
{} & $\epsilon_B$ & Fraction of shock energy carried by magnetic field \\
{} & $\epsilon_e$ & Fraction of shock energy carried by electrons
\enddata
\end{deluxetable*}

\section{Photosphere Evolution in the observer Frame and Lightcurve Reconstruction\label{app:A}}

\begin{figure}[htbp]
    \centering
    \includegraphics[width=0.70\textwidth, trim=50 140 120 60, clip]{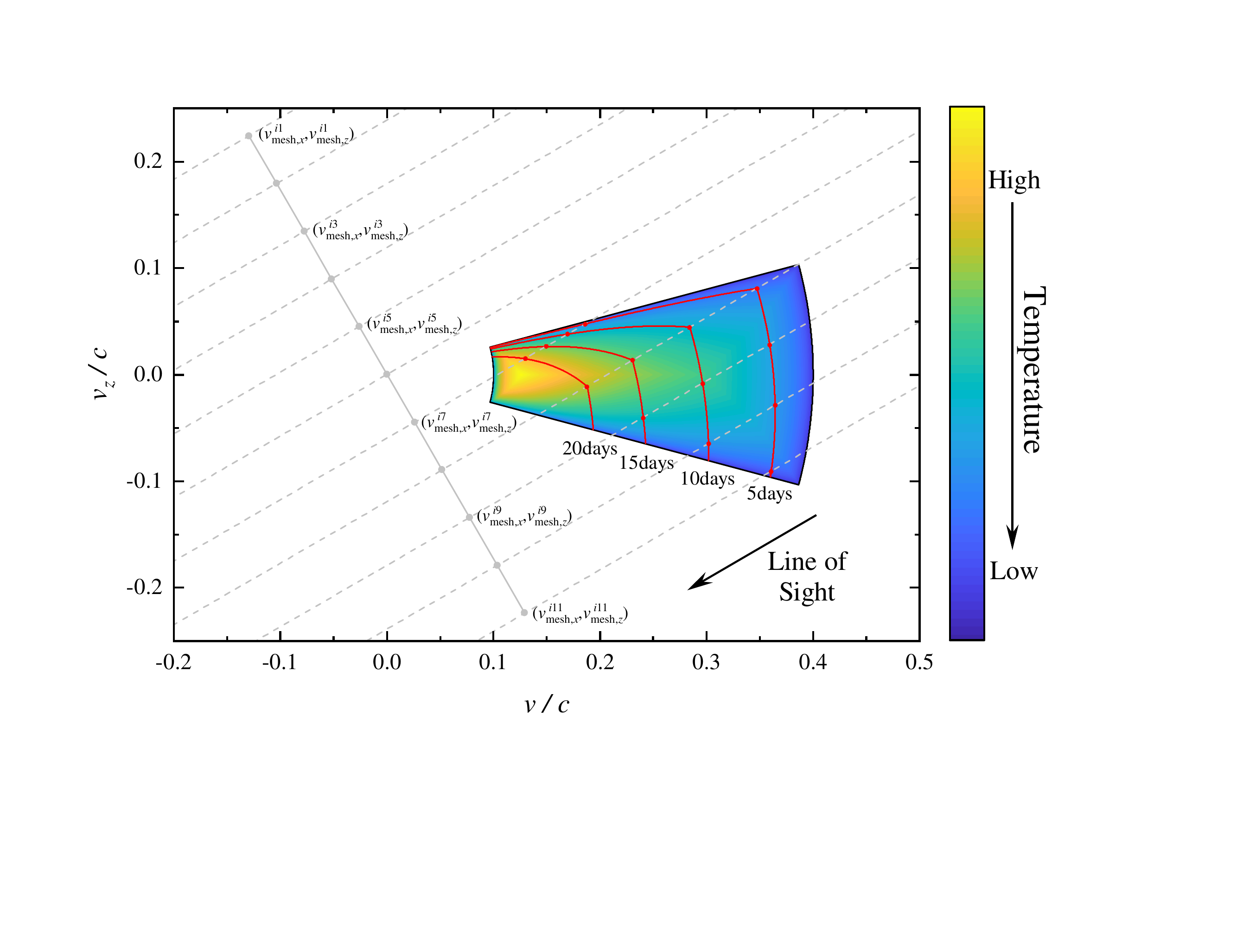}
    \caption{A schematic diagram of the evolution of photosphere in the observer frame and lightcurve reconstruction. The dynamical ejecta is taken as an example. The figure shows a sectional drawing of the $v'_xO'v'_z$ plane. The mesh grid (gray solid points) is perpendicular to the line of sight (denoted as arrow). Along the viewing direction and different mesh grid points, one can find the position of the observed photosphere (red circles) at a given time. Then, the photosphere in the observer frame (red lines) can be calculated by interpolating all points. In the case of known photosphere temperature, the emergent spectra along the viewing angle direction can be obtained after integrating over the photosphere. Here, the parameters of the dynamical ejecta are $\kappa_{\rm d} = 20\, {\rm cm}^2\,{\rm g}^{-1}$, $M_{\rm d} = 0.01\,M_{\odot}$, $\theta_{\rm d} = \pi / 12$, $\varphi_{\rm d} = \pi$, $v_{\min,{\rm d}} = 0.1\, c$ and $v_{\max,{\rm d}} = 0.4\, c$.}
    \label{fig: schematic diagram}
\end{figure}

In Section \ref{sec:2}, we have established the dynamics and temperature evolution of different components in the velocity space. In order to reconstruct viewing-angle-dependent lightcurves, one needs to solve the observed photosphere along the line of sight of an observer and integrate the projected photosphere. \cite{Rosswog2014,Grossman2014,Martin2015,Perego2017,Barbieri2019,Barbieri2020} used a semi-analytical method that divided ejecta into finite slices and summed up individual contributions of each slice to compute lightcurves with a one-dimensional viewing angle \cite[see e.g.,][used radiative transfer simulations to calculate viewing-angle-dependent lightcurve]{Darbha2020,Kawaguchi2018,Kawaguchi2020b,Korobkin2020}. This section is devoted to a numerical method of calculating the photosphere evolution and lightcurve as seen by observers with different viewing angles. We define the two-dimensional viewing angles $(\theta_{\rm view} , \varphi_{\rm view})$ within the ranges of $\theta_{\rm view} \in [0 , \pi / 2]$ and $\varphi_{\rm view} \in [0 , \pi]$, which we have shown in Figure \ref{fig:CoordinateSystem}. 

The brief steps of establishing the evolution of the photosphere in the observer frame include the following:
\begin{enumerate}
\item Comparing with the local coordinate system of all the ejecta components whose origin is $O$, we set up a three-dimensional Cartesian coordinate system with origin $O'$ and the axes $v'_x$, $v'_y$ and $v'_z$. We get a
mesh grid on the plane of $v'_xO'v'_y$ and set ${v'_z}{O}'$ as the line of sight. The points of the mesh grid are marked as $(v'^{ij}_{{\rm mesh},x} , v'^{ij}_{{\rm mesh},y} , 0)$ where the superscripts ($i$ and $j$) represent the IDs of the points at the mesh grid.
\item Rotate the density profile of all the ejecta components from the coordinate system $O$ to the coordinate system $O'$ simultaneously. In order to rotate velocity points located at the profile of the ejecta, we follow \cite{Goldstein2002} and define three Euler rotation matrices which are rotations about the $v_x-$, $v_y-$, and $v_z-$axes using the right-hand rule: 
\begin{equation}
\begin{split}
\boldsymbol{R}_{x}(\theta) =  \begin{pmatrix}
1 & 0 & 0\\ 
0 & \cos\theta & -\sin\theta\\ 
0 & \sin\theta & \cos\theta
\end{pmatrix},\\
\boldsymbol{R}_{y}(\theta) =  \begin{pmatrix}
\cos\theta & 0 & -\sin\theta\\ 
0 & 1  & 0\\ 
\sin\theta & 0 & \cos\theta
\end{pmatrix},\\
\boldsymbol{R}_{z}(\theta) =  \begin{pmatrix}
\cos\theta  & \sin\theta & 0\\ 
-\sin\theta & \cos\theta  & 0\\ 
0 & 0 & 1
\end{pmatrix}.
\end{split}
\end{equation}
Take one point $\boldsymbol{v} = [v_x , v_y , v_y]^{\rm T}$ located at the profile of the ejecta for example, the point's location at the coordinate system $O'$ after rotating is
\begin{equation}
    \boldsymbol{v}' = \boldsymbol{R}_y(\theta_{\rm view})\boldsymbol{R}_z(\varphi_{\rm view})\boldsymbol{v}.
\end{equation}

\item \label{Step:3} At one point of mesh grid $(v'^{ij}_{{\rm mesh},x} , v'^{ij}_{{\rm mesh},y} , 0)$, one can find the intersection points of the ejecta profile along the line of sight. We set $v_z$ components of each intersection point as $p^{ij}_1, p^{ij}_2, p^{ij}_3, \cdots$ from large to small, which are the velocity space distances between the intersection points and the mesh grid plane.

\item Transform the mesh grid into the coordinate system $O$. The rotation rule is
\begin{equation}
    \boldsymbol{v} = \boldsymbol{R}_z(-\varphi_{\rm view})\boldsymbol{R}_y(-\theta_{\rm view})\boldsymbol{v}'.
\end{equation}
We mark the point of the mesh grid as $(v^{ij}_{{\rm mesh},x} , v^{ij}_{{\rm mesh},y} , v^{ij}_{{\rm mesh},z})$.

\item In order to solve the photosphere in the observer frame, one needs to do the line integral to find the positions where the optical depth is $\tau_{\rm view} = 2 / 3$ along the line of sight through each mesh grid point at the coordinate system $O$. The line parameter functions through one mesh grid point $(v^{ij}_{{\rm mesh},x} , v^{ij}_{{\rm mesh},y} , v^{ij}_{{\rm mesh},z})$ are
\begin{equation}
\begin{split}
\label{Eq. Parameter function}
v_x - v_{{\rm mesh},x}^{ij} &= p\sin\theta_{\rm view}\cos\varphi_{\rm view},\\
v_y - v_{{\rm mesh},y}^{ij} &= p\sin\theta_{\rm view}\sin\varphi_{\rm view},\\
v_z - v_{{\rm mesh},z}^{ij} &= p\cos\theta_{\rm view},
\end{split}
\end{equation}
where $p$ is the parameter of the line functions. As for a given point $(v_x , v_y , v_z)$ which is located at the line, $p$ will be a velocity space distance between the given point and the mesh grid plane. The optical depth's line integral of the dynamical ejecta, neutrino-driven ejecta and viscosity-driven ejecta are respectively as follow:
\begin{equation}
\begin{split}
    \tau_{\rm view,d}^{ij} &= \int_{\rm L}\kappa_{\rm d}\rho_{\rm d}{\rm d}l \\
    &= \frac{\kappa_{\rm d}M_{\rm d}t^{-2}}{2\varphi_{\rm d}\sin\theta_{\rm d}(v_{\max , {\rm d}} - v_{\min , {\rm d}})}\int_p\frac{1}{v^2}{\rm d}p, \\
    \tau_{\rm view,n}^{ij} &= \int_{\rm L}\kappa_{\rm n}\rho_{\rm n}{\rm d}l \\
    &= \frac{35\kappa_{\rm n}M_{\rm n}t^{-2}}{64\pi(1 - \cos\theta_{\rm n})v_{\max , \rm n}}
    \int_p\frac{1}{v ^ 2}\left[ 1 - \left( \frac{v}{v_{\max , \rm n}}\right) ^ 2 \right] ^ 3{\rm d}p, \\
    \tau_{\rm view,v}^{ij} &= \int_{\rm L}\kappa_{\rm v}\rho_{\rm v}{\rm d}l \\
    &= \frac{105\kappa_{\rm v}M_{\rm v}t^{-2}}{128\pi v_{\max , \rm v}}\int_p\frac{v^2 - v_z^2}{v^4}\left[ 1 - \left( \frac{v}{v_{\max , \rm v}}\right) ^ 2 \right] ^ 3{\rm d}p,
\end{split}
\end{equation}
where $v_z = v_{{\rm mesh},z}^{ij} + p\cos\theta_{\rm view}$, $v ^ 2 = p ^ 2 + 2p(v_{{\rm mesh},x}^{ij}\sin\theta_{\rm view}\cos\varphi_{\rm view} + v_{{\rm mesh},y}^{ij}\sin\theta_{\rm view}\sin\varphi_{\rm view} + v_{{\rm mesh},z}^{ij}\cos\theta_{\rm view}) + {v_{\rm mesh}^{ij}}^2$ and ${v_{\rm mesh}^{ij}} ^ 2= {v_{{\rm mesh},x}^{ij}} ^ 2 + {v_{{\rm mesh},y}^{ij}} ^ 2 + {v_{{\rm mesh},z}^{ij}} ^ 2$. 

We have obtained the intersection points of the ejecta profile and their $v_z$-components $p^{ij}_1, p^{ij}_2, p^{ij}_3, \cdots$ in step \ref{Step:3}. At a given time, one can do the line integration along the light of sight. As for the first ejecta component that the observer will see, we calculate the entire optical depth $\tau_{\rm view}^{ij}$ by assuming the range of integration is from $p^{ij}_2$ to $p^{ij}_1$. If $\tau_{\rm view}^{ij} > 2 / 3$, the photosphere cannot penetrate this ejecta component. One can solve the photosphere's position of $\tau_{\rm view}^{ij} = 2 / 3$ and calculate its velocity space distance $p^{ij}_{\rm phot}$ between $p^{ij}_1$ and $p^{ij}_2$. On the other hand, if $\tau_{\rm view}^{ij} < 2 / 3$, the photosphere can penetrate this ejecta component and enter the next ejecta component that observer will see. $p_{\rm phot}^{ij}$ will be located at between $p_2^{ij}$ and $p_3^{ij}$. Repeat above steps until $p_{\rm phot}^{ij}$ is solved.

\item 
{However, due to the light propagation effect, photons emitted at a given time $t$ reach the observer at different arrival times $t_{\rm obs}$. We set the arrival time $t_{\rm obs} = t$ when a photon is emitted at the mesh grid plane. Therefore, the observer would see a photon emitted from the photosphere at} 
\begin{equation}
t_{\rm obs} = t\left(1 - \frac{p^{ij}_{\rm phot}}{c}\right).
\end{equation}

\item At a given observational time $t_{\rm obs}$, the photosphere location assumed as $(v_{{\rm phot}, x}^{ij} , v_{{\rm phot}, y}^{ij} , v_{{\rm phot}, z}^{ij})$ can be solved by Equation (\ref{Eq. Parameter function}) if $p_{\rm phot}^{ij}$ have been solved in the coordinate system $O'$. We have obtained the thermal temperature $T^{ij}_{\rm mesh}$ in Section \ref{sec:2} when the observational time $t_{\rm obs}$ and the observed photosphere location are known. Throughout all the points of the mesh grid, one can obtain the entire photosphere in the observer frame and the thermal temperature at the photosphere. A schematic diagram is presented in Figure \ref{fig: schematic diagram}. In order to calculate the observed flux, it is convenient to directly do the interpolation and integration by projecting the parameter information of the entire photosphere to the mesh grid if the source angular size can be ignored. The observed flux for a give frequency $\nu$ is
\begin{equation}
    F_\nu (\nu,t_{\rm obs}) \approx \frac{2}{h^2c^2D_{\rm L}^2}\iint_S\frac{\mathcal{D}(h\nu/\mathcal{D})^3}{\exp{(h\nu/\mathcal{D}k_{\rm B}T^{ij}_{\rm mesh}}) - 1}{\rm d}\sigma'^{ij},
\end{equation}
where $D_{\rm L}$ is the luminosity distance, $h$ is the Planck constant and $k_{\rm B}$ is the Boltzmann constant, and {${\rm d}\sigma'^{ij}$ is the infinitesimal projected photosphere area, i.e.,}
\begin{equation}
    {\rm d}\sigma'^{ij} = \left(\frac{t_{\rm obs}}{1 - p_{\rm phot} ^ {ij} / c} \right)^2{\rm d}v'^{ij}_{{\rm mesh},x}{\rm d}v'^{ij}_{{\rm mesh},y}.
\end{equation}
The Doppler factor is $\mathcal{D} = 1 / [\Gamma(1 - \beta\cos\Delta\theta)]$, where $\Gamma = 1 / \sqrt{1 - \beta ^ 2}$, $\beta = v / c$ and $\Delta\theta$ is the angle between the moving direction of the point located at the photosphere and the line of sight. In the coordinate system $O'$, the moving direction of the point located at the photosphere is $(v'^{ij}_{{\rm mesh},x} , v'^{ij}_{{\rm mesh},y} , p_{\rm phot}^{ij})$ while the line of sight is towards to the $v'_z-$direction. Follow the included angle formula, one can calculate $\cos\Delta\theta = p_{\rm phot}^{ij} / \sqrt{{v'^{ij}_{{\rm mesh},x}} ^ 2 + {v'^{ij}_{{\rm mesh},y}} ^ 2 + {p_{\rm phot}^{ij}} ^ 2}$. 

The conversion between frequency and wavelength is obtained from $F_\nu{\rm d}\nu = F_\lambda{\rm d}\lambda$. Therefore, the observed flux for a given wavelength $\lambda$ is
\begin{equation}
    F_\lambda(\lambda,t_{\rm obs}) \approx \frac{2}{h^4c^3D_{\rm L}^2}\iint_S \frac{\mathcal{D}^3(hc/\mathcal{D}\lambda)^5}{\exp{(hc/\mathcal{D}k_{\rm B}\lambda T^{ij}_{\rm mesh}}) - 1}{\rm d}\sigma'^{ij}.
\end{equation}

\end{enumerate}

\section{Gamma-Ray Burst Afterglow Model \label{app:B}}

In this section, we briefly describe the afterglow model we used to calculate the sGRB lightcurves along the line of sight. 

After producing the sGRB, the relativistic jet sweeps into the ISM, driving a forward shock. The swept mass per unit solid angle at a given radius $R$ is 
\begin{equation}
    \mu(R) = \frac{R^3}{3}nm_p,
\end{equation}
where $n$ is the ISM number density and $m_p$ is the proton mass. By assuming energy conservation, the Lorentz factor of the shocked material \citep{Panaitescu2000,Granot2003} is
\begin{equation}
    \Gamma(R , \theta) = \frac{\mu_0}{2\mu}\left[ \sqrt{1 + \frac{4\mu(c^{-2}{\rm d}E/{\rm d}\Omega + \mu + \mu_0)}{\mu_0^2}} - 1 \right],
\end{equation}
where $\mu_0(\theta) = ({\rm d}E/{\rm d}\Omega)/[\Gamma(0 , \theta) - 1]c^2$, ${\rm d}E/{\rm d}\Omega$ is the kinetic energy angular distribution of the jet, and $c$ is the speed of light.

For a relativistic shock propagating into a cold ISM, the physical condition of the shocked plasma is obtained from the shock-jump conditions \citep{Blandford1976}, i.e., the conservations of baryon number, energy and momentum fluxes. Based on these conservation conditions, the electron number density $n_{\rm s}$ of the shocked material is
\begin{equation}
    n_{\rm s} = \frac{\gamma_{\rm ad}\Gamma + 1}{\gamma_{\rm ad} - 1}n,
\end{equation}
while its Lorentz factor is
\begin{equation}
    \Gamma_{\rm s} = [\gamma_{\rm ad}(\Gamma - 1) + 1] \sqrt{\frac{\Gamma + 1}{\gamma_{\rm ad} (2 - \gamma_{\rm ad})(\Gamma - 1) + 2}},
\end{equation}
where $\gamma_{\rm ad}$ is the post-shock adiabatic index. We adopt $\gamma_{\rm ad}$ that is a function of $\Gamma$ by \cite{Pe'er2012}: $\gamma_{\rm ad} = (5 - 1.21937 z + 0.18203 z^2 - 0.96583 z^3 + 2.32513 z^ 4 - 2.39332 z^ 5 + 1.07136z^6)/3$, where $z = \Theta / (0.24 + \Theta)$ and 
\begin{equation}
    \Theta = \left( \frac{\Gamma \beta}{3} \right) \left( \frac{\Gamma\beta + 1.07(\Gamma\beta) ^ 2}{1 + \Gamma\beta + 1.07 (\Gamma)\beta)^2} \right).
\end{equation}
By assuming that the shock material is concentrated in a thin layer behind the shock and has a uniform radial density distribution, the thickness of the shocked layer is given by \cite{Salafia2019}
\begin{equation}
    \Delta R = \frac{R(\gamma_{\rm ad} - 1)}{3(\gamma_{\rm ad} + 1) \Gamma}.
\end{equation}

The shock surface brightness is
\begin{equation}
    I_\nu (\nu , R , \theta , \phi) = \mathcal{D}^3\Delta R'j'_{\nu'}(\nu/\mathcal{D}),
\end{equation}
where the Doppler factor is $\mathcal{D}(R , \theta , \varphi , \theta_{\rm view}) = \Gamma(R , \theta) ^ {-1}[1 - \beta(R , \theta)\cos\alpha]^{-1}$, $\beta = (1 - \Gamma^{-2})^{1 / 2}$ and $\Delta R' = \Gamma(R , \theta)\Delta R$, and $j'_{\nu'}$ is the comoving emissivity due to synchrotron emission. The synchrotron spectrum for a distribution of electrons depends on the ordering of characteristic break frequencies, i.e., the typical synchrotron frequency of the accelerated electrons with the minimum Lorentz factor $\nu'_{\rm m}$, the cooling frequency $\nu'_{\rm c}$, and the synchrotron self-absorption frequency $\nu'_{\rm a}$. The spectrum for $\nu'_{\rm m}<\nu'_{\rm c}$ is classified as the ``slow cooling" case, i.e. \citep{Sari1998}
\begin{equation}
    j'_{\nu'} = j'_{\nu',{\rm max}}\left\{
    \begin{array}{ll}
    \left(\frac{\nu'_{\rm a}}{\nu'_{\rm m}} \right)^{1/3}\left(\frac{\nu'}{\nu'_{\rm a}} \right )^{2}, & \nu'<\nu'_{\rm a}\\ \left(\frac{\nu'}{\nu'_{\rm m}} \right )^{1/3}, & \nu'_{\rm a}<\nu'<\nu'_{\rm m} \\ \left(\frac{\nu'}{\nu'_{\rm m}} \right )^{-(p - 1) / 2}, & \nu'_{\rm m}<\nu'<\nu'_{\rm c}\\ \left(\frac{\nu'_{\rm c}}{\nu'_{\rm m}} \right )^{-(p - 1) / 2}\left(\frac{\nu'}{\nu'_{\rm c}} \right )^{-p / 2}, & \nu'>\nu'_{\rm c}
    \end{array}\right.
\end{equation}
and that for $\nu'_{\rm c}<\nu'_{\rm m}$ is the ``fast cooling" case, i.e.
\begin{equation}
    j'_{\nu'} = j'_{\nu',{\rm max}}\left\{
    \begin{array}{ll}
    \left(\frac{\nu'_{\rm a}}{\nu'_{\rm c}} \right)^{1/3}\left(\frac{\nu'}{\nu'_{\rm a}} \right )^{2}, & \nu'<\nu'_{\rm a}\\ \left(\frac{\nu'}{\nu'_{\rm c}} \right )^{1/3}, & \nu'_{\rm a}<\nu'<\nu'_{\rm c} \\ \left(\frac{\nu'}{\nu'_{\rm c}} \right )^{-1 / 2}, & \nu'_{\rm c}<\nu'<\nu'_{\rm m}\\ \left(\frac{\nu'_{\rm m}}{\nu'_{\rm c}} \right )^{- 1 / 2}\left(\frac{\nu'}{\nu'_{\rm m}} \right )^{- p / 2}, & \nu'>\nu'_{\rm m}.
    \end{array}\right.
\end{equation}
In our calculations, we set $p = 2.2$. The comoving synchrotron emissivity of electrons behind the shock at the peak of spectrum is
 \begin{equation}
    j'_{\nu',{\rm max}} \approx 0.66\frac{q_e^3}{m_e^2c^4}\frac{p - 2}{3p - 1}\frac{B'\epsilon_e e}{\gamma_{\rm m}},
\end{equation}
where $e = (\Gamma - 1)n_{\rm s}m_pc^2$, $m_e$ is the mass of electron, $q_e$ is the electron charge, and $\epsilon_e$ is the fraction of internal energy that is given to electrons. The comoving magnetic field strength $B'$ is
\begin{equation}
    B' = (32\pi m_p\epsilon_B n)^{1 / 2}\Gamma c,
\end{equation}
where $\epsilon_B$ is the internal energy that goes to magnetic fields.

More specifically for each characteristic frequency, $\nu'_{\rm m}$ is the synchrotron frequency which is $\nu'_{\rm m} = \gamma_{\rm m} ^ 2 q_e B' / 2\pi m_ec$, where $m_e$ is the electron mass,
\begin{equation}
    \gamma_{\rm m} = \max\left[1 , \frac{p - 2}{p - 1}(\Gamma - 1)\frac{m_p}{m_e} \right],
\end{equation}
$\nu'_{\rm c}$ is the synchrotron frequency corresponding to the Lorentz factor $\gamma_{\rm c}$ which is 
\begin{equation}
    \gamma_{\rm c} = \frac{6\pi m_ec^2\Gamma\beta}{\sigma_{\rm T}B'^2R},
\end{equation}
where $\sigma_{\rm T}$ is the Thomson cross section. The frequency $\nu'_{\rm a}$ is synchrotron self-absorption, which we calculate based on \cite{Shen2009}:
\begin{equation}
    2k_{\rm B}T'\frac{\nu'^2_{\rm a}}{c^2} = C(p)I'_{\nu'_{\rm a}},
\end{equation}
where $k_{\rm B}$ is Boltzmann constant and $k_{\rm B}T' = \max(\gamma_{\rm a} , \min(\gamma_{\rm m} , \gamma_{\rm c}))m_ec^2$. The correction factor is
\begin{equation}
    C(p) = 
    \left\{ \begin{array}{ll}
    C_1(p) = \frac{(p + 2)(p - 1 /3)}{p+2 / 3}, &{\rm if}\ \nu'_{\rm a} < \min(
    \nu'_{\rm m} , \nu'_{\rm c}) \\
    C_2(p) = \sqrt{2}(p + 1)\frac{\Gamma\left( \frac{3p + 22}{12} \right)\Gamma\left(\frac{3p + 2}{12} \right)}{\Gamma\left(\frac{3p+19}{12}\right)\Gamma\left(\frac{3p - 1}{12} \right)}, &{\rm if}\ \min(\nu'_{\rm m} , \nu'_{\rm c}) < \nu'_{\rm a} < \max(\nu'_{\rm m} , \nu'_{\rm c}),
    \end{array}\right.
\end{equation}
where $\Gamma$ denotes the gamma function in this function. 

Due to the aberration effect, photons emitted at ($R$, $\theta$, $\varphi$) arrive at the observer with a time delay with respect to a photon emitted at $R = 0$ by
\begin{equation}
    t_{\rm obs} = \int^{R}_0\frac{{\rm d}R(1 - \beta_{\rm s}\cos\alpha)}{\beta_{\rm s}c},
\end{equation}
where $\beta_{\rm s} = (1 - \Gamma_{\rm s}^{-2})^{1 / 2}$ and $\cos\alpha = \cos\theta\cos\theta_{\rm view} + \sin\theta\sin\varphi\sin\theta_{\rm view}$. By integrating the equal-arrival time surfaces, the flux density is given by
\begin{equation}
    F_\nu(\nu, t_{\rm obs}) = \frac{2}{D_{\rm L}^2}\int^1_0{\rm d}\cos\theta\int^{\pi / 2}_{-\pi / 2}{\rm d}\varphi R^2I_\nu(\nu , R).
\end{equation}
In order to calculate the flux density from the counter-jet, one can setting $\theta_{\rm view}$ to $\theta_{\rm view} + \pi$.

\end{document}